\documentclass[a4paper,fleqn,usenatbib]{mnras}

\usepackage[T1]{fontenc}
\usepackage{ae,aecompl}
\usepackage[normalem]{ulem}

\usepackage{graphicx}
\usepackage{graphicx}	
\usepackage{amsmath}	
\usepackage{amssymb}	
\usepackage{bm}

\usepackage{newtxtext,newtxmath}

\title[PION: simulations of wind-blown nebulae]{\textsc{pion}: Simulating bow shocks and circumstellar nebulae}
\author[J. Mackey et al.]{Jonathan~Mackey,$^{1,2}$\thanks{E-mail: jmackey@cp.dias.ie (JM)}
Samuel Green,$^{1,2}$
Maria Moutzouri,$^{1,2}$
Thomas J.\ Haworth$^3$,
\newauthor
Robert D. Kavanagh$^{4}$,
Davit Zargaryan,$^{1,2,5}$
Maggie Celeste $^{1,4}$
\\
$^{1}$Dublin Institute for Advanced Studies, Astronomy \& Astrophysics Section, 31 Fitzwilliam Place, Dublin 2, Ireland\\
$^{2}$Dublin Institute for Advanced Studies, Centre for AstroParticle Physics and Astrophysics (CAPPA), DIAS Dunsink Observatory, Dunsink Lane, Dublin 15, Ireland\\
$^{3}$Astronomy Unit, School of Physics and Astronomy, Queen Mary University of London, London E1 4NS, UK\\
$^{4}$School of Physics, Trinity College Dublin, The University of Dublin, Dublin 2, Ireland\\
$^{5}$High Energy Astrophysics Laboratory, RAU, 123 Hovsep Emin St, Yerevan 0051, Armenia \\
}

\date{Submitted 09 January 2021 / Accepted 08 March 2021}
\pubyear{2021}

\begin{document}
\label{firstpage}
\pagerange{\pageref{firstpage}--\pageref{lastpage}}
\maketitle

\begin{abstract}
  Expanding nebulae are produced by mass loss from stars, especially during late stages of evolution.
  Multi-dimensional simulation of these nebulae requires high resolution near the star and permits resolution that decreases with distance from the star, ideally with adaptive timesteps.
  We report the implementation and testing of static mesh-refinement in the radiation-magnetohydrodynamics code \textsc{pion}, and document its performance for 2D and 3D calculations.
  The bow shock produced by a hot, magnetized, slowly rotating star as it moves through the magnetized ISM is simulated in 3D, highlighting differences compared with 2D calculations.
  Latitude-dependent, time-varying magnetized winds are modelled and compared with simulations of ring nebulae around blue supergiants from the literature.
  A 3D simulation of the expansion of a fast wind from a Wolf-Rayet star into the slow wind from a previous red supergiant phase of evolution is presented, with results compared with results in the literature and analytic theory.
  Finally the wind-wind collision from a binary star system is modelled with 3D MHD, and the results compared with previous 2D hydrodynamic calculations.
  A python library is provided for reading and plotting simulation snapshots, and the generation of synthetic infrared emission maps using \textsc{torus} is also demonstrated.
  It is shown that state-of-the-art 3D MHD simulations of wind-driven nebulae can be performed using \textsc{pion} with reasonable computational resources.
  The source code and user documentation is made available for the community under a BSD3 licence.
\end{abstract}


\section{Introduction}
\label{sec:intro}

Massive stars emit copious extreme-ultraviolet (EUV) photons capable of ionizing hydrogen when on the hydrogen-burning main sequence and also have line-driven stellar winds with terminal velocities $v_\infty\gtrsim1000\,\mathrm{km}\,\mathrm{s}^{-1}$ \citep{SnoMor76}, with important consequences for their surroundings \citep{Dal15}.
After the main sequence phase the outer layers of a massive star expand rapidly, and the star evolves to the upper-right part of the Hertzsprung-Russell Diagram (HRD), becoming a cool and luminous supergiant.
Such stars have extended, loosely bound envelopes, and their further evolution is determined by mass loss through winds, eruptions, or interaction with a binary companion, and by rotational mixing of nuclear-processed material from the core to the envelope \citep{Lan12, Smi14}.

After the main sequence, the dynamical time-scale of circumstellar nebulae ($\sim10^4-10^5$ yr) becomes comparable to the nuclear ($\sim10^5$ yr) and thermal ($\sim10-10^4$ yr) time-scales of a massive star.
Mass-loss rates  ($\dot{M}$) and wind velocities ($v_\infty$) can change drastically on these time-scales, meaning that the evolution of circumstellar nebulae cannot be considered in isolation from the evolution of the central star(s).
These late phases of evolution of massive stars are very uncertain because some key physical processes are poorly constrained and poorly modelled, namely convection, mass loss, rotation, and interaction with a companion \citep[for a review, see][]{Smi14}.
Significant progress is being made in understanding the radii \citep{GraLanGri18} and wind structure \citep{SanVin20} of classical Wolf-Rayet (WR) stars, winds from stars close to the Eddington limit \citep{Bes20} and potentially understanding the S-Doradus cycle of Luminous Blue Variables (LBV) \citep{GraLanMac21}.
On the other hand, we do not yet have a predictive theory of mass loss from red supergiants (RSG), for which the empirical scaling of $\dot{M}$ with stellar luminosity, mass and temperature is a subject of active debate and research \citep{BeaDavSmi20,HumHelJon20}.
Nor is there any consensus on the causes or trigger for eruptive mass-loss events such as LBV giant eruptions, but there are indications some of them could be driven by binary interaction \citep{SmiAndRes18}.
While we do not have a predictive theory for mass-loss rates across the HRD, stellar evolution calculations use mass-loss prescriptions that do cover the HRD, and so the wind-driven nebula produced around a massive star is a prediction of stellar evolution calculations.
Comparing these predictions with observations is a test of mass-loss prescriptions.

Circumstellar nebulae are complex structures, typically subject to non-linear dynamical instabilities \citep{GarLanMac96}, and must be studied with multi-dimensional radiation-magnetohydrodynamics (R-MHD), or simplifications thereof (e.g. hydrodynamics (HD) or ideal magnetohydrodynamics (MHD) with radiative heating and cooling).
This means that, while significant work has been done on modelling circumstellar nebulae (see below), its potential to test stellar evolution theory has not been exploited to the extent that it could be.

The first two-dimensional (2D) hydrodynamics simulations of  the expansion of H\,\textsc{ii} regions \citep{BodTenYor79}, wind bubbles \citep{Roz85}, superbubbles \citep{MacMcCNor89} and bow shocks \citep{MacVanWoo91} showed the importance of asymmetric ISM density and of hydrodynamical instabilities in the evolution of circumstellar nebulae.
Colliding winds in binary systems were studied by \citet{SteBloPol92}, who found that the wind collision region can be dynamically unstable and predicted that the resulting X-ray emission could vary at the level of 10\%.
A series of papers gave a quantitative understanding of the physical processes that could give rise to the structure of planetary nebulae \citep{FraMel94, Mel94}.
\citet{RagNorCan97} studied the properties of bow shocks and H\,\textsc{ii} regions around runaway stars.

By coupling stellar-evolution calculations with 2D hydrodynamics on a spherical mesh (with logarithmic radial spacing), \citet{GarMacLan96} and \citet{GarLanMac96} studied the development of nebulae around stars evolving from Main Sequence through LBV $\rightarrow$ WR, and RSG $\rightarrow$ WR phases, respectively.
They predicted lifetimes and observable properties of nebulae produced during various transitions and phases, and compared results with nebulae around a number of WR stars, finding good agreement in some cases.
The numerical methods developed have been used in many follow-up works and ported to other codes \citep[e.g.][]{VanLanGar05, ChiLanVan08, VanLanVin09, VanDecMel14}.

\citet{MeyMacLan14} implemented the wind boundary-condition and radiative heating/cooling model of \citet{MacMohNei12} into \textsc{pluto} \citep{MigZanTze12} and made 2D hydrodynamics simulations of bow shocks around massive stars moving through the Galactic plane, following this in \citet{MeyLanMac15} with simulations of supernova blastwaves interacting with the bow shocks.
This model was extended to MHD by \citet{MeyMigKui17} and also used for a number of recent studies of circumstellar nebulae \citep[e.g.][]{MeyPetPoh20}.

\citet{YorKai95} and \citet{YorWel96} developed a radiation-hydrodynamics (R-HD) solver on a multiply nested grid in 2D cylindrical coordinates ($R$ and $z$) with adaptive timesteps.
This was used by \citet{FreHenYor03,FreHenYor06} to study H\,\textsc{ii} regions \ and wind bubbles around two stars (60\,$\mathrm{M}_{\odot}$ and 35\,$\mathrm{M}_{\odot}$, respectively) for the full evolution of the star through main sequence, supergiant, and WR phases.
The same evolutionary tracks as \citet{GarMacLan96} and \citet{GarLanMac96} were used.

Three-dimensional (3D) simulations of circumstellar nebulae became possible in the past 10-15 yr.
\citet{Pit09} developed 3D hydrodynamic simulations including wind acceleration, used to study the thermal X-ray emission from binary stars in \citet{PitPar10}.
Using 3D adaptive mesh-refinement simulations, 
\citet{ParPitCor11} studied the wind-collision region of the binary system $\eta$ Carinae, and \citet{ParGos11} investigated the WR\,22 system.
H\,\textsc{ii} region expansion in turbulent clouds was investigated by a number of authors \citep{MelArtHen06, ArtHenMel11, WalWhiBis12, GeeHenTre15}.
3D simulations of bow shocks around RSGs were presented in \citet{MohMacLan12}.
The differences between 2D and 3D calculations of wind-wind interaction were investigated by \citet{VanKep12}, and 3D calculations of wind bubbles expansing in turbulent media by \citet{RogPit13}.
\citet{GeeRosBla15} and \citet{HaiWalSei18} studied combined effects of winds and H\,\textsc{ii} regions on the ISM for the full evolution of a star using R-HD, similar to previous 2D calculations by \citet{FreHenYor03}.
3D MHD calculations of wind bubbles were presented by \citet{SchBaaFic20}.

Previous work \citep{FreHenYor03,FreHenYor06} has shown the value of static mesh-refinement for simulating circumstellar structures expanding from small to large scales, motivating the work presented here.
The majority of the work cited above was performed using software that is no longer actively developed or is not freely available.
In this paper we describe the simulation code \emph{PhotoIonization Of Nebulae}, abbreviated to \textsc{pion}, a R-MHD code that has been developed with the aim of modelling nebulae around massive evolving stars.
Significant new additions to the code with respect to previous versions \citep{MacLim10, MacLim11, Mac12} are described, and the code is made available to the community under a BSD-3 licence from \href{https://www.pion.ie}{https://www.pion.ie}.

The paper is organized as follows:
section \ref{sec:algorithms} describes the numerical methods, including the wind boundary condition, static mesh-refinement, radiative transfer and MHD.
Test calculations are presented in section \ref{sec:tests} that show the strengths and weaknesses of static mesh-refinement.
Applications of the code to modelling circumstellar nebulae are presented in section \ref{sec:winds}, namely a 3D simulation of a magnetized bow shock (section \ref{sec:wind3d}), 2D simulation of the formation of ring nebulae around rotating and evolving stars (section \ref{sec:chita}), 3D R-HD simulation of the wind-wind interaction from a RSG evolving to a WR star (section \ref{sec:GS96}), and 3D MHD simulation of a wind-wind collision between two rotating stars (section \ref{sec:v444}).
In all of these cases, the results are compared with previous calculations in the literature.
Methods for postprocessing simulation snapshots are described in section \ref{sec:postprocessing}, and parallel scaling in section \ref{sec:performance}.
Conclusions are presented in section \ref{sec:conclusions}.

\section{Code description and algorithms}
\label{sec:algorithms}

\textsc{pion} is a HD and MHD grid-based simulation code that includes radiative transfer of ionizing and non-ionizing photons for R-HD \citep{MacLim10} and for R-MHD \citep{MacLim11}.
A finite-volume integration scheme was implemented that is second-order-accurate in time and space, following \citet{FalKomJoa98}.
In \citet{MacLim10,MacLim11} the formation of pillars at the boundaries of H\,\textsc{ii} regions was investigated using 3D simulations in Cartesian geometry.
Improvements to the radiative transfer and time-integration schemes were described in \citet{Mac12}.
2D simulations with axisymmetry (cylindrical coordinates in $R$ and $z$) were added following \citet{Fal91}, and a stellar wind boundary condition implemented and used in \citet{MacMohNei12} to study the nebula around Betelgeuse assuming the star was previously a blue supergiant and only recently evolved to a RSG.
This was achieved by varying the wind parameters according to results from a stellar evolution calculation.
The spherically symmetric (1D) coordinate system has also been implemented, and was used for studying the external irradiation of winds from RSG \citep{MacMohGva14, SzeMacLan18} and for modelling the D-type expansion of H\,\textsc{ii} regions \citep{BisHawWil15}.
A non-equilibrium-ionization model for the thermodynamics and ionization of the diffuse ISM was introduced in \citet{MacLanGva13} for modelling H\,\textsc{ii} regions, and a related model for molecular gas in \citet{MacGvaMoh15}, based on results from \citet{HenArtDeC09}.
These were used in \citet{MacHawGva16}, \citet{GvaMacKni17} and \citet{GreMacHaw19} for simulating circumstellar nebulae and comparing observational data with synthetic observations.

These calculations were run on a uniform rectilinear grid, decomposed into blocks for parallel code execution, using MPI for inter-process communication.
The code was shown to scale well to at least 256 cores for 2D problems, and to 1024 cores for 3D problems \citep{Mac12} in tests of strong scaling (i.e.\ fixed problem size, variable number of MPI processes).
\textsc{pion} has proven to be a useful code for studying circumstellar nebulae and expanding bubbles driven by photoionization and winds, but most applications have been two-dimensional because of the limitations of the uniform computational grid.

\subsection{Stellar-wind boundary condition}
\label{sec:stellar-wind}

A stellar wind is modelled as a source of mass, momentum and energy within a sphere of user-specified radius on 1D-spherical, 2D-cylindrical and 3D-Cartesian grids.
There are three options in \textsc{pion} specified by wind-type parameter 0, 1 or 2 in the input parameter-file.
Type 0 is a spherically symmetric wind that is constant in time; type 1 is a spherically symmetric and time-varying wind with properties specified by a text file containing the time evolution; and type 2 is a latitude-dependent and time-varying wind.
These are described in the following subsections and demonstrated in sections \ref{sec:wind3d}, \ref{sec:GS96} and \ref{sec:chita}, respectively.
There is no limit on the number of wind sources that can be included in a simulation.

For all wind-boundary types, it is possible to specify chemical element abundance fractions (by mass) as passive scalar variables that are advected across the simulation domain \citep[see also e.g.][]{GeoWalFol13}.
These can be constant in time or with time-varying values read in from a text file.
We implemented the consistent multi-species advection (sCMA) scheme of \citet{PleMul99} for tracking the fractional abundances of these chemical elements.
This ensures that the non-uniform elemental abundances are tracked accurately as they expand outwards and mix with fluid elements that have (potentially) different abundances.

\subsubsection{Constant wind}
The simplest wind boundary (type 0) is spherically symmetric and constant in time, and the wind is injected at the terminal velocity, i.e., it is assumed that the wind boundary region is significantly bigger than the star.
The boundary region is specified by a position and a radius, both in cm, and the physical properties of the wind are specified by the mass-loss rate, $\dot{M}$, the wind terminal velocity, $v_\infty$, stellar radius, $R_\star$, temperature, $T_\mathrm{eff}$, equatorial rotation velocity, $v_\mathrm{rot}$, surface split-monopole magnetic field strength, $B_\star$, and the mass fractions of any chemical elements tracked.
Generally the wind boundary region should be 10-20 grid cells in radius to suppress grid-related artefacts in the expanding flow.

If $\left|v_\mathrm{rot}\right|>0$ then the spherical symmetry is broken for multi-dimensional simulations, because the azimuthal component of velocity and magnetic field are non-zero.
The magnetic field is taken to be weak (dynamically) and to follow a split monopole swept into a Parker spiral at large distance from the stellar surface.
Both toroidal and poloidal field components are included, and for simplicity it is assumed that the rotation and magnetic axes are coincident.
The rotational component of the wind velocity decays with distance, $r$, from the star as $r^{-1}$ and is generally negligible.
The boundary condition follows closely the methods commonly used for MHD modelling of the Solar Wind and Heliosphere \citep[e.g.][]{PogZanOgi04}, also similar to the recent implementation on a spherical coordinate grid by \citet{SchBaaFic20}, and it is demonstrated in sections \ref{sec:wind3d} and \ref{sec:v444}.

\subsubsection{Time-varying wind}
Wind type 1 is an extension of type 0 for time-varying sources that are specified through a tab-separated text file containing the evolution of the star in question.
The columns in this file are: time, mass, luminosity, temperature, mass-loss rate, rotation velocity, critical rotation velocity, $v_\mathrm{crit}\equiv v_\mathrm{esc}/\sqrt{2}$ (where $v_\mathrm{esc}$ is the surface escape velocity), wind terminal velocity, and mass-fractions of any chemical elements tracked.
All values are assumed to be in cgs units and can be modified output from a stellar evolution calculation \citep[e.g.][]{MacMohNei12} or an ad-hoc model \citep[cf.][]{LanGarMac99}.
The evolving stellar wind module was previously used in \citet{MacMohNei12} to study the hydrodynamics of the nebula produced when a blue supergiant evolves redward to a RSG, and follows similar algorithms from the earlier literature \citep{GarLanMac96, VanLanGar05}.
Here the module is demonstrated in section \ref{sec:GS96} for the nebula produced when a RSG evolves to a WR star.

\subsubsection{Latitude-dependent and time-varying wind}
Wind type 2 provides a prescription for latitude-dependent winds from rotating stars, and the option to read time-evolution of stellar-wind and radiation properties from a text file.
The latitude-dependent wind is modelled following \citet{LanGarMac99}, who introduced a mathematical model of the focusing of stellar wind towards the equator as the star approaches the so-called $\Omega$-limit \citep{Lan97}, defined as the equatorial surface rotation speed, $v_\mathrm{rot}$, for which the net acceleration on the surface layers is zero.
The critical rotation velocity is used to define the rotation parameter, $\Omega\equiv v_\mathrm{rot}/v_\mathrm{crit} <1$.
Eqs. 3-5 in \citet{LanGarMac99} are used to calculate the latitude dependence of the wind density and velocity as a function of $\Omega$.
This algorithm is based on the theory of \citet{BjoCas93} and it produces many of the observed features of bipolar nebulae \citep{LanGarMac99, ChiLanVan08, VanLanYoo08}, particularly for stars that reach critical rotation in the temperature range of $6000-10\,000$\,K when embarking on a blue loop.
This module is demonstrated in section \ref{sec:chita} where the results are compared with previous literature results.

\subsection{Upgraded Magnetohydrodynamics implementation}

\textsc{pion} has a MHD implementation presented in \citet{MacLim11}, which is effective for simulating the magnetohydrodynamics of H\,\textsc{ii} regions \citep{MacLanGva13}.
This uses a modified version of the \citet{DedKemKro02} mixed-GLM divergence-cleaning algorithm for mitigating against the growth of magnetic monopoles.
It uses either the linear MHD solver described by \citet{FalKomJoa98}, or the Roe solver in conserved variables of \citet{CarGal97}, following \citet{StoGarTeu08}.

Neither of these MHD Riemann solvers is robust enough for the high-Mach-number shocks encountered in stellar-wind bubbles around hot stars.
We implemented the HLL solver in HD \citep{HarLaxVan83} and MHD \citep{Jan00} following \citet{MigZanTze12}, and also the more accurate HLLD solver \citep{MiyKus05} for MHD. 
The HLLD solver is also not sufficiently robust for the high-Mach-number flows in stellar wind simulations, because it is not positive definite in gas pressure \citep{MigZanTze12}.
Following \cite{MigZanTze12} we implemented a shock detection scheme and a switch that locally replaces the HLLD with the HLL scheme, which is positive definite.
This improves the code stability, but for some problems the simpler HLL scheme should be used everywhere.

Following \citet{DerWinGas18} we included the Powell source terms \citep{PowRoeLin99}, added the source terms for the \citet{DedKemKro02} $\psi$ field, which we re-scaled as in \citet{DerWinGas18}, and included $\psi$ in the total energy density.
This introduces very small changes in the solution to test calculations, and some improvements in the robustness of the scheme. 
We did not implement the full scheme of \citet{DerWinGas18} with their entropy-stable Riemann solver, and so the rest of the MHD implementation is as in \citet{MacLim11}.

\subsection{Static Mesh-Refinement}

A number of implementations of static mesh-refinement have been described in the literature, typically arranged as a multiply nested grid that is centred on a region of interest.
\citet{FreHenYor03,FreHenYor06} used a 2D nested grid with axisymmetry described in \citet{YorKai95} and \citet{YorWel96} to study expanding nebulae.
This has one advantage over a spherical grid with logarithmically spaced radial cells, in that the latter has a global timestep for all cells whereas the nested grid can have adaptive timestepping.
This efficiency comes at the cost that all radial columns away from a point source are not equal -- angle-dependent numerical viscosity and grid-artefacts are inevitably introduced, as can be seen by comparing results from \citet{GarLanMac96} and \citet{FreHenYor06}.
In particular with 2D simulations, the symmetry axis has a coordinate singularity that affects results, also seen in bow-shock simulations \citep{GreMacHaw19}.
For 3D this is less of a problem, but the viscosity of the numerical scheme for expansion along grid axes remains different from expansion at an angle to the grid.

Recently \citet{StoTomWhi20} desribed the implementation of static and adaptive mesh-refinement algorithms in the \textsc{Athena++} software framework, again demonstrating the dramatic improvements that can be obtained with these techniques.
The advantages in computational efficiency of a nested grid compared with a uniform grid are clear: for 2D calculations with a uniform grid, doubling the resolution everywhere increases the computational cost by a factor of 8; for 3D calculations it is a factor of 16.
Adding a nested grid that is a factor of 2 smaller than the coarse grid in each dimension, but that retains the same number of zones, increases the computational load by a factor of 3 (the fine grid requires the same amount of computation as the coarse grid \emph{per step}, but must take twice the number of steps), and this is independent of dimensionality.
Adding a third level requires 7 times more computation than just a single level, whereas for a uniform grid the cost of quadrupling the resolution would be $64\times$ (2D) or $256\times$ (3D) more work.
A nested grid also has a modest efficiency advantage over spherical-coordinate grids with a cell size that increases with radius, in that adaptive timesteps can be used.
On a spherical grid all cells must use the same timestep, usually dictated by the smallest cells close to the origin.
For $N$ refinement levels, the computational saving using adaptive timestepping compared with a global timestep on all levels approaches a factor of $N/2$ for large $N$.

There are three additions to a uniform-grid algorithm required for a nested grid:
\begin{enumerate}
\item
  The refined grid should obtain its external boundary data from its parent (coarser) grid, by interpolating the coarse-grid zones to the zone-centres of the boundary data on the refined grid.  This interpolation should be done to the same order of accuracy as the spatial reconstruction used, and should conserve the total mass, momentum and energy of the coarse-grid zone.  This is known as \emph{prolongation} \citep[e.g.][]{TotRoe02}.
\item
  The coarse grid should update its zones by obtaining averaged data from any finer-level grid (where applicable).  This is known as \emph{restriction} \citep[e.g.][]{TotRoe02}.
\item
  The flux entering/leaving a finer-level grid should be recorded and sent to the coarser-level parent grid to ensure that this flux is consistent across all grid levels \citep{BerCol89}.
  This is required so that conserved quantities are indeed conserved; otherwise mass, momentum and energy can disappear because of inconsistencies between levels.
\end{enumerate}

All of these are well-established techniques, but they are described below because the implementation depends on the time-integration scheme adopted as well as the parallelisation strategy.

\subsubsection{Coarse-to-fine interpolation (Prolongation)}
We follow the scheme used for MPI-AMRVAC \citep{MelKepCas07,KepTeuXia20} on a cell-by-cell basis, and for a grid with $D$ spatial dimensions.
For a scheme that is first-order accurate in space, we can simply copy the coarse-grid values to the $2^D$ fine-grid cells.
For a second-order scheme, linear interpolation and correction are applied as follows:
\begin{enumerate}
\item
  For each coarse-grid cell, $i$, with cell volume $V_i$ and cell-centred vector of primitive variables $\bm{P}_i$, calculate slopes, $\bm{m_k}$, of the primitive variables in each dimension $k$.
\item
  Send these data to the finer grid, and the finer grid receives the data.
\item
  Using the slopes $\bm{m_k}$ interpolate $\bm{P}_i$ to the cell centres of the $2^D$ fine-grid cells contained within the coarse-grid cell $i$, assigning primitive variable data $\bm{P}_j$ to these cells.
  Depending on grid dimensionality, this uses linear, bilinear, or trilinear interpolation.
\item
  The conserved quantities $\bm{U}_i$ and $\bm{U}_j$ are calculated from $\bm{P}_i$ and $\bm{P}_j$, respectively.
\item 
  The difference vector $\bm{\Delta} = \bm{U}_i V_i - \sum_j \bm{U}_j V_j$ is calculated, to ensure that the conserved quantities have consistent values within the same volume in both levels.
\item
  The fine-grid cells $j$ are corrected by adding $1/2^D$ of this difference to each $\bm{U}_j$ (also dividing out the total volume):
  \begin{equation}
  \bm{U}_j \rightarrow \bm{U}_j + \frac{\Delta}{2^D V_i}
  \end{equation}
\item
  The fine-grid primitive vectors $\bm{P}_j$ are obtained from the corrected $\bm{U}_j$ vectors, for each fine-grid cell.
\end{enumerate}
This ensures that conserved quantities are conserved when a coarse grid cell is prolongated on to the finer grid.

\subsubsection{Fine-to-coarse averaging (Restriction)}
This is much more straightforward than prolongation, and also independent of the spatial order of accuracy of the scheme.
\begin{enumerate}
\item
  For each set of $2^D$ fine-grid cells $j$, contained within the coarse-grid cell $i$, we calculate the average of the conserved quantities contained within the volume $V_i$ of cell $i$:
  \begin{equation}
  \bm{U}_i = \frac{\sum_j \bm{U}_j V_j}{\sum_j V_j}
  \end{equation}
\item
  The list of $\bm{U}_i$ vectors is sent to the coarse grid, and the coarse grid receives data.
\item
  Vectors $\bm{U}_i$ are converted to a primitive vector and assigned to each coarse-grid cell $i$.
\end{enumerate}

\subsubsection{Flux correction on coarse grid zones abutting a fine grid boundary}

\citet[][hereafter BC89]{BerCol89} describe a method to ensure consistent fluxes across cell boundaries at different levels of refinement, with the assumption that the most accurately calculated flux is at the finest level.
This finest-level flux is then propagated to coarser levels as needed, and the coarse-cell fluxes are corrected to agree with the finest-level flux.
The \textsc{pion} implementation is described here for 2 levels, which is the only case that arises for a nested grid arrangement.
It is assumed that the coarse and fine grids are assigned to different MPI processes, although the update algorithms do not make the MPI calls if the grids are on the same process.

The correction is not needed for the half-step in the second-order scheme, because this is only an approximate time-centred state used to calculate fluxes that are accurate to second order.
This means that the full-step fluxes over two fine-grid steps must be added together and sent to the coarse grid \emph{after} the full-step coarse fluxes have been calculated but \emph{before} the coarse grid cells have their state advanced in time.

When the coarse and fine grids are set up, the edge cells of the refined levels are identified, as well as the coarse grid cells that share the same edge.
Any fine-grid cells whose outer face is the edge of the fine level have up to $D$ (for edge/corner cells) extra state vectors allocated and initialized to zero, to record the full-step fluxes as they are calculated.

In addition, the fine grid allocates up to $2D$ vectors of C-style \emph{structs} (one for each outward normal direction on the grid).
Each element in each vector represents a cell face on the coarse grid for which the flux will be corrected by the fine grid.
The structs contain a list of pointers to the fine-grid cells contributing to this coarse-cell face, a vector of corresponding areas of the cell faces (not all identical for curvilinear grids), and a state vector to hold the time-integrated flux through the fine-cell faces over the two time steps.
The coarse grid allocates a similar vector of structures to record the uncorrected fluxes, but there is only one coarse cell and one face area in each struct.

The scheme is as follows, shown only for one coordinate direction:
\begin{enumerate}
  \item At the start of an even-numbered fine grid timestep, reset BC89 fluxes to zero.
  \item Record fluxes, $\bm{F}_j$ across all fine-grid boundary cells $j$ during the timestep.
  \item Calculate time- and area-integrated flux, $\bm{\Delta U}_i^f$, through surfaces of the $2^{D-1}$ cells, $j$, on the fine grid that correspond to the surface of coarse-grid cell, $i$.
  Add these to the vector of structs on fine grid by summing the fluxes:
\begin{equation}
  \bm{\Delta U}_i^f = \Delta t^f \sum_{j=0}^{2^{(D-1)}} \bm{F}_j A_j^f
\end{equation}
    where $A_j^f$ is the surface area of the face of cell $j$, and $\Delta t^f$ is the fine-grid timestep.
  \item On the coarse grid record fluxes, $\bm{F}_i$, through cell faces that map on to the edge of fine grid, and calculate $\bm{\Delta U}_i^c = \Delta t^c \bm{F}_i A_i^c$ (where $\Delta t^c=2\Delta t^f$ is the coarse-grid time step and $A_i^c$ is the area of the face of cell i).
  \item Complete the odd-numbered fine-grid step by repeating steps (ii) and (iii), adding to $\bm{\Delta U}_i^f$ as before.
  \item Send array of $\bm{\Delta U}_i^f$ values from MPI process of fine grid to MPI process of coarse grid.
  \item Correct $\bm{\Delta U}_i^c$ values on the coarse grid so they agree with $\bm{\Delta U}_i^f$, and modify fluxes accordingly.
\end{enumerate}

\subsection{Time-integration scheme}

The coarse-to-fine update can only be applied once every full step of the coarse grid for a fully consistent solution, and so the finer-level grid must calculate two timesteps between updates, following the algorithm above.
The boundary region should be six cells deep in order to complete two full steps on the finer level without updating the boundary conditions (for a second-order scheme), compared with four cells if the update was every step.
The fine-to-coarse boundaries are updated every full step on the fine-level grid.

The following time-integration scheme was implemented, based on the uniform-grid scheme of \citet{FalKomJoa98} and using adaptive timesteps on nested grids.
We update level $l$ by one step and level $l+1$ by two steps, and the algorithm is recursive.
\begin{enumerate}
  \item Begin timestep level $l$, to advance current time, $t_0$, by $\Delta t_l$.
  \item If an even step, receive coarse-to-fine external boundary data from level $l-1$.
  \item Update any other external boundary conditions (including from domain decomposition).
  \item Send coarse-to-fine data to $l+1$.
  \item Advance level $l+1$ by one step.
  \item Calculate fluxes on level $l$ and calculate the time-centred state at $t_0+0.5\Delta t_l$.
  \item Update internal boundary conditions (e.g.~stellar wind properties).
  \item Receive fine-level data from $l+1$ and replace level $l$ states with these data (including optical depths, if raytracing).
  \item Update external boundary conditions except for coarse-to-fine level boundary.
  \item Do raytracing on time-centred state to calculate optical depths for full step.
  \item Calculate level $l$ fluxes for full step, using the time-centred state, saving fluxes needed for BC89 correction.
  \item Advance level $l+1$ by one step.
  \item Receive BC89 fluxes summed over two steps from $l+1$ and correct level $l$ fluxes accordingly.
  \item Update state vector on level $l$ to $t=t_0+\Delta t_l$.
  \item Update internal boundary conditions (e.g.~stellar wind properties), receive fine-level data from $l+1$ (including optical depths) and replace level $l$ states with these data.
  \item Raytrace level $l$ to calculate optical depths for next (half) step.
  \item If an even-numbered step, send BC89 fluxes to $l-1$.
  \item Send fine-to-coarse averaged data to $l-1$ (including optical depths).
  \item return.
\end{enumerate}

After a full step on the coarsest grid, the timestep is re-calculated on all grids and a new step is started.
The refined grids use the same timestep for the duration of the coarse step, and so a safety factor is included in addition to the usual CFL condition.

\subsection{Radiative Transfer}
Raytracing is implemented in \textsc{pion} using the method of short characteristics with the On-The-Spot approximation, i.e. scattered radiation is locally re-absorbed and so only direct radiation from point sources needs to be transported across the grid \citep{MelIliAlv06,Mac12}.
If radiation sources are always on the most refined grid, then raytracing on a nested grid proceeds on the finest level exactly as for a uniform grid.
On the next coarser level the stored quantities (whether optical depth, column density, or radiation density) are mapped on to the coarser grid cells from the finer grid, and raytracing proceeds through the rest of the level.
This procedure is repeated for all coarse levels.

Raytracing must be performed on all levels when calculating the timestep on the coarsest level (because the photoionization and recombination time-scales also limit the timestep), and then twice each timestep per level for the second-order scheme \citep{Mac12}.
So for a grid with 4 levels, the finest level (level 3)  has 17 raytracings per coarse-grid step, level 2 has 9 raytracings, level 1 has 5, and level 0 has 3.

\subsection{Summary}
The upgraded MHD algorithms with static mesh-refinement are implemented in first-order and second-order schemes, and the code was run on different numbers of MPI processes to check for consistency.
The results for the first-order scheme for HD, MHD and R-MHD were shown to be byte-for-byte identical, independent of the number of MPI processes.
For the second-order scheme, HD and MHD results are byte-for-byte identical when run on different numbers of MPI processes, and R-MHD results show very small differences (relative difference $\approx10^{-4}$ in primitive variables) at the end of a simulation, arising because of the adaptive timestepping algorithm in the implicit solver for ionization and heating/cooling.
The conservation of mass, momentum and energy were also checked and found to be consistent with roundoff error.

\section{Test Calculations}
\label{sec:tests}

\begin{figure*} 
\centering
\includegraphics[height=7.5cm]{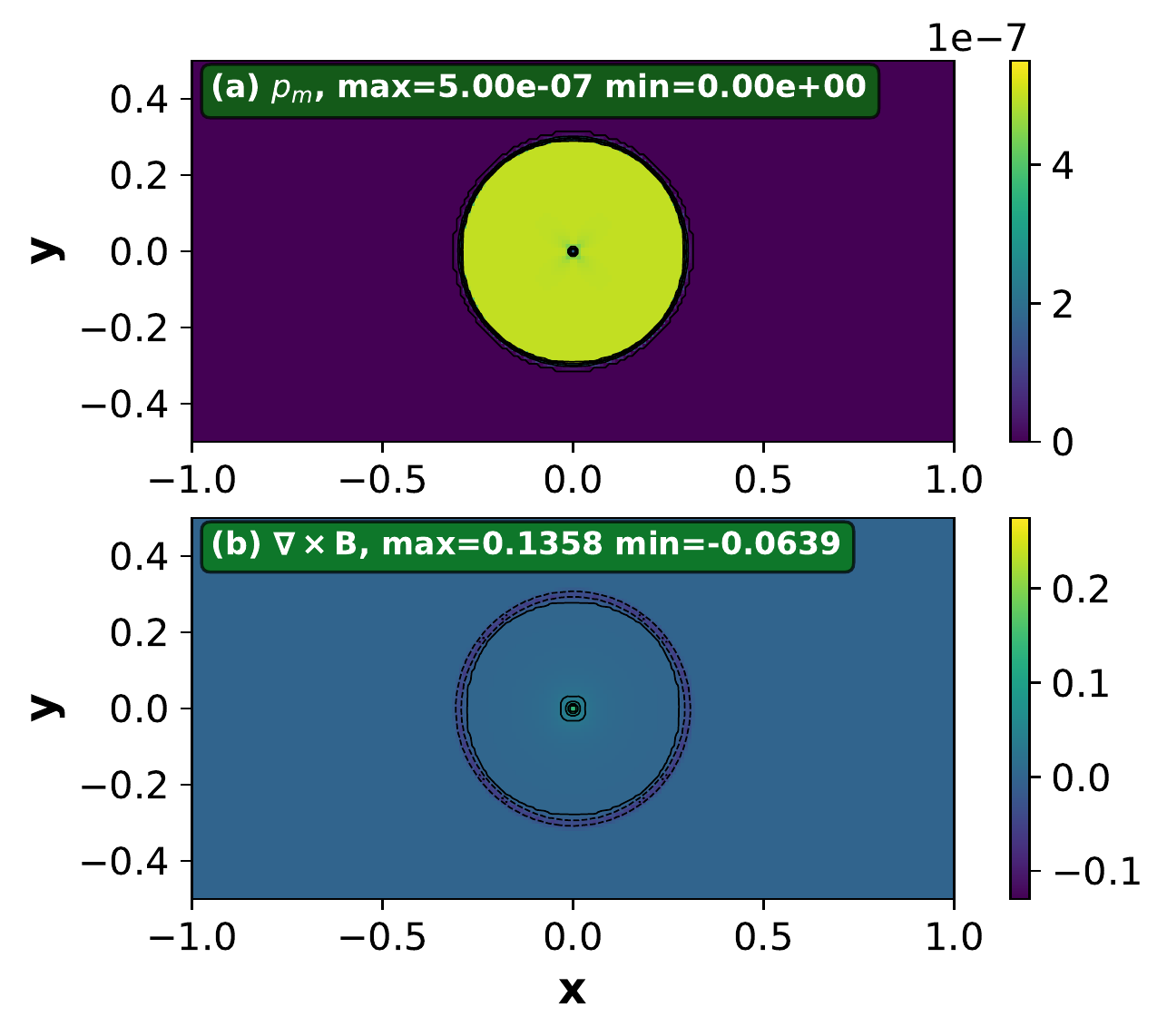}
\includegraphics[height=7.5cm]{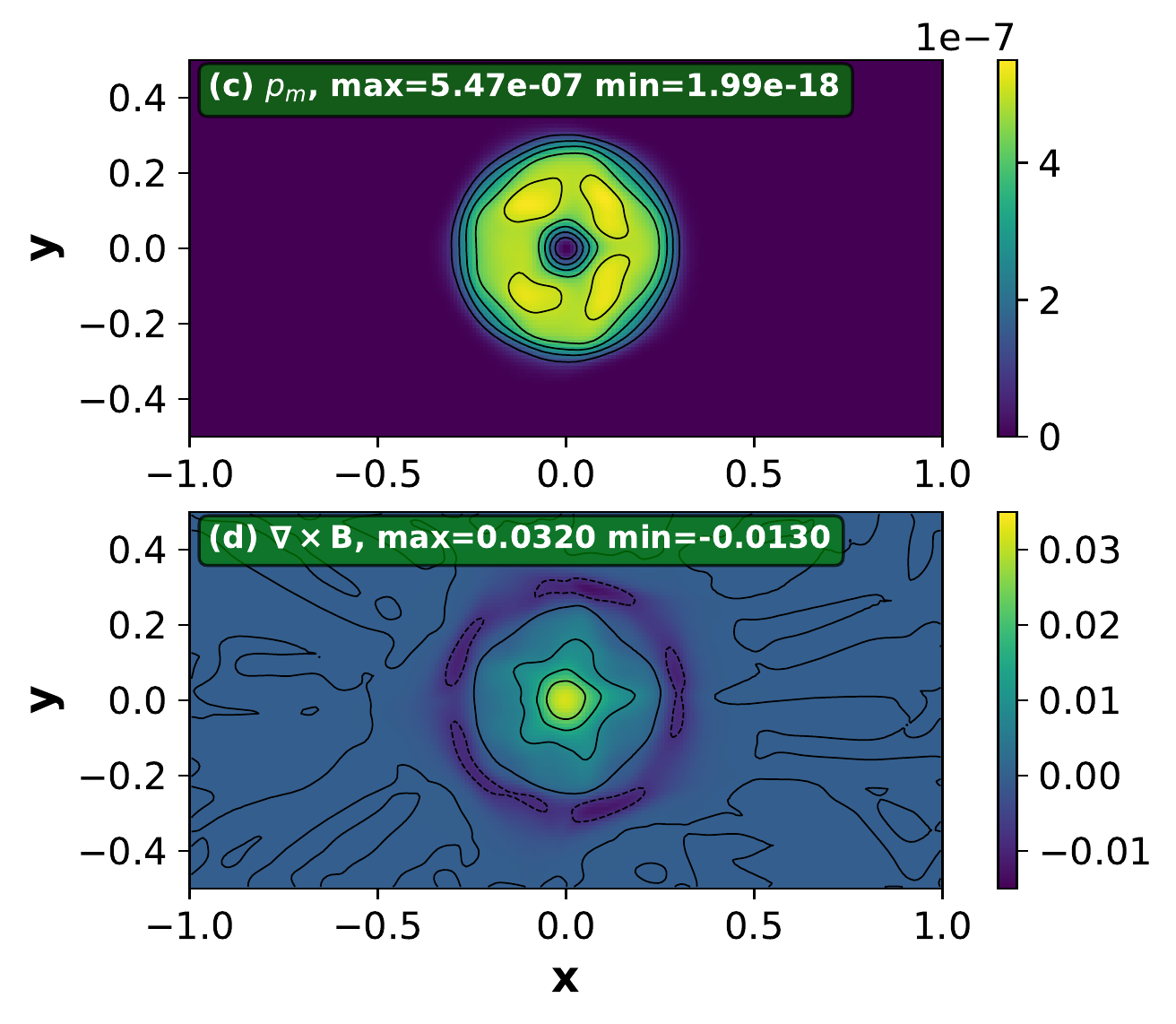}
\caption{
  Field-loop advection test, showing the initial conditions (panels a,b) and the final state at $t=2$ after advecting twice across the domain (panels c,d).
  The magnetic pressure $p_\mathrm{m}=|\bm{B}|^2/(8\pi)$ is plotted in the upper panels, and the current density in the lower panels, using a linear colour scale as indicated.
  In the upper panels, contours of magnetic pressure are shown from $p_\mathrm{m}=1\times10^{-7}$ to $p_\mathrm{m}=5\times10^{-7}$, linearly spaced in steps of $1\times10^{-7}$.
  For panel (b) the current density contours are $\bm{\nabla\times B}=[-0.06,-0.03,0,0.03,0.06,0.09,0.12]$,
and for panel (d)  $\bm{\nabla\times B}=[-0.008,0,0.008,0.016,0.024,0.032]$, using broken lines for negative contours.
  }
\label{fig:fl_ug}  
\end{figure*}

\begin{figure*} 
\centering
\includegraphics[height=7.5cm]{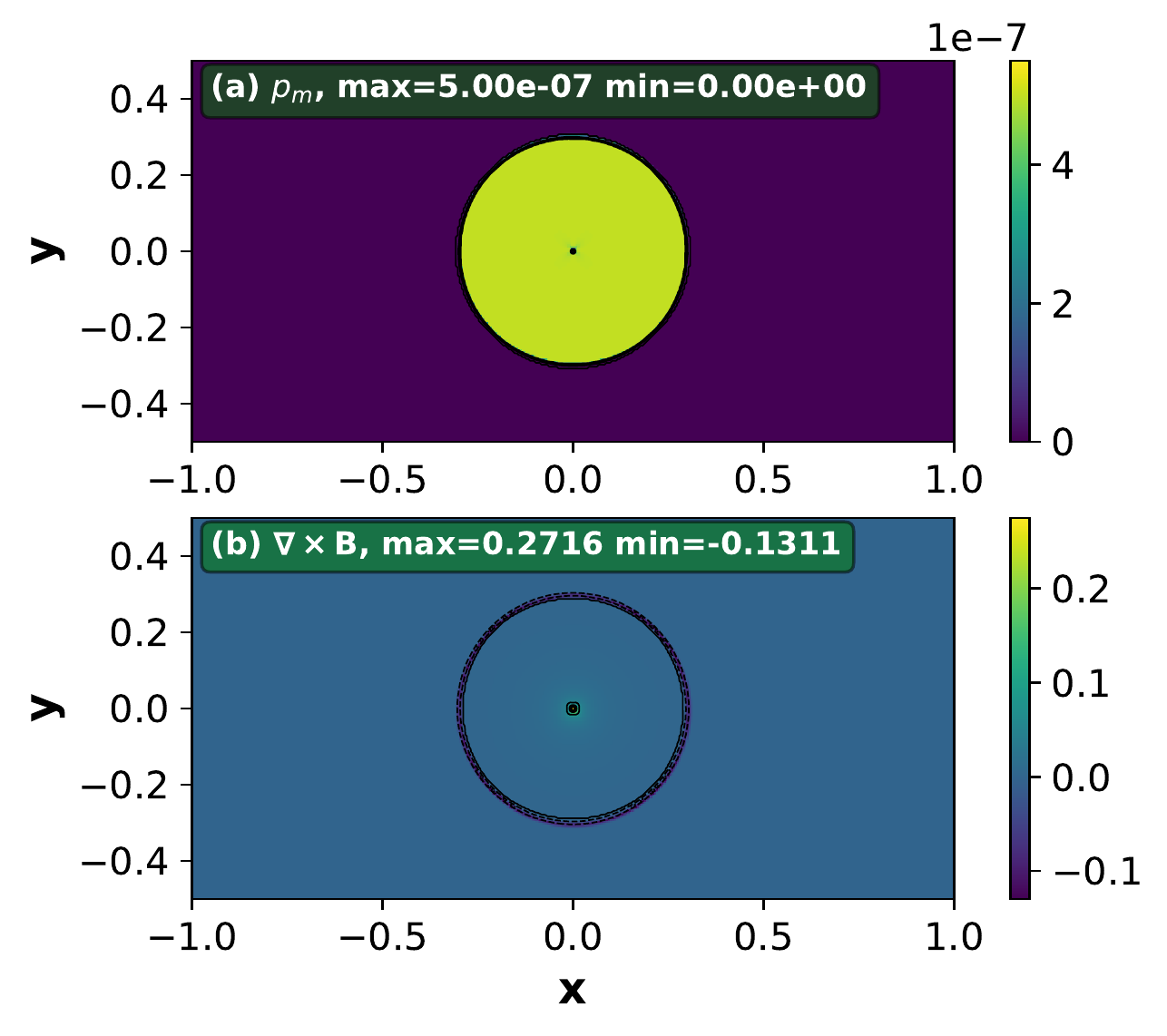}
\includegraphics[height=7.5cm]{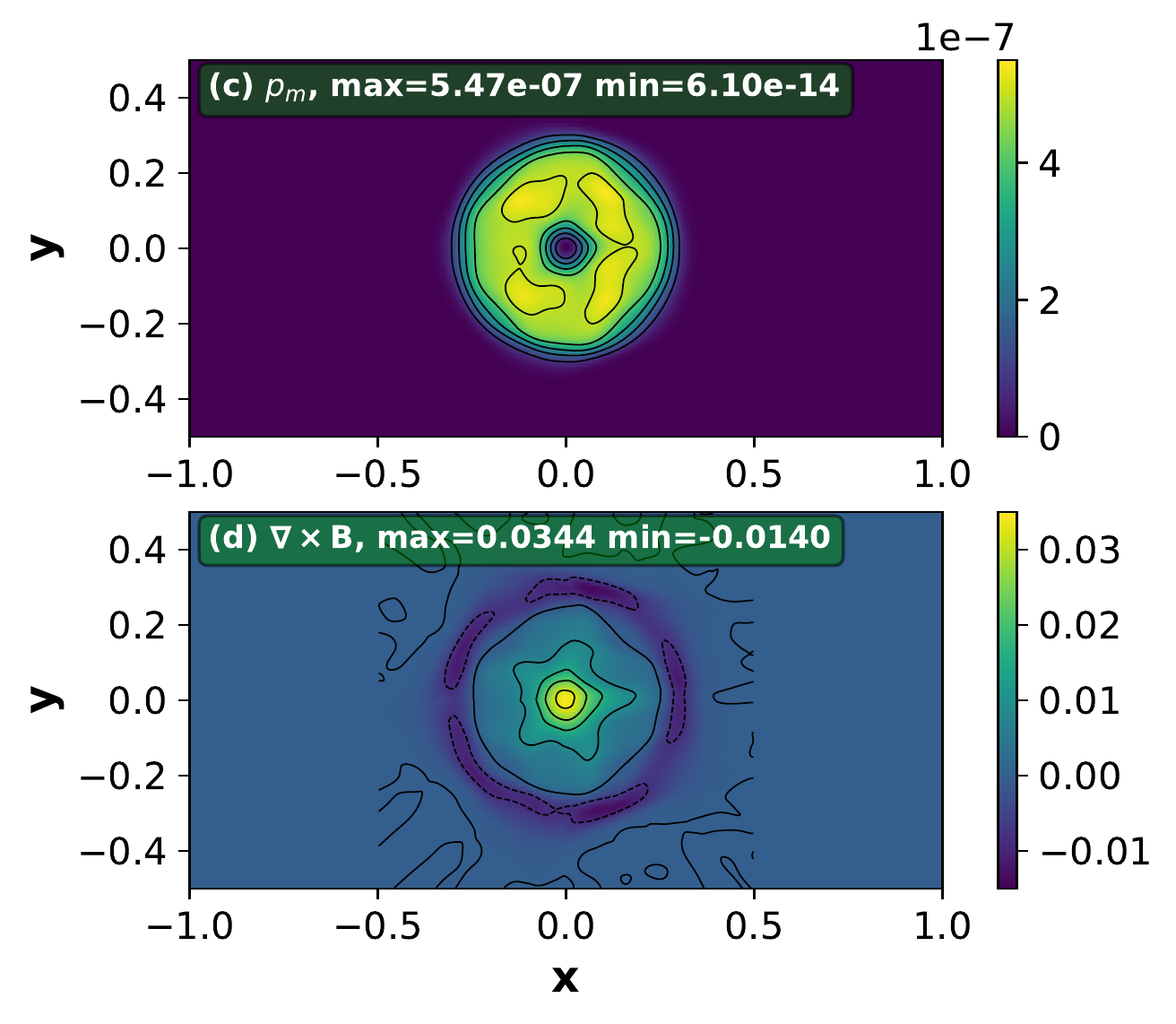}
\caption{
  As Fig.~\ref{fig:fl_ug} but using a nested grid with 2 levels, centred on $[0,0]$, again plotting magnetic pressure above and current density below.
  The left panels show the initial conditions and the right panels the results at $t=2$ after advecting twice across the domain.
  In the upper panels, contours of magnetic pressure are shown as before, but only for the refined grid.
  For panel (b) the current density contours are $\bm{\nabla\times B}=[-0.12,-0.06,0,0.06,0.12,0.18,0.24]$,
and for panel (d)  $\bm{\nabla\times B}=[-0.008,0,0.008,0.016,0.024,0.032]$, using broken lines for negative contours.
  }
\label{fig:fl_ng}  
\end{figure*}

\subsection{Advection of a Magnetic Field Loop}
\label{sec:advection}

Advection across refinement boundaries can verify that the refinement has been implemented correctly, and can show that the accuracy of the nested-grid integration algorithm is the same as that of the uniform-grid.
This is demonstrated with 2D test problems using periodic boundaries, where the whole domain is advected twice through the domain and back to its starting location.

The advection of a magnetic field loop is a good test of the diffusivity of an MHD scheme \citep[e.g.][]{GarSto05, StoGarTeu08}.
A weak magnetic field loop is set up in the $x$-$y$ plane using the vector potential $\bm{A}=[0,0,A_z]$, with $z$-component
\begin{equation}
  A_z = 
    \begin{cases}
    \sqrt{4\pi}A_0(R_0-r) & (r=\sqrt{x^2+y^2})<R_0 \\
    0          & r\geq R_0 \;,
    \end{cases}
\end{equation}
using $A_0=0.001$ and $R_0=0.3$.
This generates a constant circular magnetic field of strength $\sqrt{4\pi}A_0$ within $r<R_0$, a current sheet at $r=R_0$, and a current spike at $r=0$ whose amplitude increases with increasing numerical resolution.
The initial uniform density is $\rho=1$, thermal pressure is $p_\mathrm{g}=1$, and velocity is $\bm{v}=[2,1,0]$, using an adiabatic equation of state with $\gamma=5/3$.
The magnetic pressure, $p_\mathrm{m}\equiv B^2/8\pi=5\times10^{-7}$, is therefore negligible and the field is advected with the flow.

For the uniform-grid simulation, a 2D domain with $x\in[-1,1]$ and $y\in[-0.5,0.5]$ is used, and for the nested-grid simulation the domain is $x\in[-1,1]$ and $y\in[-1,1]$ (so that the field loop fits entirely in the first refined level centred on $[0,0]$).
In both cases the HLLD solver is used, and $200\times100$ grid cells per level.
Results from the uniform-grid simulation are plotted in Fig.~\ref{fig:fl_ug}, showing the initial conditions (left) and the final state at $t=2$ (right), by which time the loop has advected twice across the domain at an angle $30^\circ$ to the positive $x$-axis.
These results are similar to those shown for the previous version of \textsc{pion} in \citet{MacLim11} using the linear solver of \citet{FalKomJoa98}.
The decay of magnetic pressure very closely follows the results presented in \citet{MacLim11} (because the integration scheme is essentially unchanged) and is not shown here.

Results for the simulation with one level of refinement, and a refined grid on $x\in[-0.5,0.5]$ and $y\in[-0.5,0.5]$, are plotted in Fig.~\ref{fig:fl_ng}, where now contours are only plotted for the refined grid.
This simulation has a larger domain from $[-1,1]$ to $[1,1]$.
The loop still advects twice across the domain, crossing both the refined and coarse levels, but spending most time on the coarse level.
Results are very similar to the uniform-grid case, except that the extrema of $\bm{\nabla\times B}$ are slightly more pronounced, meaning that the initial conditions are marginally better preserved.
No artefacts are introduced by advecting the field loop across refinement levels.

\begin{figure*} 
\centering
\includegraphics[width=0.32\textwidth]{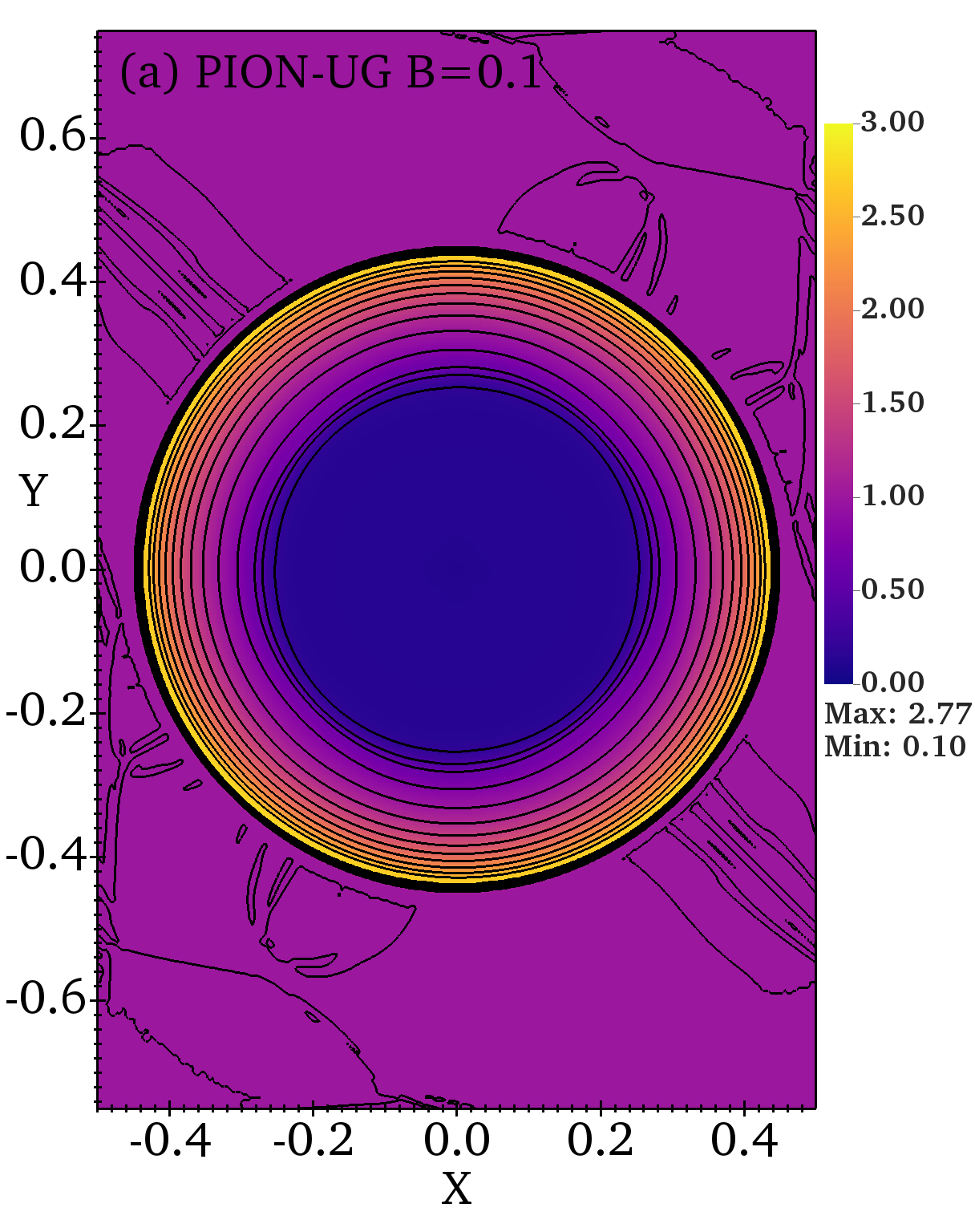}
\includegraphics[width=0.32\textwidth]{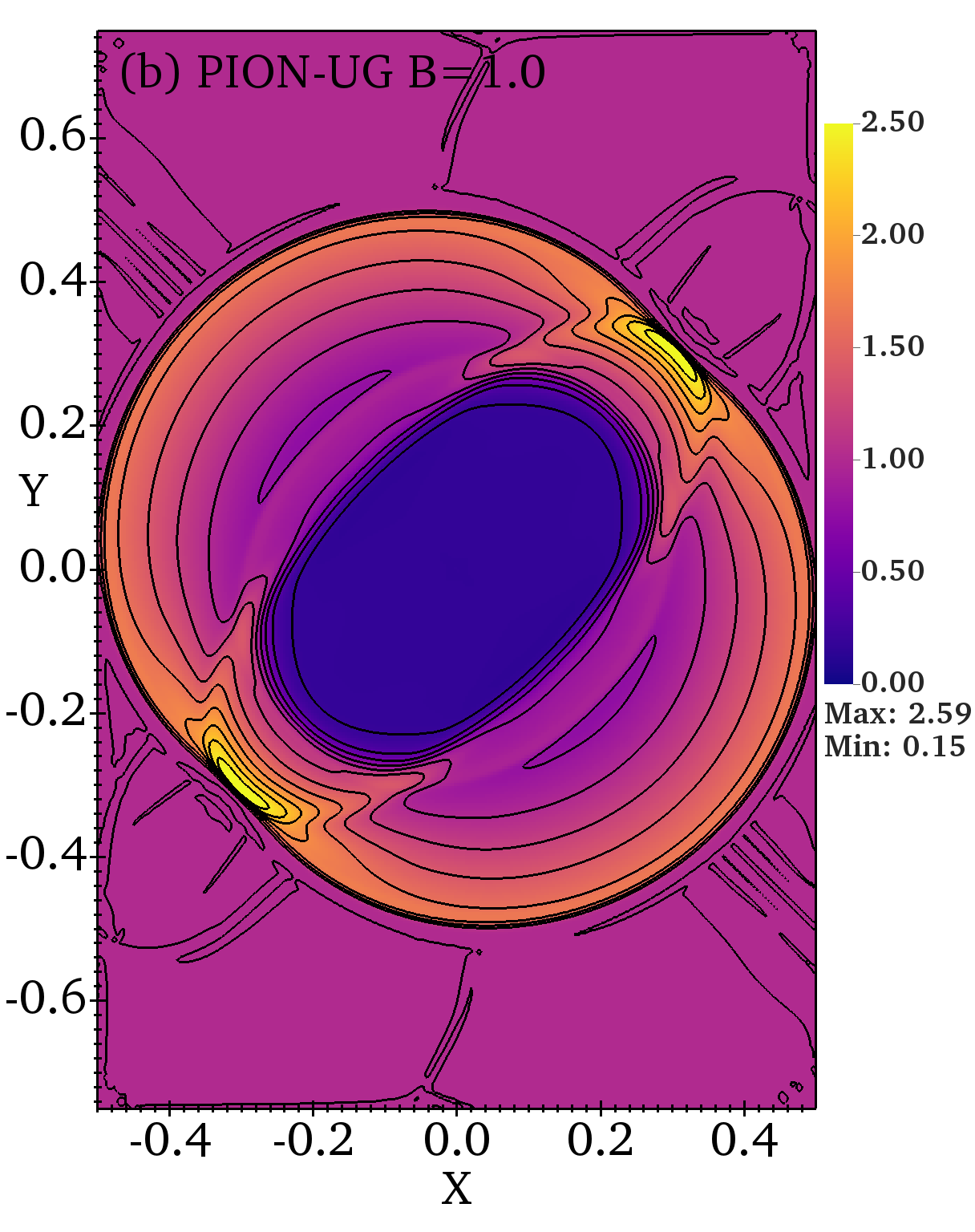}
\includegraphics[width=0.32\textwidth]{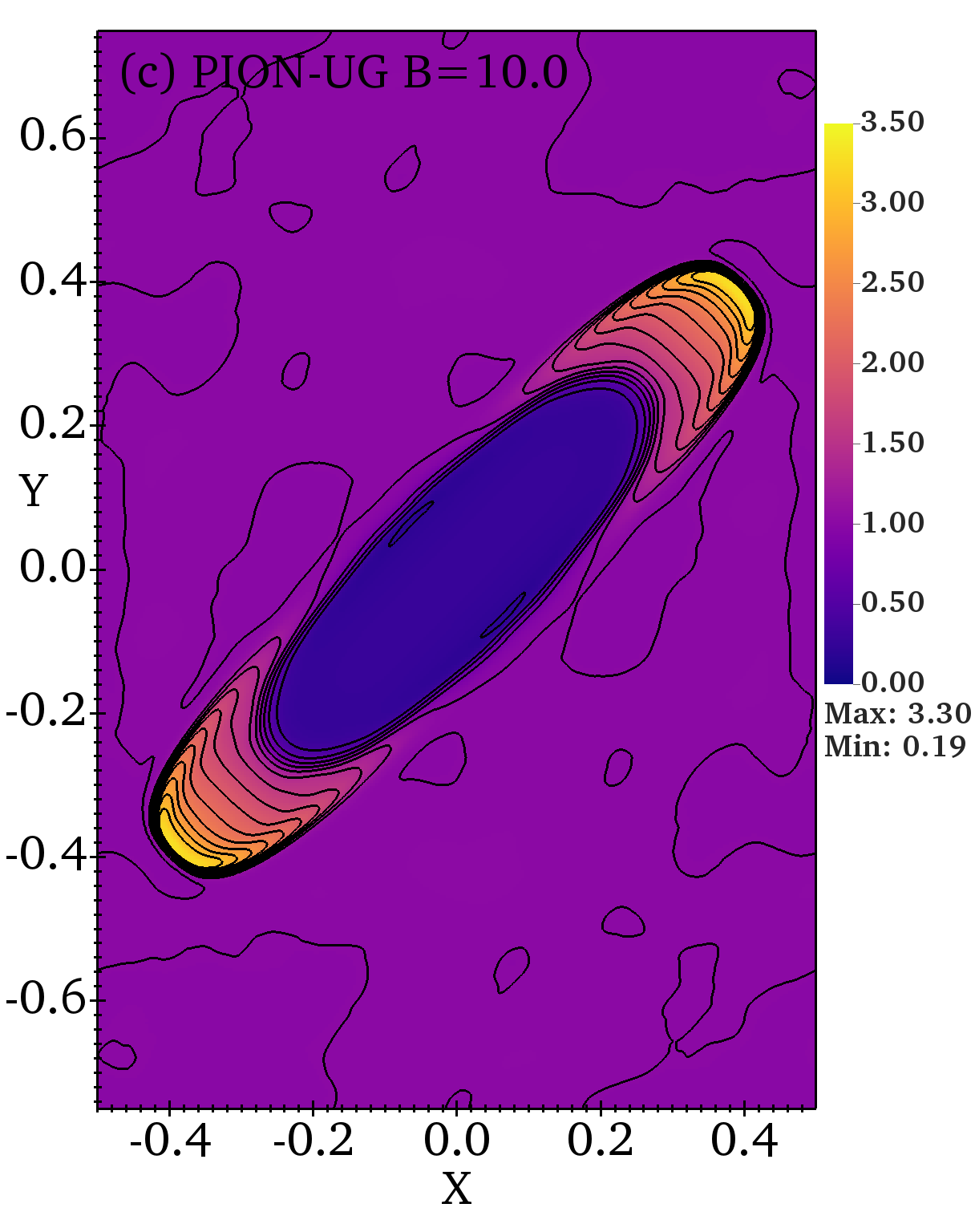}
\caption{
  MHD Blastwave test calculation in 2D using a uniform grid with $256\times384$ cells, calculated with \textsc{pion}, for a background magnetic field of strength $B=0.1$ (a), $B=1$ (b), and $B=10$ (c), at an angle of 45$^\circ$ to the positive $x$-axis.
  Gas density is plotted at $t=0.2$ using the indicated linear colour scale.
  Contours of density are plotted on a linear scale starting from $\rho=0$ separated by $\Delta\rho=0.2$.
  }
\label{fig:mhdblast-ug}  
\end{figure*}

\subsection{MHD Blast Wave in 2D}
\label{sec:MHDBW2D}

The expansion of a blastwave in a 2D Cartesian domain is a standard test problem, \citep[e.g.][]{StoGarTeu08}.
Here we set up the problem as in \citet{StoGarTeu08} and \citet{MacLim11}: the domain is $x\in[-0.5,0.5]$, $y\in[-0.75,0.75]$, resolved by $256\times384$ cells, with a uniform background density $\rho=1$, pressure $p_\mathrm{g}=0.1$, and magnetic field strength of 0.1, 1.0, or 10.0 (in units where factors of $4\pi$ do not appear, so e.g., $B=1$ corresponds to $B=\sqrt{4\pi}$ in CGS units).
The field is oriented at an angle of 45$^\circ$ to the $x$-axis and the medium is initially at rest.
A circle of radius 0.1 is filled with gas at pressure $p_\mathrm{g}=10$ and the system is allowed to evolve to $t=0.2$.
Periodic boundary conditions are imposed on all sides, although they are not relevant to the dynamics until $t>0.2$.

We calculate the three problems (weak, medium and strong magnetic field) using the HLLD solver with a CFL number of 0.24, initially on a uniform grid.
The results for all three cases are shown in Fig.~\ref{fig:mhdblast-ug} at $t=0.2$, and are comparable to results obtained with \textsc{Athena} \citep{StoGarTeu08}.
The features visible in the contour at $\rho=1$, outside the outer shock, arise from the diffusion of divergence errors by the $\psi$ field of the GLM-MHD scheme.
Apart from this, the symmetry of the blast wave and contact discontinuity are maintained well, and the HLLD solver is at least as good the solver presented in \citet{MacLim11} while being significantly more robust.

\begin{figure*} 
\centering
\includegraphics[width=0.32\textwidth]{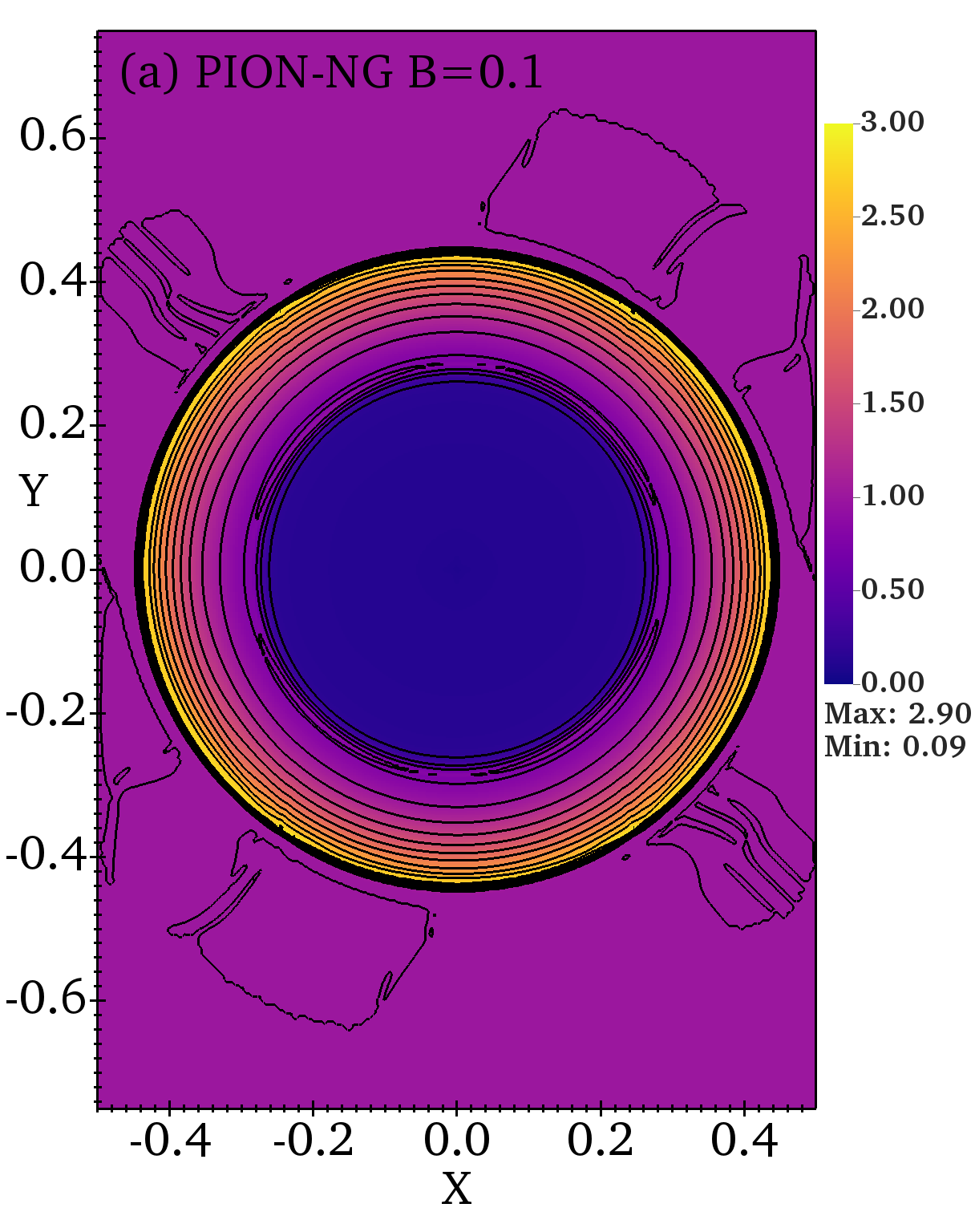}
\includegraphics[width=0.32\textwidth]{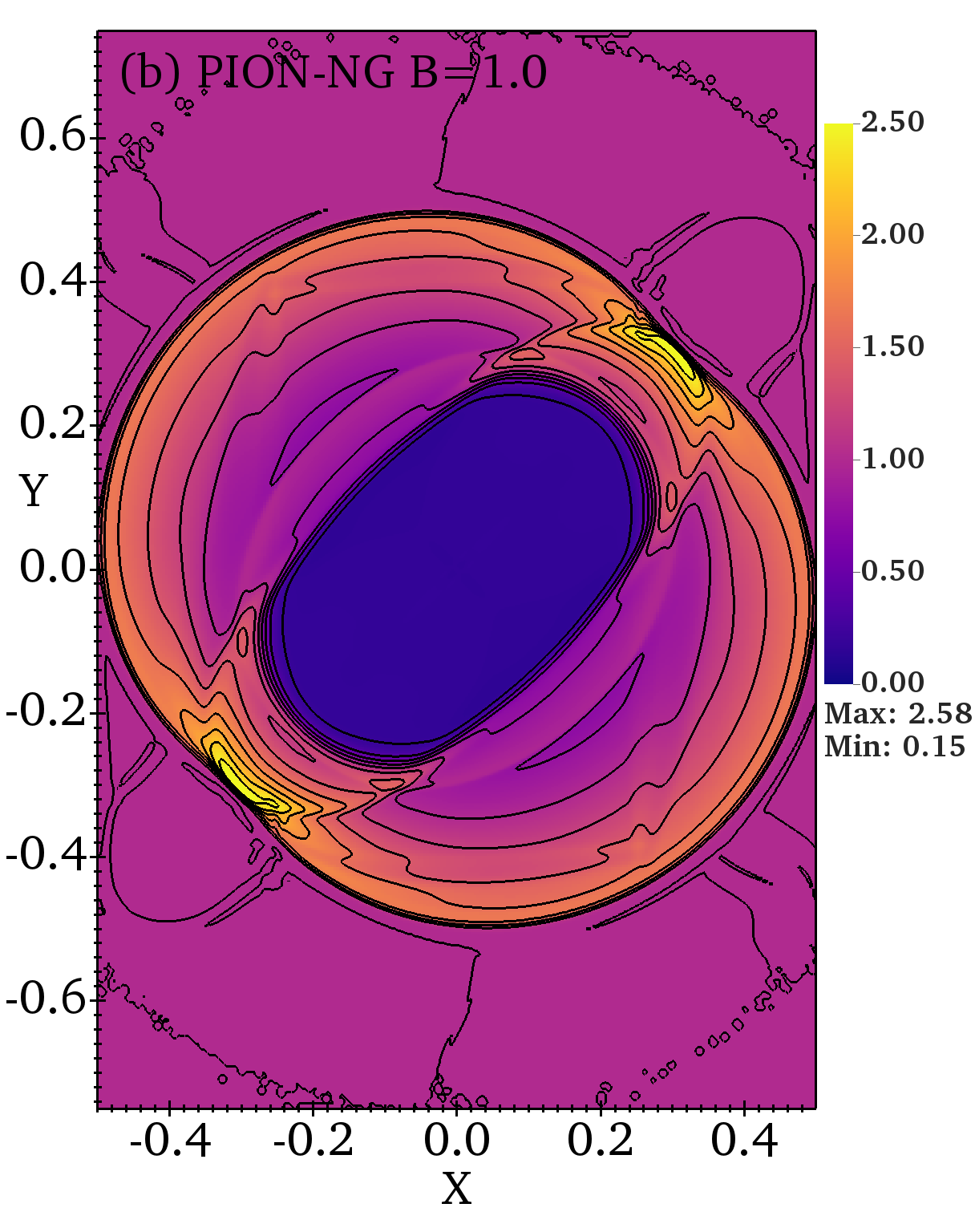}
\includegraphics[width=0.32\textwidth]{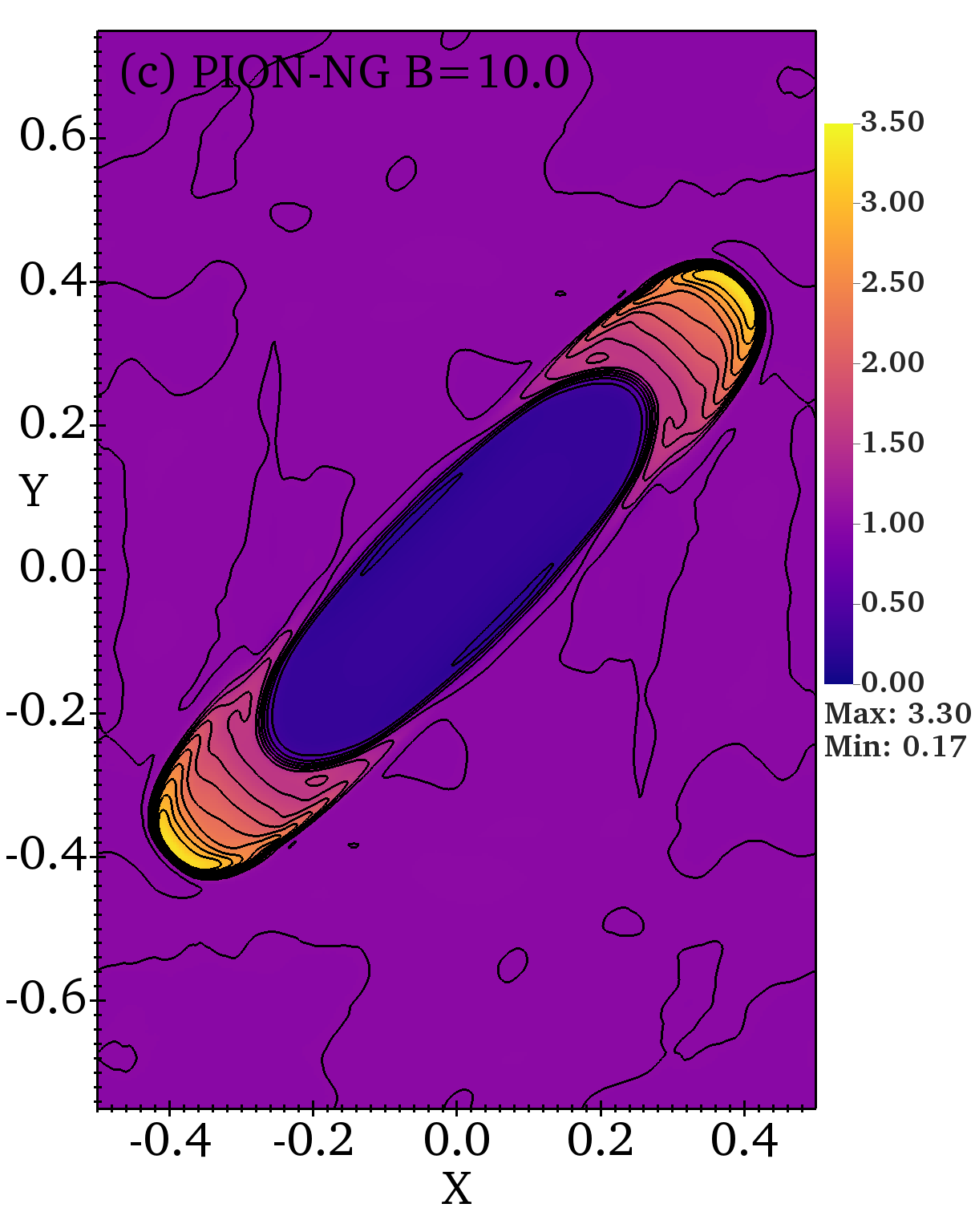}
\caption{
  As Fig.~\ref{fig:mhdblast-ug} but using a nested grid with 2 levels, centred on the origin.
  }
\label{fig:mhdblast-B}  
\end{figure*}

For comparison, the results using a nested grid with two levels, centred on the origin, for all three magnetic field strengths are plotted in Fig.~\ref{fig:mhdblast-B}, using the same colour scale and contours as Fig.~\ref{fig:mhdblast-ug}.
The inner part of the solution is solved  more accurately, as expected because of the higher resolution, but the most obvious difference from Fig.~\ref{fig:mhdblast-ug} is that the refinement boundary has left an imprint in the form of waves trailing the forward shock in an approximate parallelogram shape in the middle panel ($B=1$).
For the case where $B=0.1$ this grid-refinement-boundary effect is much less noticeable than for $B=1$.
The effect is almost absent in HD calculations, similar to panel (a) in Fig.~\ref{fig:mhdblast-B} for which $p_\mathrm{g}$ in the hot region is much larger than the magnetic pressure, $p_\mathrm{m}\equiv B^2/2$ (in these units), i.e.~the plasma $\beta\equiv p_\mathrm{g}/p_\mathrm{m} \gg1$.
In panel (c) of Fig.~\ref{fig:mhdblast-B} the grid effect is visible in that the blastwave is no longer mirror-symmetric along its axis, but the error is not worse than in panel (b).
In panel (b) the initial conditions have $\beta\ll1$ in the undisturbed medium, and $\beta=2$ in the hot region of the initial conditions.
Panel (c) has $\beta<1$ in all regions of the initial conditions.

\begin{figure*} 
\centering
\includegraphics[width=0.32\textwidth]{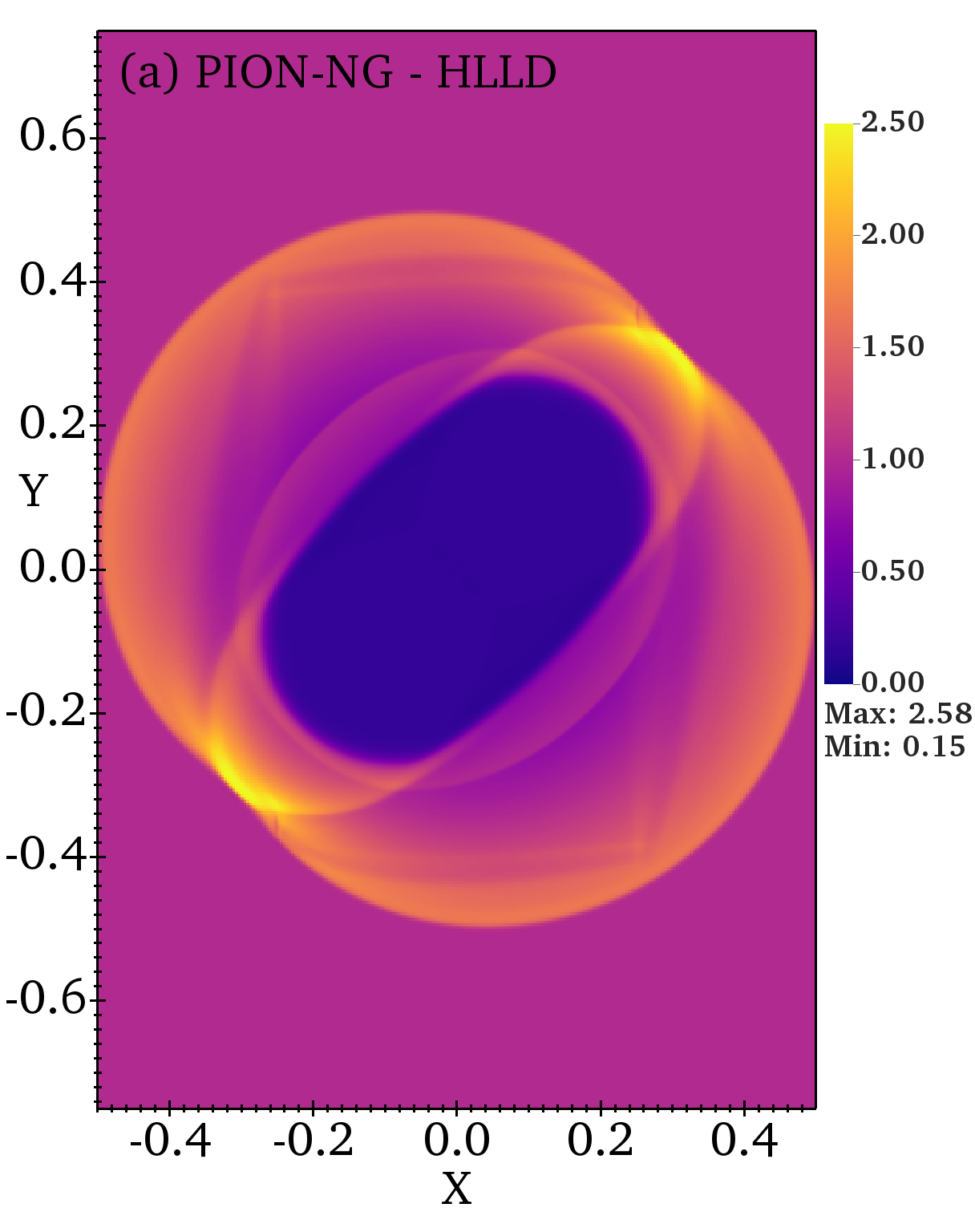}
\includegraphics[width=0.32\textwidth]{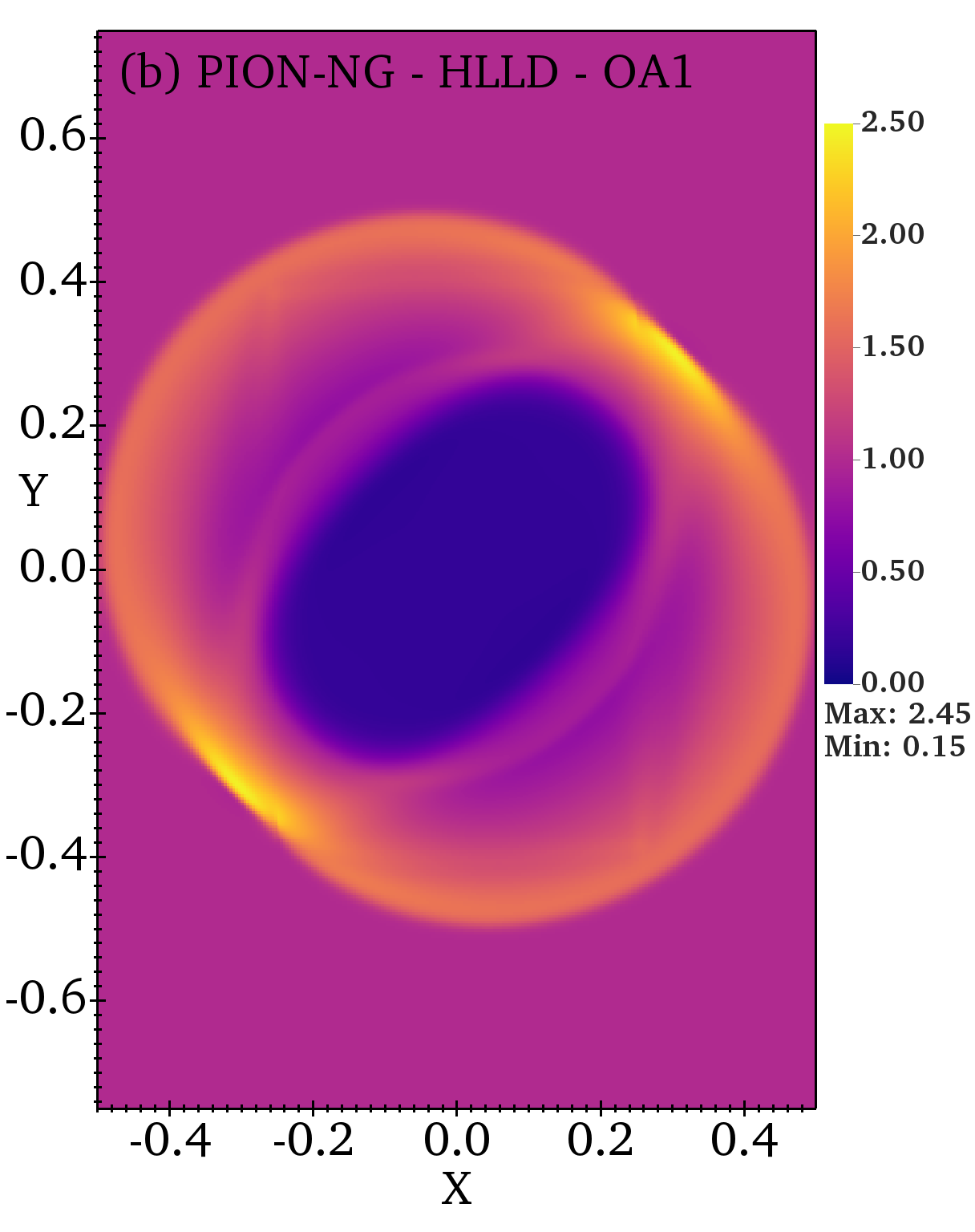}
\includegraphics[width=0.32\textwidth]{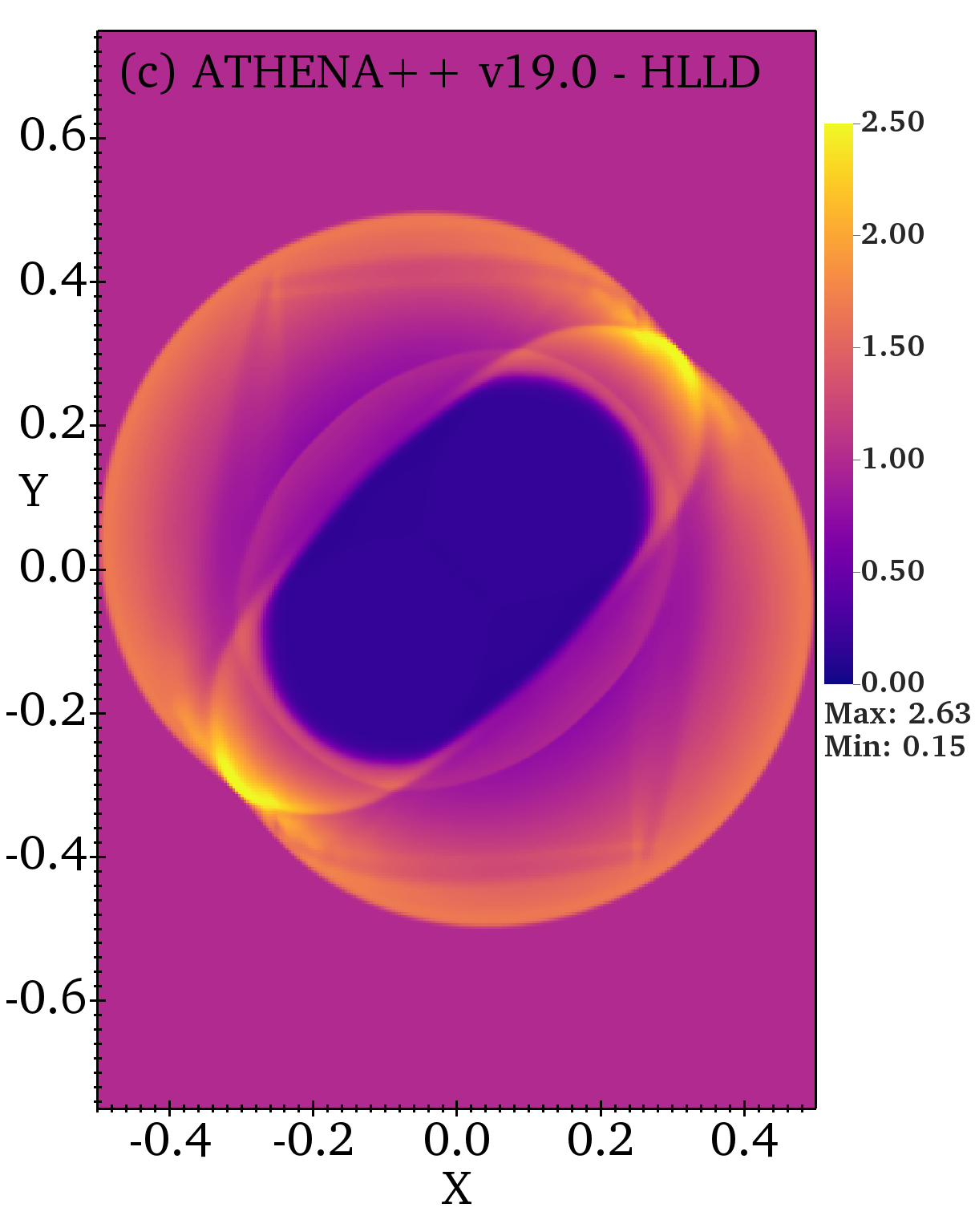}
\caption{
  The MHD Blastwave test calculation in 2D using static mesh-refinement and 2 levels, for the case $B=1$, with the finer level centred on the origin and extending to $x=\pm0.25$ and $y=\pm0.375$ calculated using (a) \textsc{pion} with the second-order scheme, (b) \textsc{pion} with the first-order scheme, and (c) \textsc{Athena++} verson 19.0.
  Gas density is plotted at $t=0.2$ using the same colour scale, as indicated.
  }
\label{fig:mhdblast-ng}  
\end{figure*}

The medium-field case ($B=1$) was investigated in more detail by running with a first-order integration scheme and using \textsc{Athena++} version 19.0 \citep{StoTomWhi20} with the same resolution and static mesh-refinement.
Results are plotted in Fig.~\ref{fig:mhdblast-ng} using \textsc{pion} with the HLLD solver and the second-order scheme (a), the HLLD solver and the first-order scheme (b) and using \textsc{Athena++} (c).
Both codes show basically the same result, with small differences because \textsc{Athena++} uses a different integration scheme, especially with regard to integration of the magnetic field.
The first-order scheme is more diffusive for all waves, and so features are not as sharp, but the imprint of the refinement boundary remains.

The error appears to be related to an inconsistency introduced in the discretized equations when an oblique shock crosses the refinement boundary, for strongly magnetized plasma with $\beta\lesssim1$.
Significant effort was made to characterise and eliminate the issue, but no satisfactory solution was found.
The features are basically the same whether one uses ideal MHD, ideal MHD with Powell source terms, or ideal MHD with the GLM-MHD divergence cleaning method, although there are small differences in each case.
Removing the \citet{BerCol89} flux correction also does not remove the error (or change the solution to any great extent).
It is worth noting that \textsc{Athena++} uses a constrained transport scheme to eliminate $\bm{\nabla\cdot B}$ \citep{GarSto05}, completely different from the methods used here, and so the issue is not caused by the divergence-cleaning implementation.

Features introduced to the flow by waves crossing refinement boundaries are also discussed in \citet[][figs.~38 and 39]{StoTomWhi20}, where a simulation of the relativistic and magnetized Kelvin-Helmholtz instability is run with a uniform grid and with AMR.
There are noticeable differences, with the uniform-grid simulation showing much smoother and more symmetric flow.
It is unavoidable that refinement boundaries introduce some numerical errors to the solution, but the results in Fig.~\ref{fig:mhdblast-ng} appear to be a worst-case scenario in terms of the refinement boundary having an effect on the overall solution.
In particular, the results presented below for 3D simulation of magnetized bow shocks and H\,\textsc{ii} region expansion have almost indiscernible artefacts in the flow variables at refinement boundaries, even though in some cases $\beta\sim1$ in the post-shock medium.

\subsection{Expansion of a D-type Ionization Front}
\label{sec:ionization}

The accuracy of \textsc{pion} in tracking ionization fronts propagating at various speeds from D-type to R-type across a uniform grid was presented in \citet{Mac12}.
The implementation in the nested-grid is very similar, in particular the calculation of optical depths and timestepping restrictions, applied on a level-by-level basis.
In this section we calculate the D-type expansion of an H\,\textsc{ii} region using the parameters and initial conditions of the ``Early phase'' calculation of \citet{BisHawWil15}.

A source of Lyman-continuum photons emits at a rate $Q_0=10^{49}\,\mathrm{s}^{-1}$ from the origin, into a uniform neutral ISM of density $\rho_0=5.21\times10^{-21}\,\mathrm{g\,cm}^{-3}$ composed purely of hydrogen.
The gas has a two-temperature-isothermal equation of state, where neutral gas has temperature $T_0=100$\,K and sound speed $c_0\approx0.91\,\mathrm{km\,s}^{-1}$, and ionized gas has temperature $T_\mathrm{i}=10^{4}$\,K and sound speed $c_\mathrm{i}\approx12.85\,\mathrm{km\,s}^{-1}$, with the temperature in partially ionized cells linearly interpolated between these values.
The Str\"omgren radius is $R_\mathrm{s}=0.314$\,pc ($0.97\times10^{18}$\,cm), and the stagnation radius (where the H\,\textsc{ii} region is in pressure equilibrium with the undisturbed ISM) is $R_\mathrm{stag}=(c_\mathrm{i}/c_0)^{4/3}R_\mathrm{s} = 10.72$\,pc, approximately $34\times$ larger.

The Early-Phase test calculation in \citet{BisHawWil15} was evolved for $0.141$\,Myr, on a grid that extends to $4R_\mathrm{s}\approx3.9\times10^{18}$\,cm in each dimension.
We evolve this solution out to a larger extent of $8\times10^{18}$\,cm in each dimension, so that we can test the adaptive resolution effectively (the D-type expansion only begins at $r\approx R_\mathrm{s}$ and the effects of numerical resolution only become clear once a shock and swept-up shell can form).
We calculate the radius of the ionization front, $R_\mathrm{IF}$, from the ionized volume as follows
\begin{equation}
R_\mathrm{IF} = \left( \frac{3}{4\pi} \sum_i y_i(\mathrm{H}^+) V_i \right)^{1/3} \;,
\end{equation}
where the sum is over all cells, $i$, on the domain with cell volume $V_i$ that have H$^+$ fraction $y_i(\mathrm{H}^+)>0.01$.
This allows a consistent solution even when the shocked shell becomes distorted in multi-dimensional calculations \citep[cf.][]{BisHawWil15}.

\begin{figure} 
\centering
\includegraphics[width=0.4\textwidth]{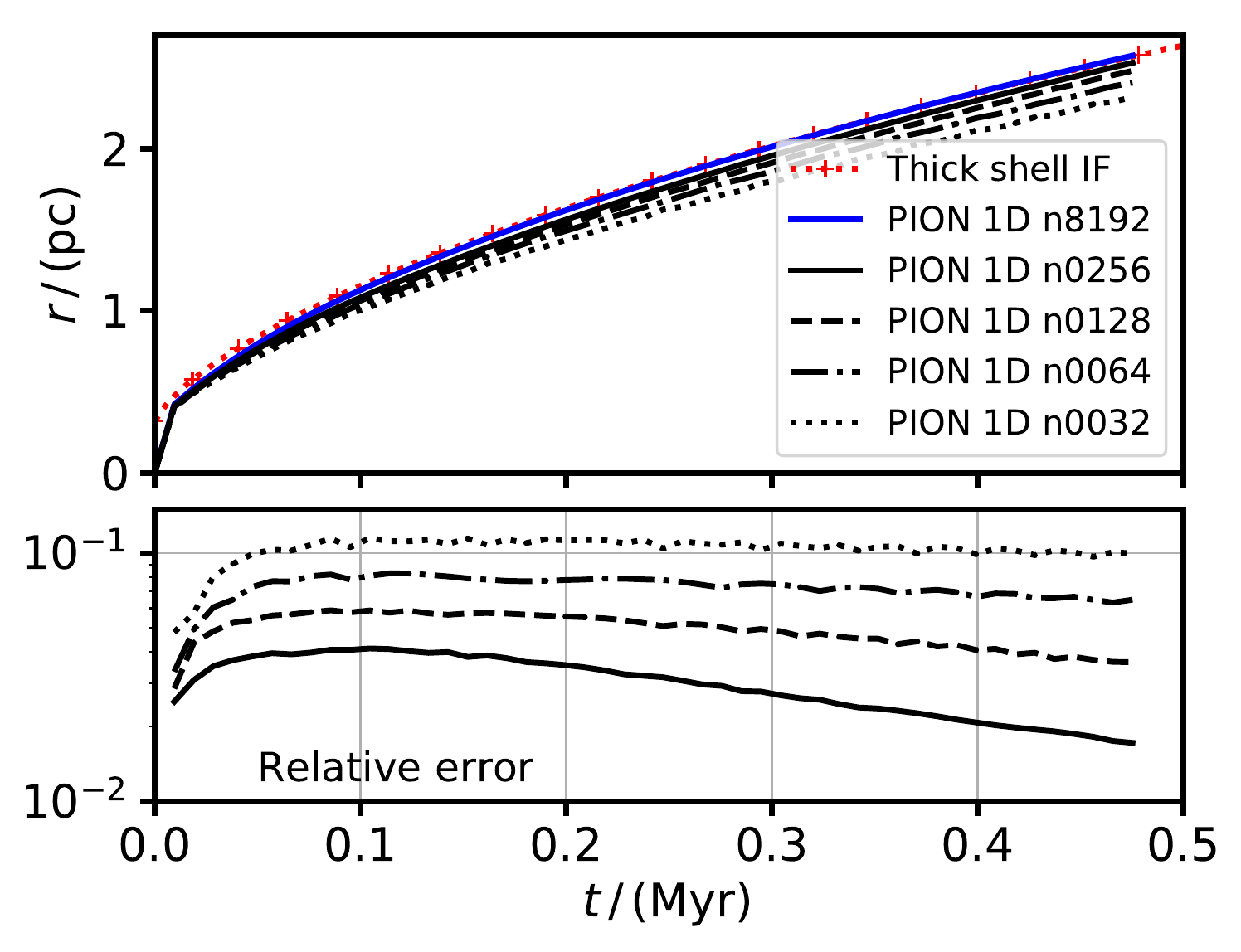}
\caption{
  Expansion of an H\,\textsc{ii} region simulated in 1D with spherical symmetry, showing how the accuracy improves with increasing numerical resolution.
  \emph{Upper panel:} the radius of the ionization front as a function of time, for different resolutions, compared with the `thick-shell' solution of \citet{WilBisHaw18}.
  \emph{Lower panel:} the relative difference between the radius for a given resolution and the radius when run with 8192 grid cells (absolute value).  Line styles are the same as in the upper panel.
  }
\label{fig:DTE1D}  
\end{figure}

The results for a series of 1D calculations with different spatial resolution are shown in Fig.~\ref{fig:DTE1D}, including the relative difference between the low resolution calculations and the calculation with 8192 cells.
Here we show results using uniform grids with 32, 64, 128, 256, and 8192 cells on the domain $r\in[0,8\times10^{18}]$\,cm.
These are compared with the `thick-shell' solution of \citet{WilBisHaw18}, which was found to be an excellent analytic solution for the early phase of expansion.
Both the ionization front and shock front are effectively discontinuities, for which the order of accuracy is reduced to first order by the slope limiter, and so the relative error of the solution improves approximately proportional to the resolution.

\begin{figure} 
\centering
\includegraphics[trim = 0mm 9mm 0mm 2mm, clip, width=0.4\textwidth]{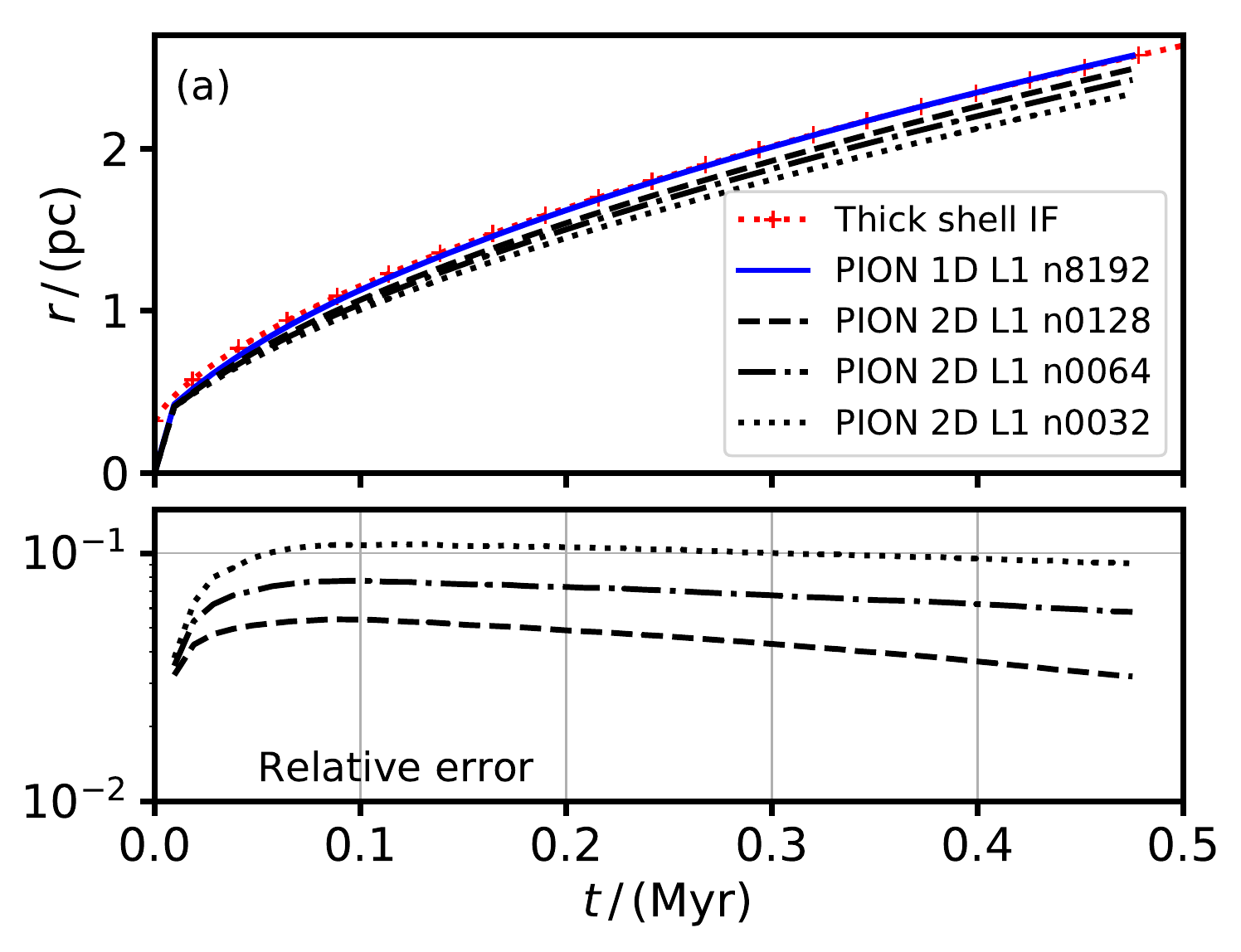}
\includegraphics[trim = 0mm 9mm 0mm 2mm, clip, width=0.4\textwidth]{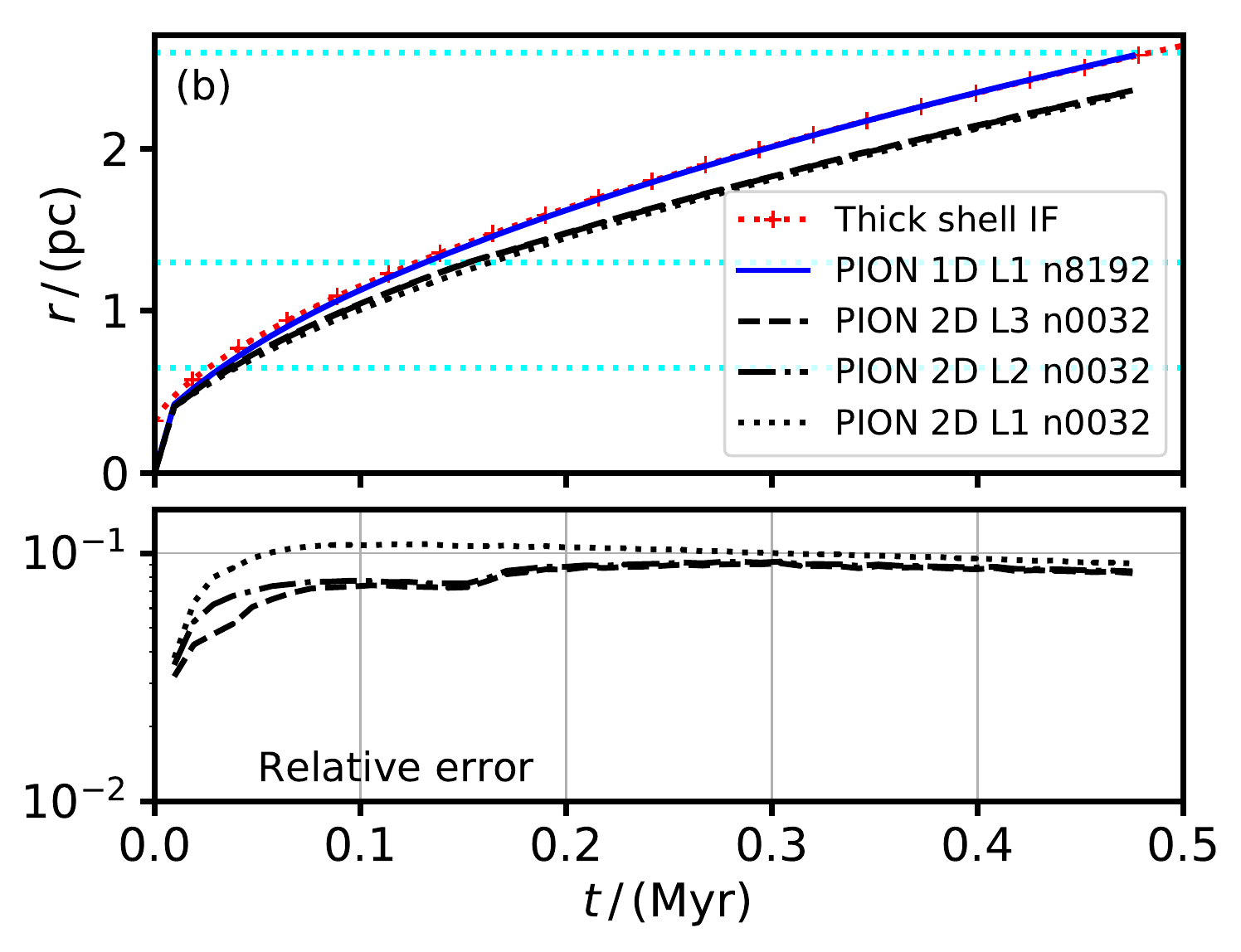}
\includegraphics[trim = 0mm 9mm 0mm 2mm, clip, width=0.4\textwidth]{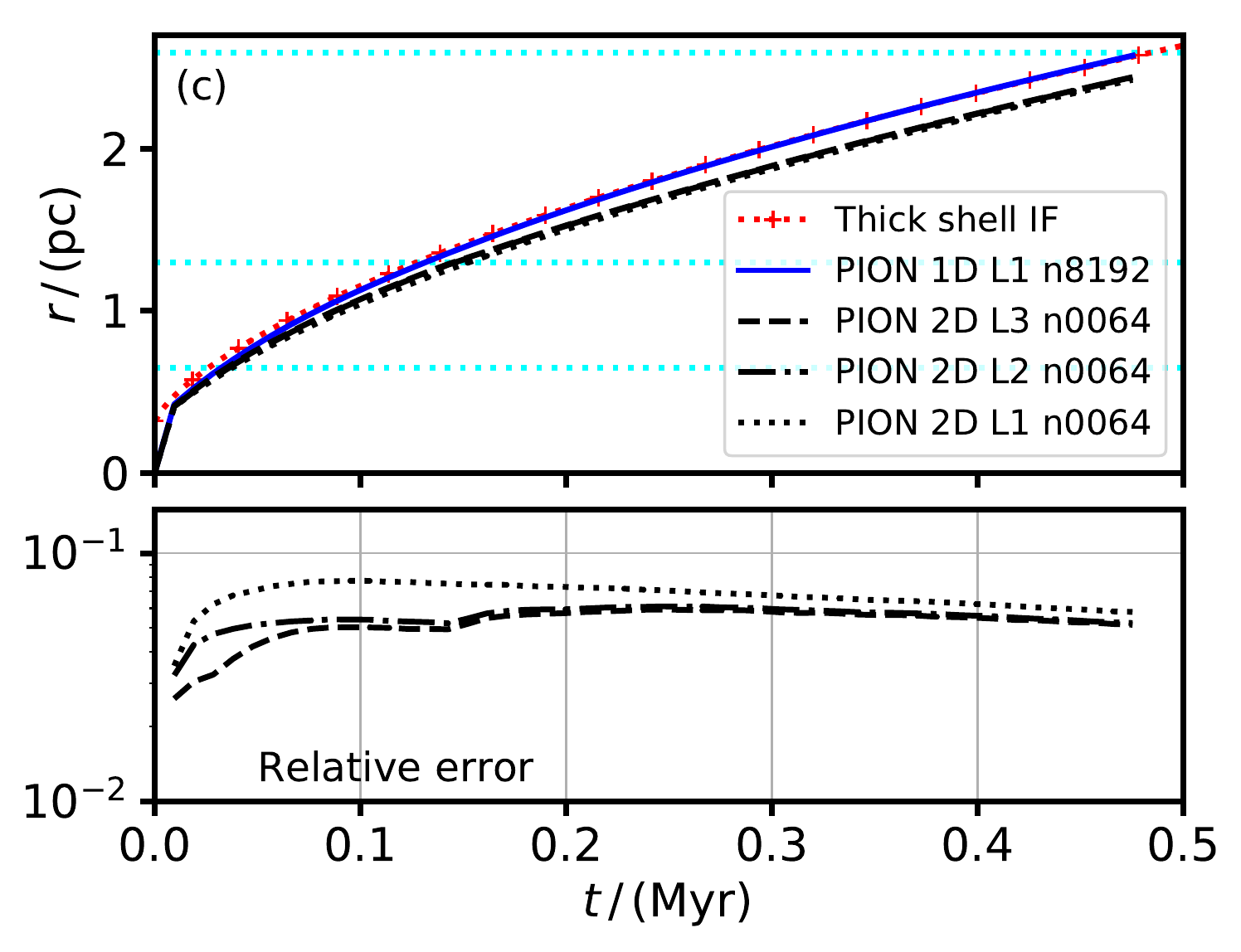}
\includegraphics[width=0.4\textwidth]{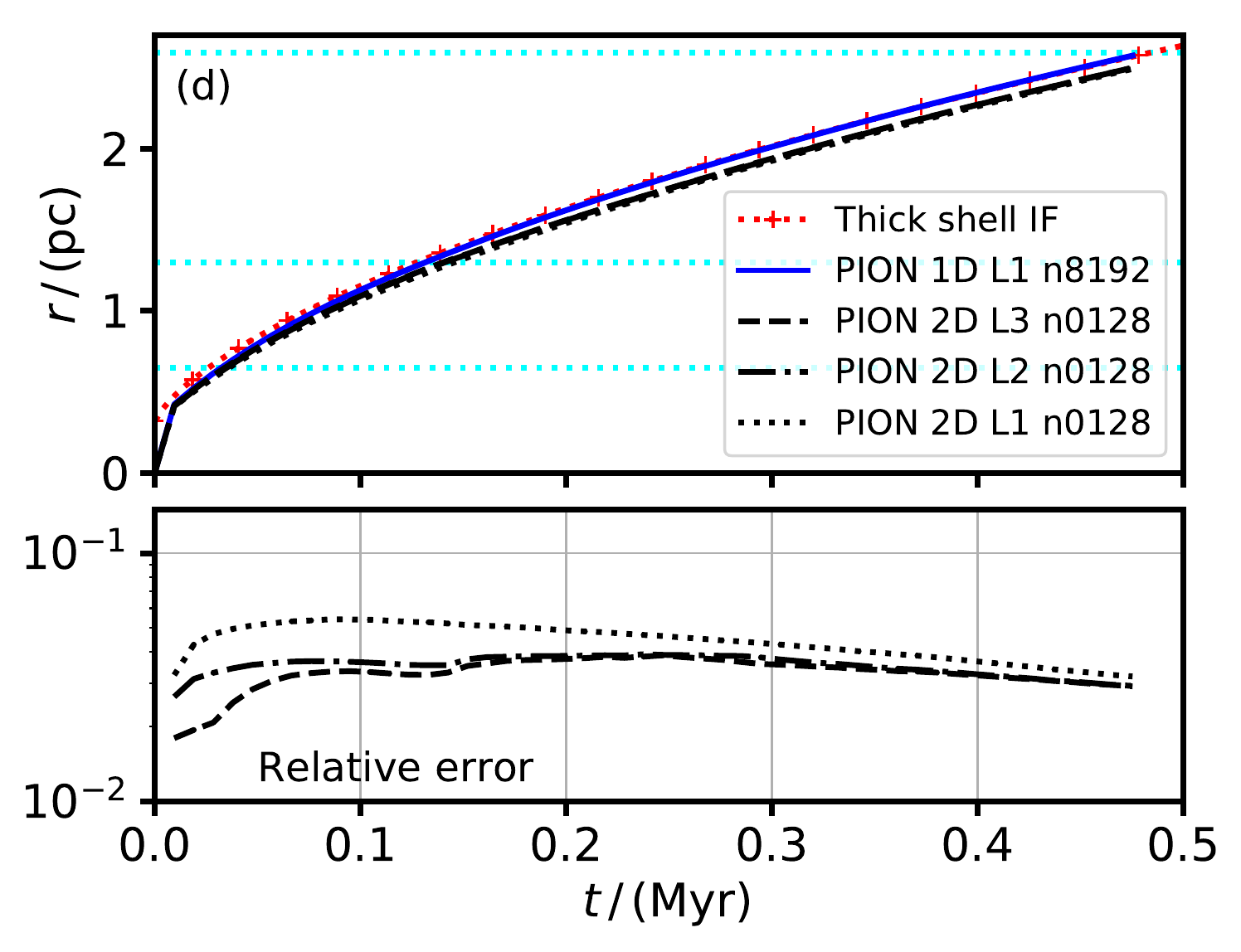}
\caption{
  Each panel is as for Fig.~\ref{fig:DTE1D}, except now the calculation is for a 2D simulation of one quadrant with different grid resolutions and numbers of refinement levels.
  \textbf{(a)} uniform-grid results with different resolutions.
  \textbf{(b)} grid resolution $32^2$ with 1, 2, and 3 refinement levels.
  Here the cyan horizontal lines show the boundaries of the 3 levels.
  \textbf{(c)} grid resolution $64^2$ with 1, 2, and 3 refinement levels.
  \textbf{(d)} grid resolution $128^2$ with 1, 2, and 3 refinement levels.
  }
\label{fig:DTE2D}  
\end{figure}

Fig.~\ref{fig:DTE2D} shows the same information but for 2D simulations with uniform and nested grids with up to 3 levels of refinement and  different grid resolutions (per level) from $32^2$ to $128^2$.
The H\,\textsc{ii} region crosses the finest level-boundary at $t\approx0.05$\,Myr, and the second level-boundary at $t\approx0.15$\,Myr (shown by the cyan dotted lines).
Again the solutions are compared with the 1D calculation using 8192 cells.
The accuracy of the 2D uniform-grid solutions is almost the same as in 1D with the same resolution, as expected.
For simulations with refined grids the solution is always better than the unform-grid solution at the same resolution, but approaches this solution at large radius.
When the ionization front is within a refined grid, the accuracy of the solution is comparable to that of a uniform grid with the same cell-diameter.
After the ionization front crosses a refinement boundary the solution accuracy begins to degrade to that of the equivalent uniform grid with the coarser resolution.
The 3D results, simulated on one octant with reflection symmetry, are indistinguishable from 2D calculations at the same resolution, and are therefore not shown.

\begin{figure} 
\centering
\includegraphics[width=0.4\textwidth]{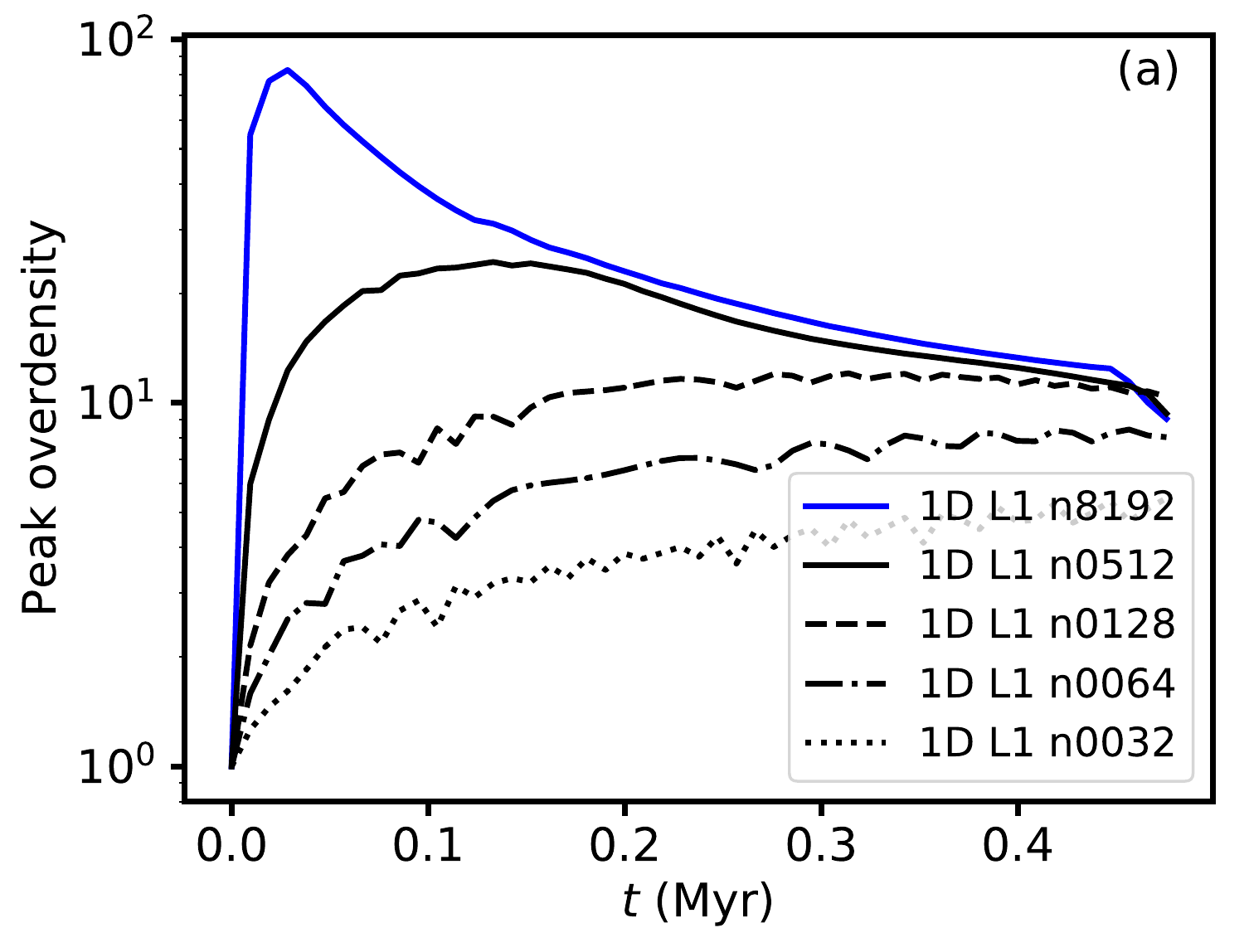}
\includegraphics[width=0.4\textwidth]{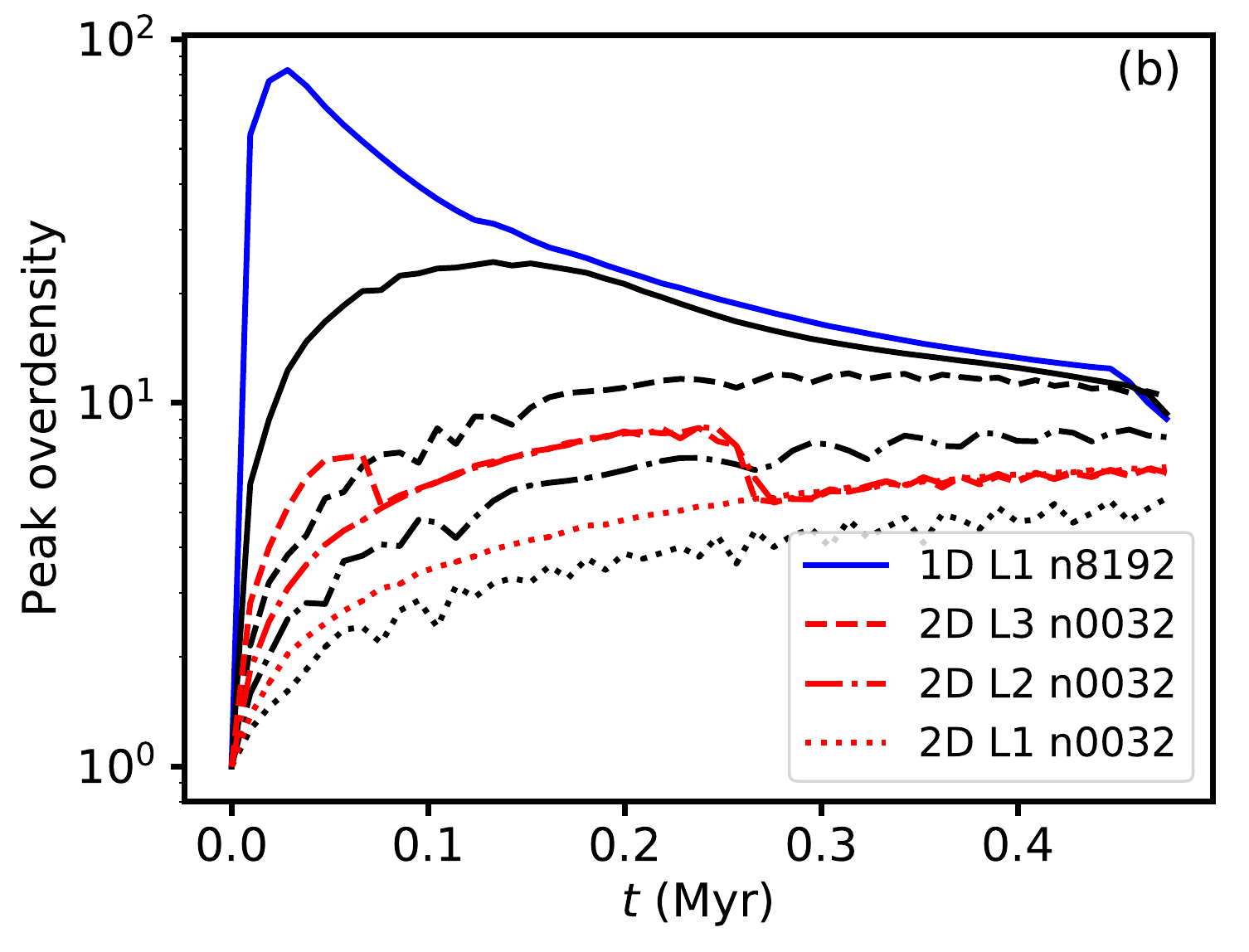}
\caption{
  \textbf{(a)} Maximum shell density as a function of time for 1D simulations of D-type expansion of an H\,\textsc{ii} region, run with differing spatial resolutions from 32 cells to 8192 cells, as indicated in the legend.
  \textbf{(b)} As above, but with 2D results overplotted in red, for nested grids of $32^2$ cells with 1, 2, and 3 levels, as indicated in the legend.
  }
\label{fig:DTE1D_maxd}  
\end{figure}

In all cases the error in ionization-front position is about 3 cell-diameters, comparable to the numerical resolution of the scheme.
This error arises because the shocked shell must be a few cells thick in order to resolve the shock and ionization front, whereas the physical shell thickness is such that the shell is unresolved at all radii for the low-resolution multi-dimensional simulations shown here.
This is shown in Fig.~\ref{fig:DTE1D_maxd} (a), where the maximum overdensity in the shocked shell is plotted as a function of time for various different grid resolulions in 1D simulations, from 32 cells to 8192 cells.
This is a reasonable proxy for whether or not the shell is numerically resolved, although not sufficient to show that all properties of the shell are correct (e.g.~thickness).
The simulations using grids with 32 and 64 cells are clearly unresolved at all radii, whereas the grid with 128 cells approaches the correct peak overdensity for $t>0.4$\,Myr, but is increasingly underresolved at earlier times.
The overdensity decreases sharply for the high-resolution calculation at the last snapshot because the shocked shell is starting to leave the domain.
In Fig.~\ref{fig:DTE1D_maxd} (b) 2D results for grids with $32^2$ cells and 1, 2 and 3 refinement levels are shown.
The results are slightly better than the 1D peak overdensity for an equivalent resolution, but at no stage is the shell resolved.

These results show that the accuracy of the expansion of H\,\textsc{ii} regions for multi-dimensional simulations with mesh refinement is comparable to that of the equivalent 1D simulation with the same spatial resolution.
This means that the dynamics of expanding nebulae can be resolved to approximately the same degree at all stages of expansion.
This has advantages for certain applications such as expanding WR nebulae \citep{FreHenYor06} and Planetary Nebulae.

\section{Applications to stellar-wind bubbles}
\label{sec:winds}

Four examples are presented here of simulations of winds from massive stars: a constant wind from an O star driving a bow shock, the nebula produced by a RSG that evolves on a blue loop and spins up to critical rotation, the nebula produced by a RSG evolving to a WR star when it loses its envelope, and the wind-wind collision of two stars in close proximity to each other.
The simulations are mainly chosen for ease of comparison with the previous literature on these topics.

\begin{figure*}
\includegraphics[width=0.85\linewidth]{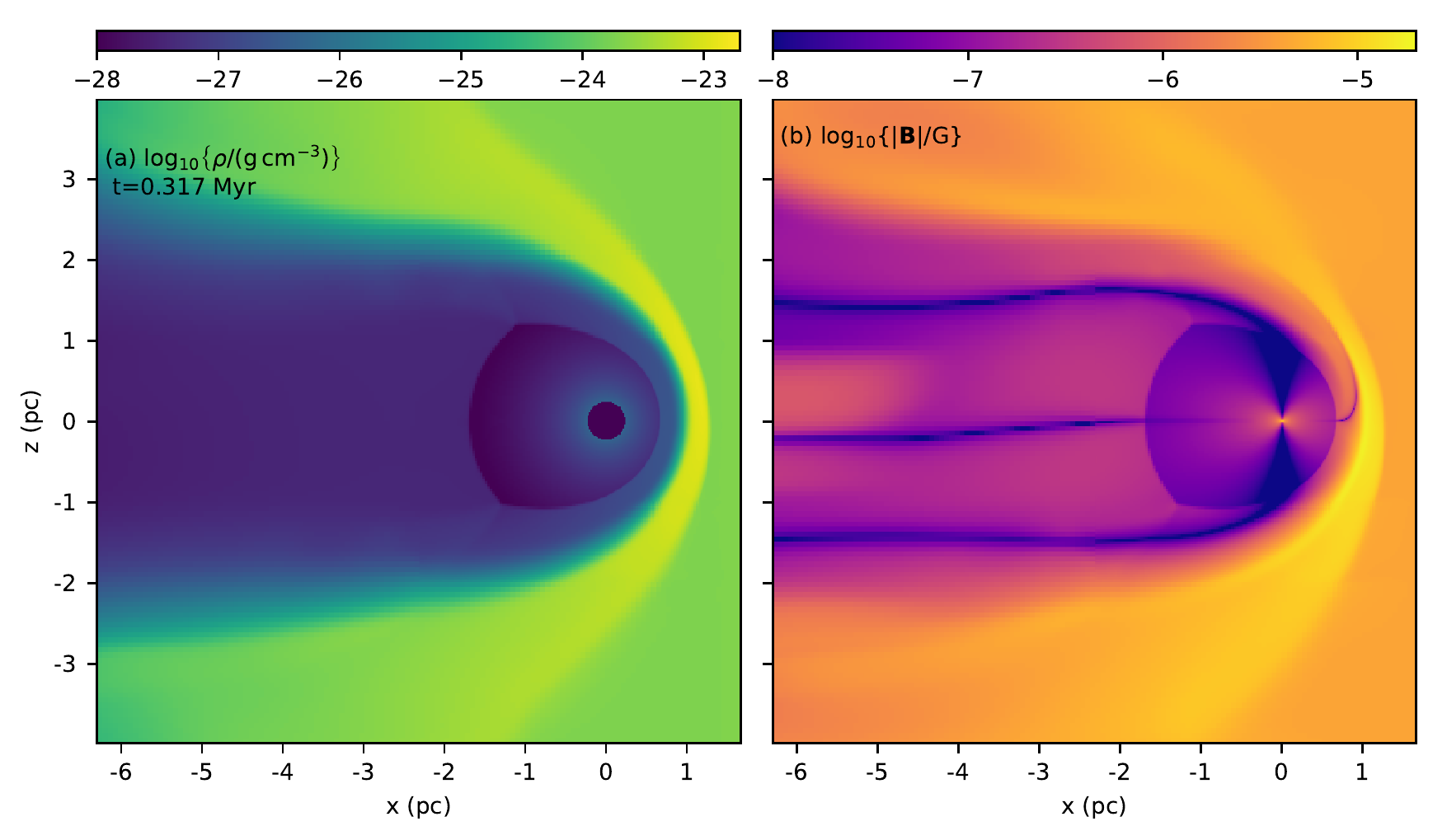}
\caption{\label{fig:b3d_DB}  (a) Gas density, $\log_{10}\left\{\rho/\mathrm{(g\,cm}^{-3})\right\}$, and (b) magnetic field magnitude, $\log_{10}(|\bm{B}|/\mathrm{G})$, in the $x$-$z$ plane through $y=0$ are plotted on a logarithmic scale as indicated, for a 3D MHD simulation of a bow shock produced by a massive star.  The star is at the origin and moving in the $+\hat{x}$ direction.
The magnetic axis of the star is $\hat{z}$, the stellar surface field is $B=10$\,G, and the upstream ISM field is $\bm{B}_0=[4,1,1]\times10^{-6}$\,G.}
\end{figure*}

\subsection{Stellar-wind bubble in 3D with MHD}
\label{sec:wind3d}

\begin{table}
\caption{\label{tab:wind3d}Stellar Wind and ISM parameters for the 3D MHD simulation of a bow shock in section~\ref{sec:wind3d}.}
\begin{center}
\begin{tabular}{ll}
\textbf{Parameter}&\textbf{Value}\\
\hline
Stellar mass-loss rate, $\dot{M}$ & $10^{-7}\,\mathrm{M}_\odot\,\mathrm{yr}^{-1}$ \rule{0pt}{2.6ex} \\
Stellar wind terminal velocity, $v_\infty$ & 2000\,km\,s$^{-1}$ \\
Stellar surface rotation  (equator), $v_\mathrm{rot}$  & 100\,km\,s$^{-1}$ \\
Surface split-monopole field strength, $|\bm{B}|$ & 10\,G \\
Surface temperature, $T_\mathrm{eff}$ & 35\,000\,K \\
\hline
ISM density, $\rho_0$ & $2.0\times10^{-24}\,\mathrm{g\,cm}^{-3}$  \rule{0pt}{2.6ex} \\
ISM pressure, $p_\mathrm{g}$   & $2.9\times10^{-12}\,\mathrm{dyne\,cm}^{-2}$ \\
ISM velocity, $\bm{v}$ & $[-30,0,0] \,\mathrm{km\,s}^{-1}$ \\
ISM B-field, $\bm{B}_0$  & $[4,1,1]\times10^{-6}$\,G \\
\hline
\end{tabular}
\end{center}
\end{table}

Here we introduce the standard wind module in \textsc{pion}, using a spherically symmetric, constant (in time), hypersonic wind from a slowly rotating star (wind type 1 from section \ref{sec:stellar-wind}).
We demonstrate for the first time with \textsc{pion} the implementation of a magnetized wind from a rotating star, a preliminary version of which was presented in \citet{MacGreMou20}.

\subsubsection{Simulation setup and initial conditions}
We set up a 3D MHD simulation of the bow shock produced by a star moving with $v_\star=30\,\mathrm{km}\,\mathrm{s}^{-1}$ through the diffuse ISM, with 3 grid levels.
The parameters of the stellar wind and the ISM are in Table~\ref{tab:wind3d}, and are typical of a late O star of mass $M\approx30\,\mathrm{M}_{\odot}$.
The ISM pressure corresponds to a gas temperature of $T\approx7\,800$\,K, appropriate for photoionized gas.
The standoff distance of the bow shock is defined by
\begin{equation}
R_\mathrm{SO} \equiv \sqrt{\frac{\dot{M}v_\infty}{4\pi\rho_0 (v_\star^2+c_s^2)}} \;,
\end{equation}
where $\dot{M}$ is the mass-loss rate of the star, $v_\infty$ is the terminal wind velocity, $\rho_0$ is the background ISM density, $v_\star$ is the space velocity of the star, and $c_s\equiv \sqrt{\gamma p_\mathrm{g}/\rho_0}$ is the sound speed of the background medium ($\gamma$ is the adiabatic index of the gas).
For these parameters, $R_\mathrm{SO}  \approx0.70$\,pc, and we expect to find the wind termination shock with this separation from the star.
The location of the contact discontinuity and the forward shock depend on the compressibility of the gas through the shocks and on radiative cooling efficiency \citep{SchBaaFic20}.

The radiative heating and cooling prescription follows \citet{GreMacHaw19} and the shocked wind is almost adiabatic whereas the shocked ISM is effectively isothermal.
The ISM is assumed to be fully ionized, and to consist of hydrogen with mass fraction 0.714, and helium with abundance one tenth that of H by number, and Solar metallicity \citep[cf.][]{GreMacHaw19}.
There is no non-equilibrium chemistry, and cooling flag 8 is used, corresponding to photoionized gas that is heated by photoionizations.
The heating rate is the product of the local recombination rate $n_e n_\mathrm{H} \alpha_\mathrm{B}$ and a heating per photoionization of 5\,eV, appropriate for an O star.
The recombination rate, $\alpha_\mathrm{B}$, is from \citet{Hum94}.
Cooling is the sum of Bremsstrahlung assuming H and He are fully ionized \citep{Hum94, RybLig79},  and metal-line cooling.
For metal-lines we take the maximum of the \citet{WieSchSmi09} collisional ionization-equilibrium (CIE) cooling curve (metals only) and equation A9 from \citet{HenArtDeC09} for forbidden-line cooling from photoionized CNO ions, which would not be present in CIE because the relevant elements would be at a lower ionization stage at $10^4$\,K.

The simulation was initialised with a coarse grid of $128^3$ grid cells and size $2.4576\times10^{19}$\,cm ($\approx 8$\,pc) in each dimension (each cell has diameter $\Delta x=1.92\times10^{17}$\,cm).
The simulation extents in the $x$-direction are $x\in[-19.432\times10^{18},5.144\times10^{18}]$\,cm, and $\{y,z\}\in[-12.288\times10^{18},12.288\times10^{18}]$\,cm; the focal point of the nested grids is at $[5.144\times10^{18},0,0]$\,cm; and the star is at the origin.
We added two levels of refinement to the coarse grid, giving a finest level cell-diameter $\Delta x= 4.8\times10^{16}$\,cm.
We set the wind inner boundary radius to $9.6\times10^{17}$\,cm, corresponding to 20 grid cells on the most refined level.

\begin{figure*}
\includegraphics[width=0.85\linewidth]{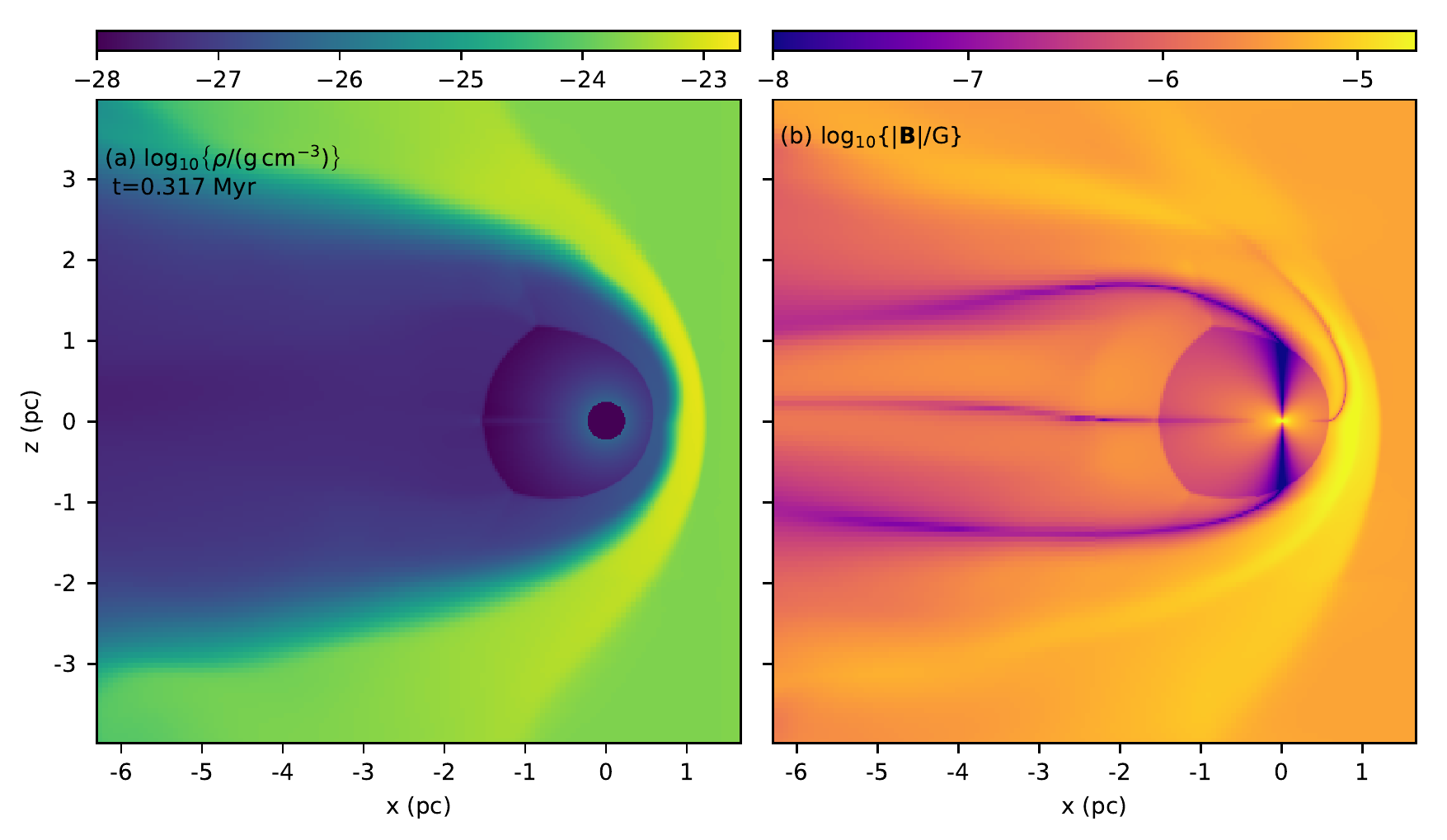}
\caption{\label{fig:b3d_B100} As Fig.~\ref{fig:b3d_DB} but for a simulation with a $10\times$ stronger stellar surface field of $B=100$\,G.}
\end{figure*}

We used the standard second-order-accurate integration scheme (in time and space) described above, together with the HLL MHD Riemann solver, to evolve the simulation for $10^{13}$\,s, about 1.2 times the crossing time for the star to cross the simulation domain, and nearly 14 times the dynamical time-scale of the bowshock ($\tau_\mathrm{dyn}=R_\mathrm{SO}/v_\star\approx7.2\times10^{11}$\,s).
The simulation takes about 3\,300 CPU-hours to run to completion.
For reference, a simulation with $256^3$ and 3 levels would take about $16\times$ longer (50\,000 CPU-hours) and $512^3$ with 3 levels would take $\approx$800\,000 CPU-hours.

\subsubsection{Results}
Results at $t=10^{13}$\,s are plotted in Fig.~\ref{fig:b3d_DB}, with gas density in panel (a) and the magnitude of the magnetic field in panel (b).
A slice through the 3D domain at $y=0$ is shown, with the star at the origin.
The inner 75 per cent of the wind boundary region is set to have very low density, hence the dark circle in panel (a) around the origin.
The typical features of the bow shock are seen: the termination shock of the wind, closer to the star in the upstream direction because of the asymmetric ram pressure of the ISM; the strong contact discontinuity where density changes by a factor of $\gtrsim10^3$ and the shocked ISM in the bow-shaped arc.
The rotation and magnetic axis of the star is $\hat{z}$; this has no effect on the gas density profile because the field is relatively weak compared with the ram pressure of the wind, but it can be seen clearly in panel (b).

The magnetic structure of the wind is the classical \citet{Par58} Spiral, where the stellar field is wound up by the star's rotation and drops off as $\left|\bm{B}\right|\propto r^{-1}$ near the equator and $\left|\bm{B}\right|\propto r^{-2}$ at the poles.
The field switches direction at the equator, producing a current sheet similar to the Heliospheric Current Sheet in the Solar wind.
Downstream the current sheet is swept into the wake behind the star and remains near the $z=0$ plane, but in the upstream direction this sheet is swept to large $z$ and back with the flow of the bow shock along the contact discontinuity between wind and ISM.
The weakly magnetized polar regions are also swept back in the shocked wind to the wake behind the star.
These are the typical features also seen in MHD simulations of the Heliosphere \citep[e.g.][]{PogZanOgi06}, but on larger scales because of the stronger wind.
In general the magnetic and rotation axes may be misaligned (as is the case for the Sun), but this requires a significantly more complicated inner boundary condition  \citep[e.g.][]{PogSueBor13, DalSteUdD19} and is deferred to future work.

The changing resolution is most clearly visible at the contact discontinuity, for which the thickness of the transition layer is mediated by numerical diffusion and therefore scales with the cell diameter.
For the HLL solver used here the transition is spread over approximately 10-15 cells, because the Riemann solver does not contain a contact discontinuity.
Shocks are resolved by 2-3 cells, by contrast, and so the effect of resolution is not as obvious.
Artefacts such as reflected waves introduced at the resolution boundaries are not visible in Fig.~\ref{fig:b3d_DB} in the way that they were for the MHD blastwave in Fig.~\ref{fig:mhdblast-ng}; the resolution effects are primarily related to the numerical diffusion.

The shocked ISM is asymmetric in the sense that the bowshock is thinner and has higher density in the upper half-plane, although the effect is weak.
This is because the ISM magnetic field is closer to the shock normal direction in the upper half-plane than the lower half-plane, and so the magnetic field is less compressed through the forward shock and hence provides less pressure support.
The angle between the shock normal and the star's velocity vector is also larger in the upper half-plane; a geometric measurement of the symmetry axis of the bow shock would therefore be a (in this case only slightly) biased estimator of the relative velocity between star and ISM.

\subsubsection{Stronger stellar magnetic field}
A simulation was also run with a stellar magnetic field 10 times stronger, and the results are plotted in Fig.~\ref{fig:b3d_B100} in the same way as for Fig.~\ref{fig:b3d_DB}.
Such a surface field (100\,G) is allowed by observational upper limits for most O stars \citep{FosCasSch15}, although here the Alfv\'enic Mach number of the wind, \begin{equation}
\mathcal{M}_\mathrm{A}\equiv v_\infty \sqrt{4\pi\rho}/ \left|\bm{B}\right| \;,
\end{equation}
has a value $\mathcal{M}_\mathrm{A}\approx10$ (in the freely expanding wind this is independent of radius near the equator because both $\sqrt{\rho}$ and $\left|\bm{B}\right|$ decrease as $r^{-1}$).
An Alfv\'enic Mach number much smaller than this would lead to a non-spherically symmetric wind and would require a more complicated inner boundary condition.

Here there is some accumulation of wind material at the equatorial current sheet (panel (a) of Fig.~\ref{fig:b3d_B100}), on account of the reduced magnetic pressure in this region compared with the neigbouring regions.
The contact discontinuity also has a feature near $z=0$ in the upstream direction, apparently from the sweeping of the current sheet to the upper half-plane.
This could be related to similar features seen in the Heliosphere for ideal MHD simulations \citep{WasTan01}, for which a deeper understanding requires a multi-fluid description of the flow and/or kinetic theory \citep{PogZanOgi06}.
This shows approximately where we expect to reach the limits of applicability of our imposed boundary condition and the single-fluid approximation, i.e., as long as $\mathcal{M}_\mathrm{A}\gtrsim10$ the wind prescription is reasonable.

\begin{figure} 
\centering
\includegraphics[width=0.45\textwidth]{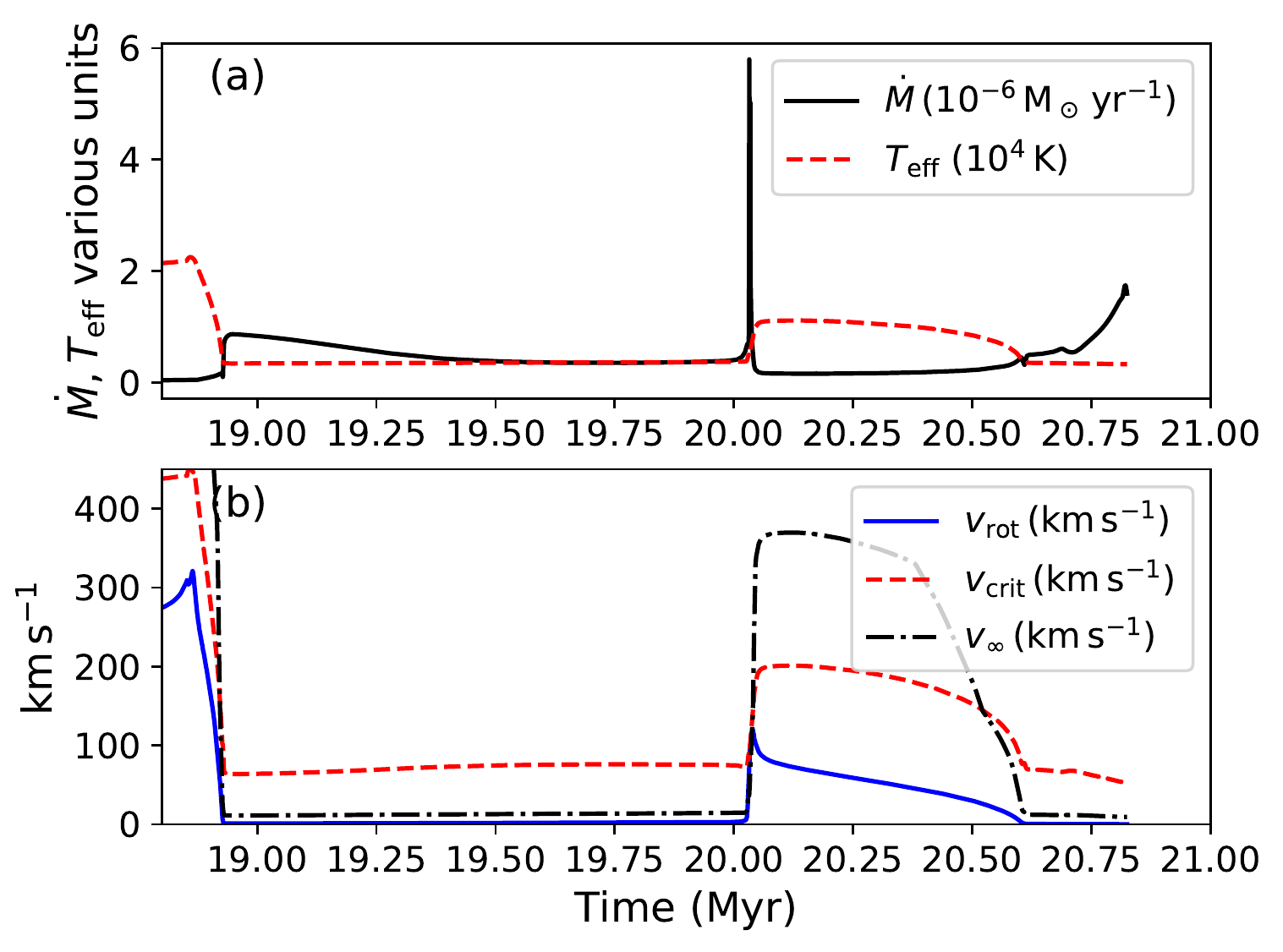}
\caption{
  Wind properties for the simulation in Section~\ref{sec:chita}: (a) Mass-loss rate, $\dot{M}$, and Effective Temperature, $T_\mathrm{eff}$, and (b) rotation velocity, $v_\mathrm{rot}$, critical velocity, $v_\mathrm{crit}$, and wind terminal velocity, $v_\infty$, as a function of time (since the birth of the star).
  The RSG phase begins at 18.9\,Myr, the phase of critical rotation at 20.035\,Myr as the star evolves to hotter surface temperatures on a blue loop.
  }
\label{fig:f12b}    
\end{figure}

\subsection{Time-varying stellar wind for a rotating star}
\label{sec:chita}

\citet[][hereafter CVL08]{ChiLanVan08} studied the formation of ring nebulae around blue supergiants (BSGs) as a result of single star evolution.
A RSG embarks on a blue loop, spins up to critical rotation as a result of contraction of the envelope, ejects an equatorially enhanced wind, resulting in a ring nebula around a BSG.
We use their work as a test calculation against which to compare results obtained with \textsc{pion} using the latitude-dependent and time-varying wind module (type 2 in section \ref{sec:stellar-wind}).
For this test calculation we use the stellar evolution model F12B calculated by \citet{HegLan00} and whose circumstellar medium was simulated by CVL08.

\subsubsection{Stellar evolution calculation and initial conditions}
Fig.~\ref{fig:f12b} shows the wind evolution for the evolutionary track F12B, starting at the end of the main sequence at 18.8\,Myr, through a RSG phase lasting just over 1\,Myr, a blue loop lasting $\approx0.5$\,Myr and finishing with a second RSG phase lasting $\approx0.25$\,Myr.
The beginning of the blue loop is marked by a spike in mass-loss rate driven by the star reaching the $\Omega$-limit as it contracts.

\begin{figure} 
\centering
\includegraphics[width=0.45\textwidth]{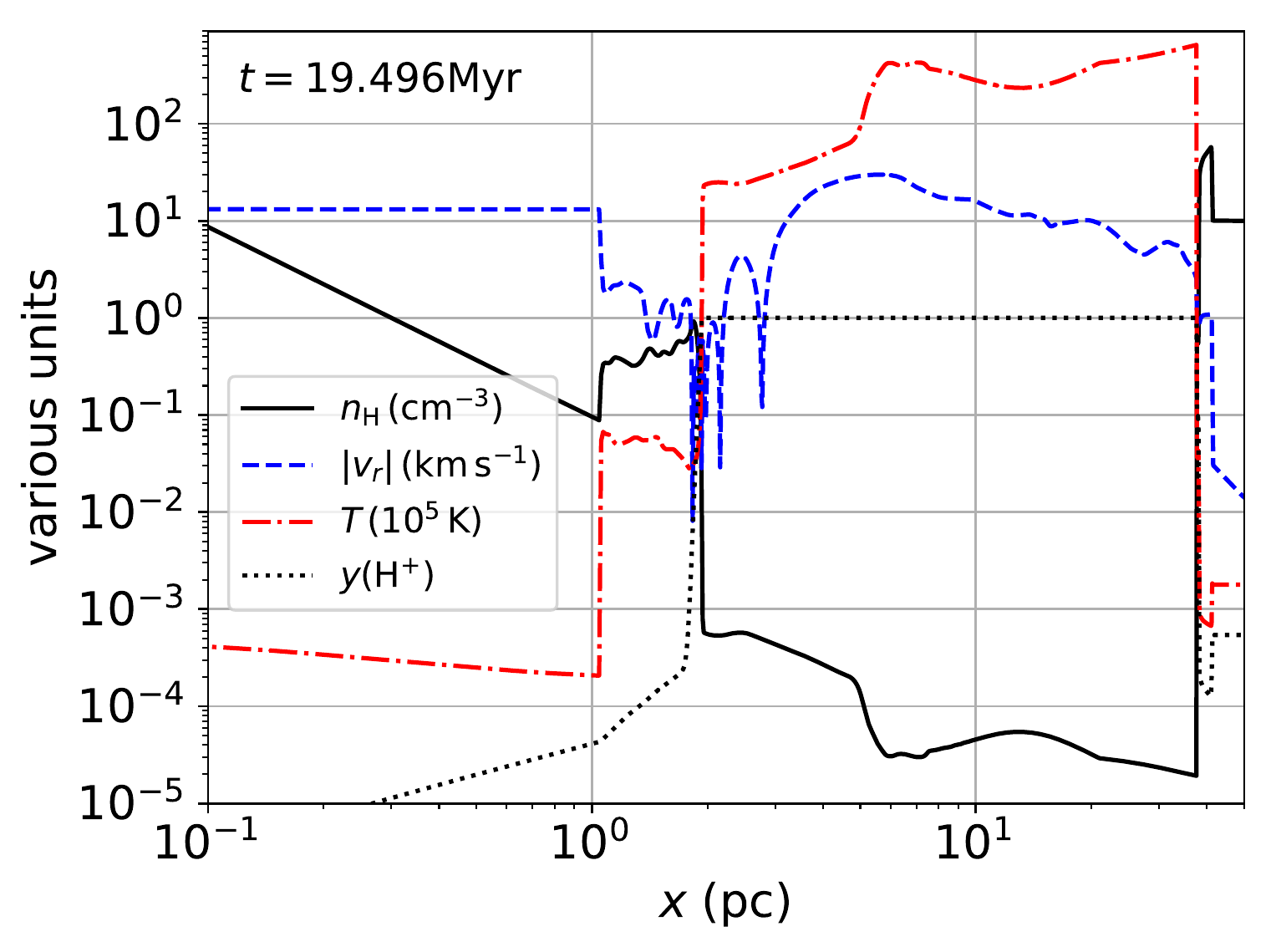}
\caption{
  Circumstellar medium produced by the F12B stellar evolution calculation in a 1D spherically symmetric R-HD simulation, with the snapshot taken midway through the RSG phase of evolution.
  This snapshot is mapped on to the 2D grid and used as the initial conditions for modelling the later blue loop.
  }
\label{fig:f12b-1d}
\end{figure}

The circumstellar medium was first simulated in 1D with spherical symmetry, up to a point mid-way through the RSG phase, at $t\approx19.5$\,Myr.
This is plotted in Fig.~\ref{fig:f12b-1d}, where we show H number density, temperature, radial velocity, and H ionization fraction, $y(\mathrm{H}^{+})$.
This calculation was performed on a simulation domain extending from the origin to $r\approx52$\,pc, resolved by 4096 uniformly spaced grid-cells.
The shocked RSG wind forms a diffuse shell at $r\approx1pc$ with $n_\mathrm{H}\sim1\,\mathrm{cm}^{-3}$, confined externally by a hot and low-density wind bubble from the main-sequence phase.
This simulation snapshot was then mapped on to a 2D cylindrical grid with 10 refinement levels and $512\times256$ cells on each level.
The coarsest grid extends to $z\in[-38.9,38.9]$\,pc and $R\in[0,38.9]$\,pc with rotational symmetry in the azimuthal coordinate, and the inner wind boundary has a radius of 0.025\,pc (84 grid cells on the finest level).
The radial resolution on the nested grid is comparable to that used by CVL08: they used 900 zones to $r=2$\,pc, and we have $256+5\times128=896$ cells to the same radius.
The angular resolution of CVL08 is superior at the wind inflow boundary (they use 200 cells for $\theta=\pi/2$ and we have 132, i.e., our angular resolution at the boundary is 0.68$^\circ$).

Photoionization plus gas heating and cooling is solved used the ``MPv3'' microphysics module that was used for modelling H~\textsc{ii} regions in \citet{MacLanGva13} and photoionization-confined shells around RSGs in \citet{SzeMacLan18}.
Radiative transfer of ionizing photons is included although the stellar temperature only reaches 11\,000\,K in the BSG phase and so the EUV output of the star is very weak.
Collisional heating from shocks is the main heating process active in the simulation (cf.~CVL08).

\subsubsection{Results}
The 2D simulation presented below starts at 19.5\,Myr in the middle of the RSG phase and ends at 20.1\,Myr, encompassing most of the RSG phase, the period of rapid rotation, and the first 50\,000 yr of the BSG phase.
The outer boundary conditions are outflow, but the boundary condition for the RSG wind bubble is effectively the confining thermal pressure of the hot and low-density bubble from the main sequence that was modelled in 1D, and which is not shown in the plots below.

\begin{figure}
\centering
\includegraphics[trim = 0mm 8mm 0mm 2mm, clip, width=0.4\textwidth]{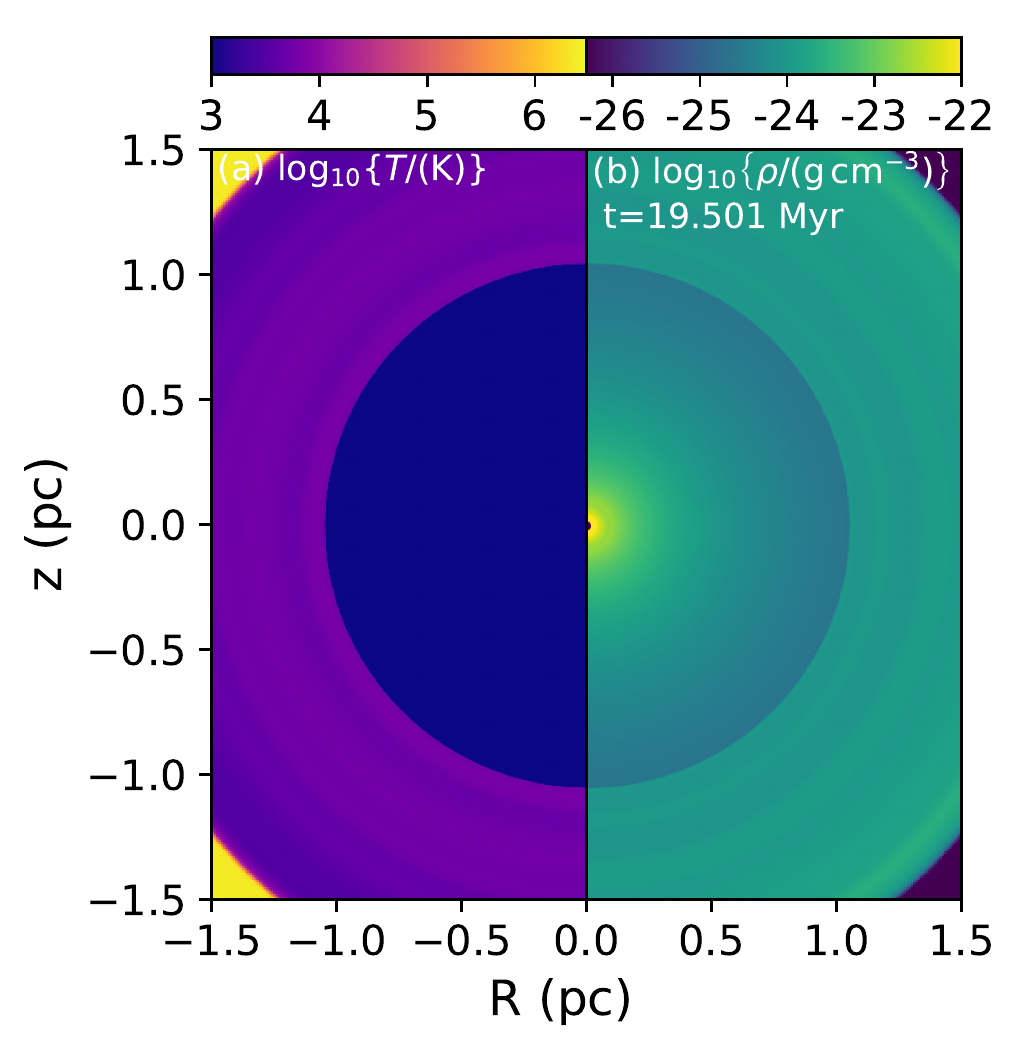}
\includegraphics[trim = 0mm 8mm 0mm 13.5mm, clip, width=0.4\textwidth]{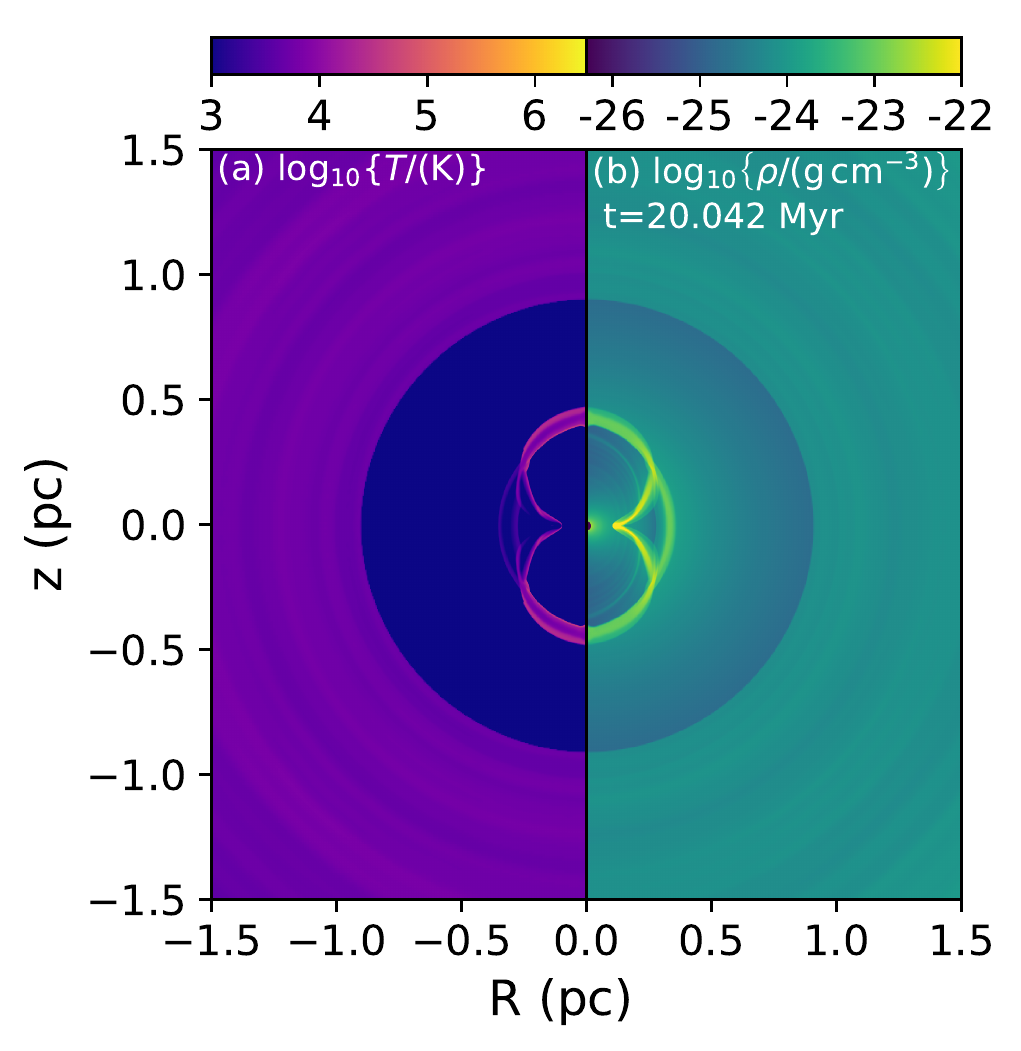}
\includegraphics[trim = 0mm 2mm 0mm 13.5mm, clip, width=0.4\textwidth]{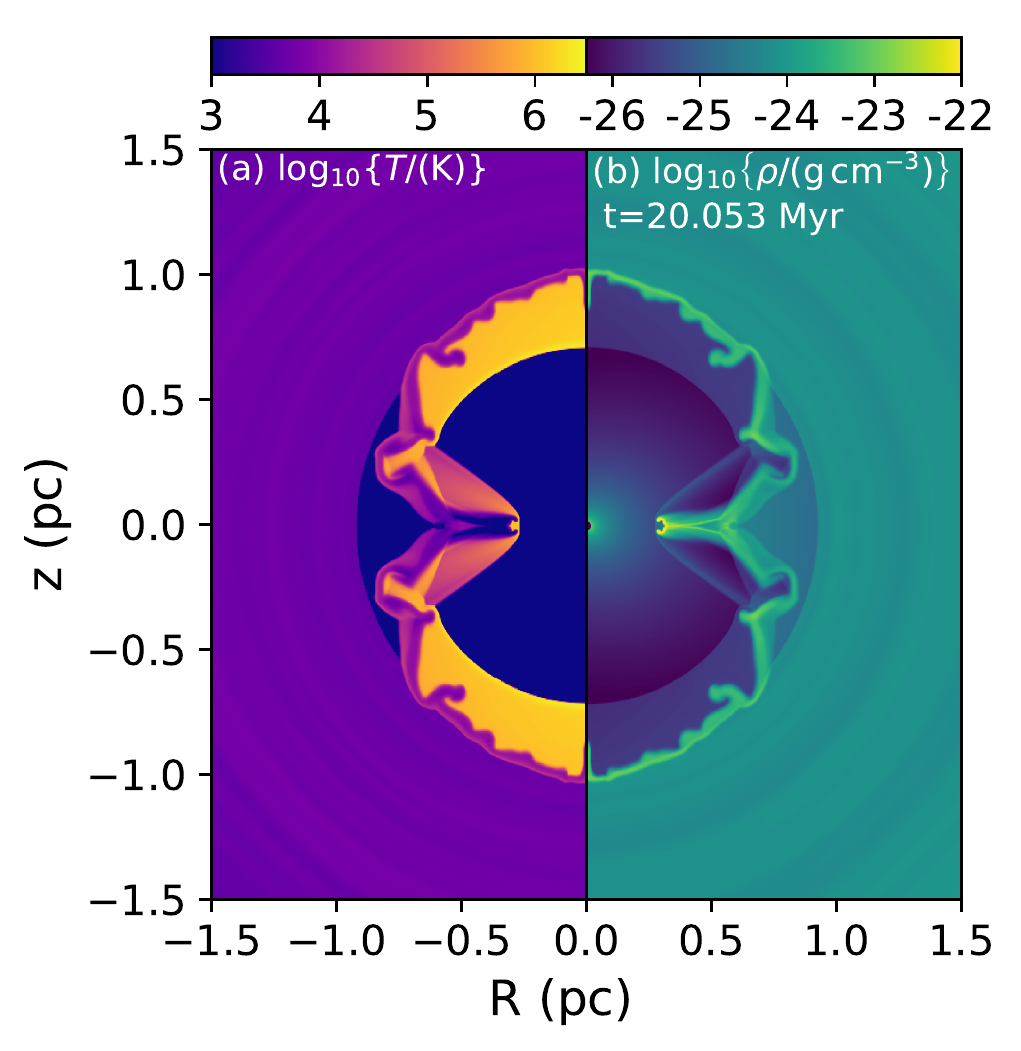}
\caption{
  Density and Temperature for the 2D simulation described in Section~\ref{sec:chita}: the wind bubble during the RSG phase shortly after mapping the 1D simulation on to the 2D nested grid (top panel), just after the phase of critical rotation (middle), and when the swept up wind reaches the termination shock of the RSG wind (bottom).
  The left half-plane shows $\log_{10}T$ in K and the right half-plane shows $\log_{10}\rho$ in g\,cm$^{-3}$. 
  Not all of the simulation domain is shown (the full domain extends to 39\,pc).
  }
\label{fig:Chita_DT}
\end{figure}

Fig~\ref{fig:Chita_DT} shows snapshots of results (a) at 19.5\,Myr in the middle of the RSG phase, (b) just after the phase of critical rotation at 20.04\,Myr, and (c) 20\,000 yr after onset of critical rotation, at approximately the same time as the final panel of fig.~A.4 of CVL08.
There are differences in our results for the structure of the RSG wind, compared with CVL08: the termination shock is approximately at the same radius, but there is significantly more post-shock radiative cooling and compression in the CVL08 calculation.
This can be traced back to differences in the wind velocity during the RSG phase: we find $v_\infty=[10-15]\,\mathrm{km}\,\mathrm{s}^{-1}$ whereas CVL08 calculate $v_\infty=[30-40]\,\mathrm{km}\,\mathrm{s}^{-1}$.
This appears to be related to a typo in eq.~2 of \citet{EldGenDai06}, where the wind multiplier $\beta_\mathrm{w}$ is erroneously multiplying $v_\mathrm{esc}^2$ instead of $v_\mathrm{esc}$.
This would introduce an error of a factor of up to $\sqrt{8}$ for RSGs which can explain the difference.

This modelling difference means that the RSG wind density is different by the same factor at a given distance from the star, and that the interaction of the BSG wind with the swept-up RSG wind proceeds somewhat differently.
Nevertheless the middle panel of Fig.~\ref{fig:Chita_DT} shows that the early part of the wind-wind interaction is very similar in the two studies.
A dense equatorial ring is expanding slowly away from the star, and two polar lobes are expanding rapidly, driven by the (now almost spherical) BSG wind with $v_\infty\approx350\,\mathrm{km}\,\mathrm{s}^{-1}$, sweeping up the slow wind from the RSG phase.
Once $T_\mathrm{eff}>10^4$\,K the parameter $\beta_\mathrm{w}=1.3$, and so our wind prescription gives $v_\infty$ larger than that of CVL08 by a factor of $\sqrt{1.3}$.
Fig.~\ref{fig:f12b} shows that our $v_\infty$ peaks at $v_\infty\approx370\,\mathrm{km}\,\mathrm{s}^{-1}$ whereas CVL08 have a peak value of $\approx320\,\mathrm{km}\,\mathrm{s}^{-1}$; the difference is consistent with $\sqrt{1.3}\times$.

The bottom panel of Fig.~\ref{fig:Chita_DT} can be compared with the lower panels of fig.~A.4 of CVL08:
the BSG wind has swept up a shell to $r\approx1$\,pc in the polar direction, and the shell is thin, radiative, and weakly unstable.
In the equatorial plane a dense ring has expanded to $r\approx0.25$\,pc and this slowly receding ring creates a bow shock in the BSG wind.
The termination shock of the BSG wind is well-separated from the swept-up shell in all directions except the equatorial plane.
These results are all consistent with CVL08 except that they find the equatorial ring has expanded almost 3 times as far in this time.
This is probably related to the initial injection velocity of the ring, because at all latitudes the wind velocity is multiplied by the parameter $\beta_\mathrm{w}$.

\subsubsection{Including magnetic fields}
\begin{figure}
\centering
\includegraphics[width=0.4\textwidth]{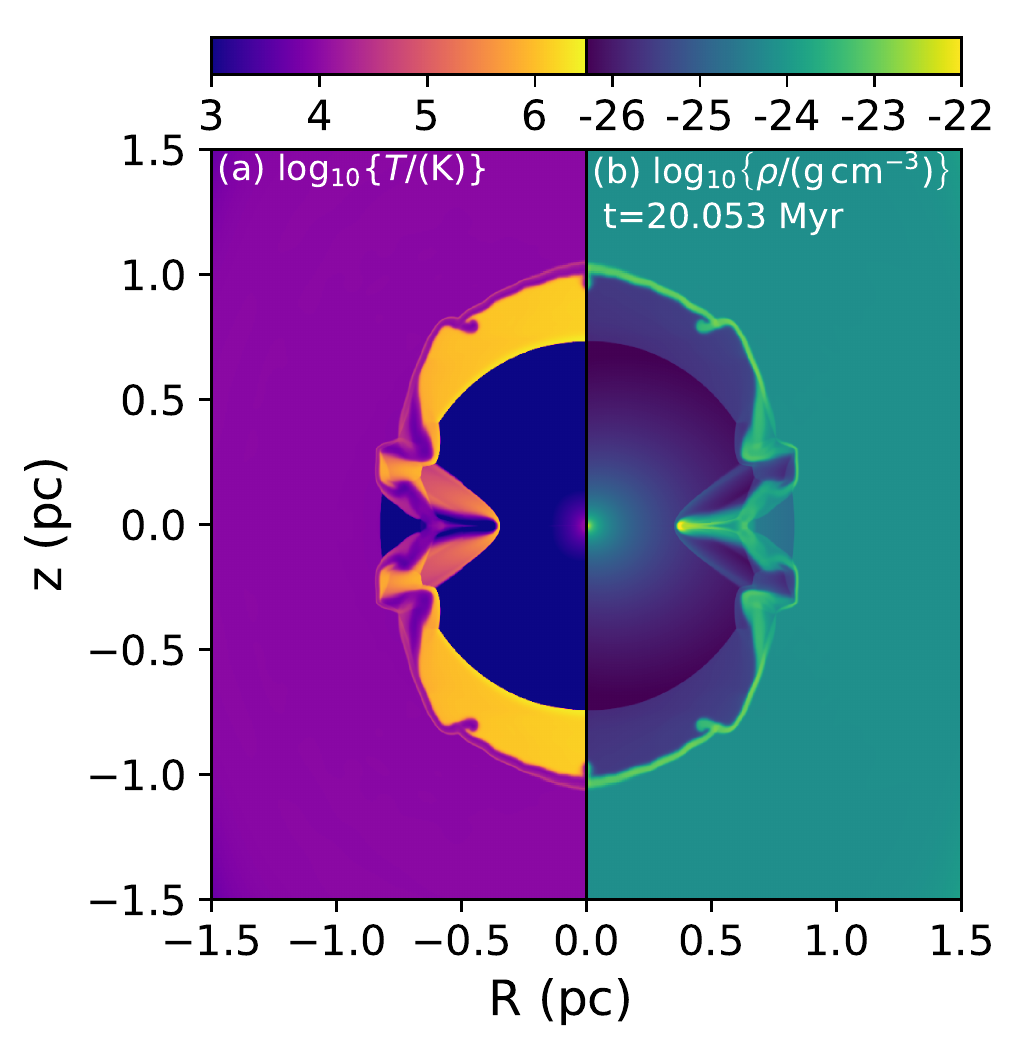}
\caption{
  As Fig.~\ref{fig:Chita_DT} but for an MHD simulation with a simple prescription for the stellar magnetic field.
  The left half-plane shows $\log_{10}T$ in K and the right half-plane shows $\log_{10}\rho$ in g\,cm$^{-3}$, with colour mapping indicated in the colourbars.
  Not all of the simulation domain is shown (the full domain extends to 2\,pc in all directions)
  }
\label{fig:chita_mhd}
\end{figure}

A stellar magnetic field can be easily included given a prediction or assumption for the time-variation of the surface field of the star.
As a simple example, Fig.~\ref{fig:chita_mhd} shows the density and temperature field for a similar simulation, but this time assuming a stellar surface magnetic field of 
\begin{equation}
B(t) = \left( \frac{10\, \mathrm{R}_\odot}{R_\star(t)} \right)^2 10\,\mathrm{G} \;,
\end{equation}
where $R_\star(t)$ is the time-dependent stellar radius.
Again we assume that the surface magnetic field is a split monopole swept into a Parker spiral by the stellar rotation, as in Sect~\ref{sec:wind3d}.
We assume an ISM magnetic field strength $B_z = 10^{-6}$\,G.
In this case we could not map the 1D simulation on to a 2D grid because the stellar and ISM magnetic fields break the spherical symmetry, and so we started the simulation at the beginning of the RSG phase, expanding into a uniform ISM with gas density $\rho_0=2.338\times10^{-24}$\,g\,cm$^{-3}$ and pressure $p_0=1.318\times10^{-12}$\,dyne\,cm$^{-2}$.
This sets the stagnation radius of the RSG wind bubble, which we chose to be approximately consistent with CVL08.

This simulation used a much smaller domain as a result of the simpler boundary condition: the largest grid extends to $z\in[-2,2]$\,pc and $R\in[0,2]$\,pc with rotational symmetry in the azimuthal coordinate.
We use 6 refinement levels with $640\times320$ grid cells in each level, and the inner wind boundary has a radius of 0.025\,pc (128 grid cells on the finest level, resolving the angular direction by 402 cells over 180$^\circ$).

The results are very similar to the R-HD calculation, except that the termination shock of the RSG wind is at slightly smaller radius in the MHD simulation and has been overrun at nearly all latitudes by the BSG wind.
While the magnetic field here was deliberately chosen to be sufficiently weak that it has little dynamical impact on the nebula, the field configuration and strength is still useful when predicting non-thermal emission such as synchrotron radiation \citep[cf.][]{DelPoh18}.

This module can be used, when coupled with results from stellar evolution calculations of single rotating stars, to study the ring nebulae produced by, e.g., spin-up to critical rotation due to stellar contraction, thought to be a potential explanation for the polar caps of Sher 25 and other BSGs \citep[][CVL08]{GvaKniBes15} and for the bipolar structure of nebulae around some Luminous Blue Variables \citep{LanGarMac99}.
The presented R-HD calculation takes $\approx9000$ CPU-hours running on 32 cores; an equivalent high-resolution 3D simulation would take $\gtrsim10^6$ CPU-hours.
The runtime could be significantly reduced by mapping from 1D at the end of the RSG phase instead of at the middle.
The calculations are more demanding than the constant-wind simulations of Section~\ref{sec:wind3d} because of the radiative transfer and the requirement for high angular resolution at the wind boundary.

\subsection{3D R-HD simulation of an expanding WR Nebula}
\label{sec:GS96}
\begin{figure}
\centering
\includegraphics[width=0.4\textwidth]{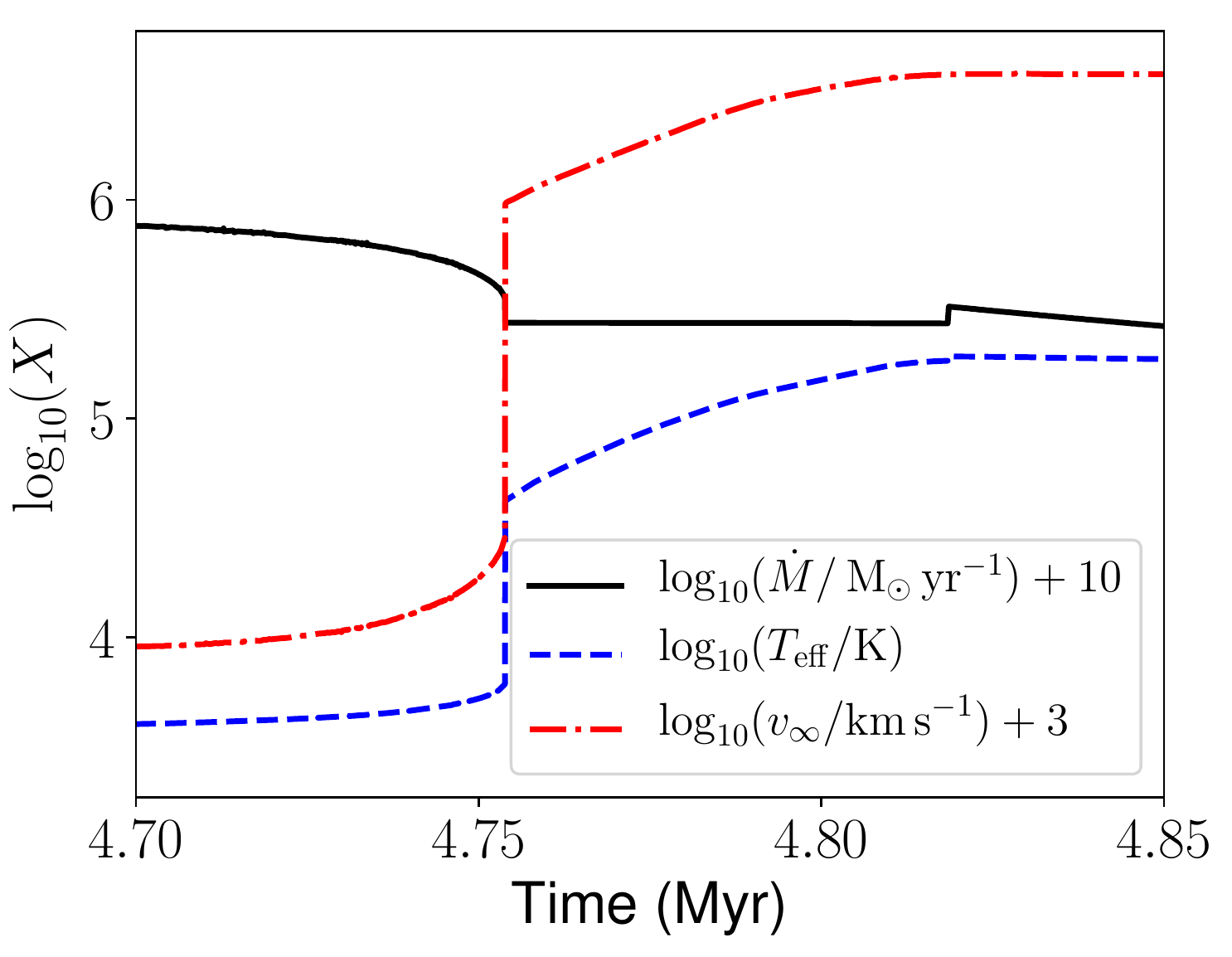}
\caption{
  Evolution of stellar and wind papameters for the $35\,\mathrm{M}_\odot$ evolutionary model used in section~\ref{sec:GS96}, from \citet{GarLanMac96}.
  The mass-loss rate, effective temperature and wind terminal-velocity are plotted around the evolutionary transition from RSG to WR.
  }
\label{fig:wr-wind}
\end{figure}

\begin{figure*}
\centering
\includegraphics[trim = 0mm 6mm 0mm 2mm, clip, width=0.8\textwidth]{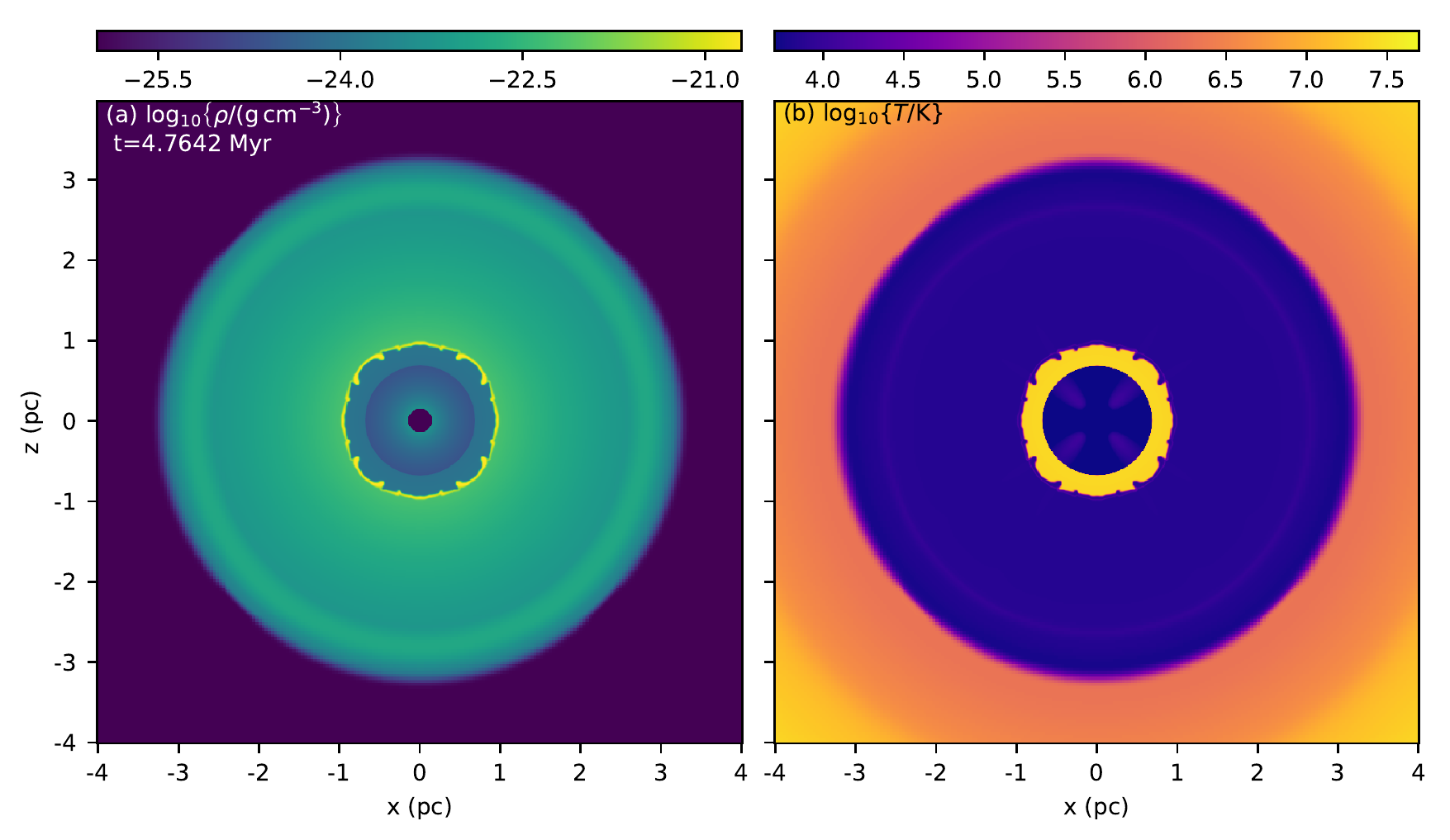}
\includegraphics[trim = 0mm 6mm 0mm 12mm, clip, width=0.8\textwidth]{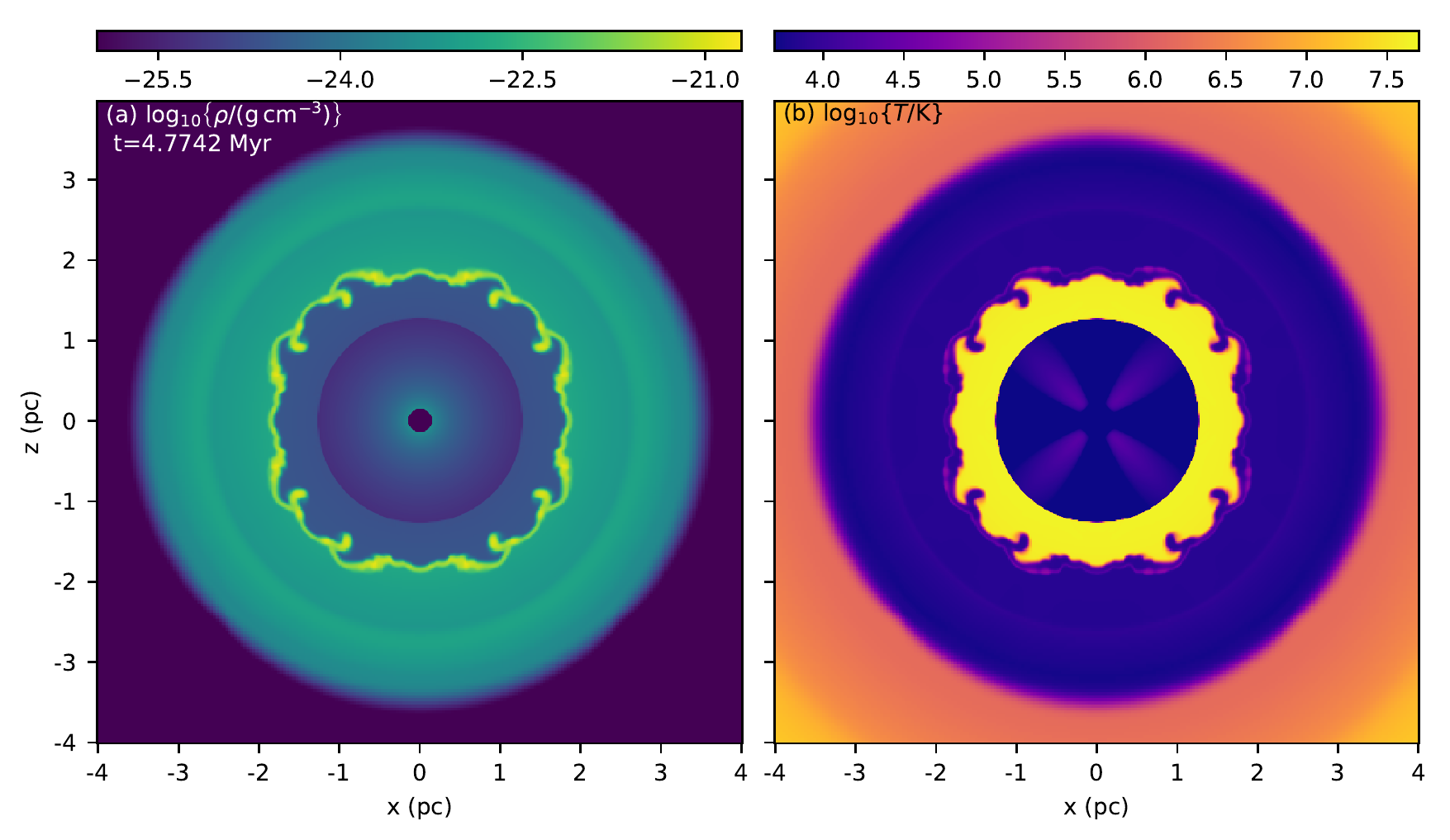}
\includegraphics[trim = 0mm 6mm 0mm 12mm, clip, width=0.8\textwidth]{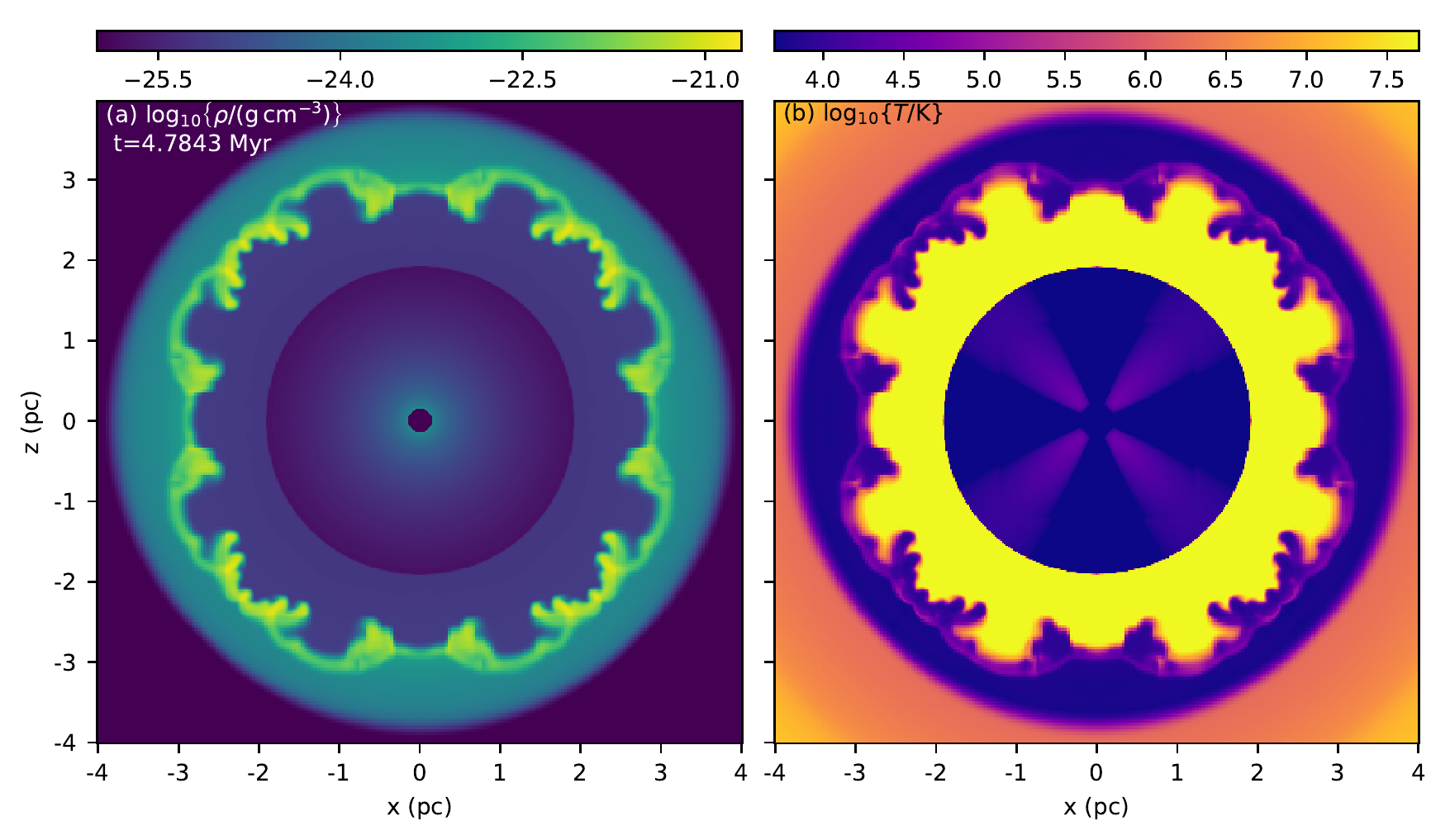}
\caption{
  Log of gas density (left) and gas temperature (right) for a 3D simulation of an expanding WR Nebula.
  In both panels a slice through the $x$-$z$ plane at $y=0$ is shown, with the star at the origin.
  The axes show the domain in parsecs.
  The first row shows the CSM 10\,000 yr after the RSG$\rightarrow$WR transition; the second row after 20\,000 yr; the third after 30\,000 yr.
  }
\label{fig:wr-3d}
\end{figure*}

\begin{figure*}
\centering
\includegraphics[trim = 0mm 6mm 0mm 2mm, clip, width=0.8\textwidth]{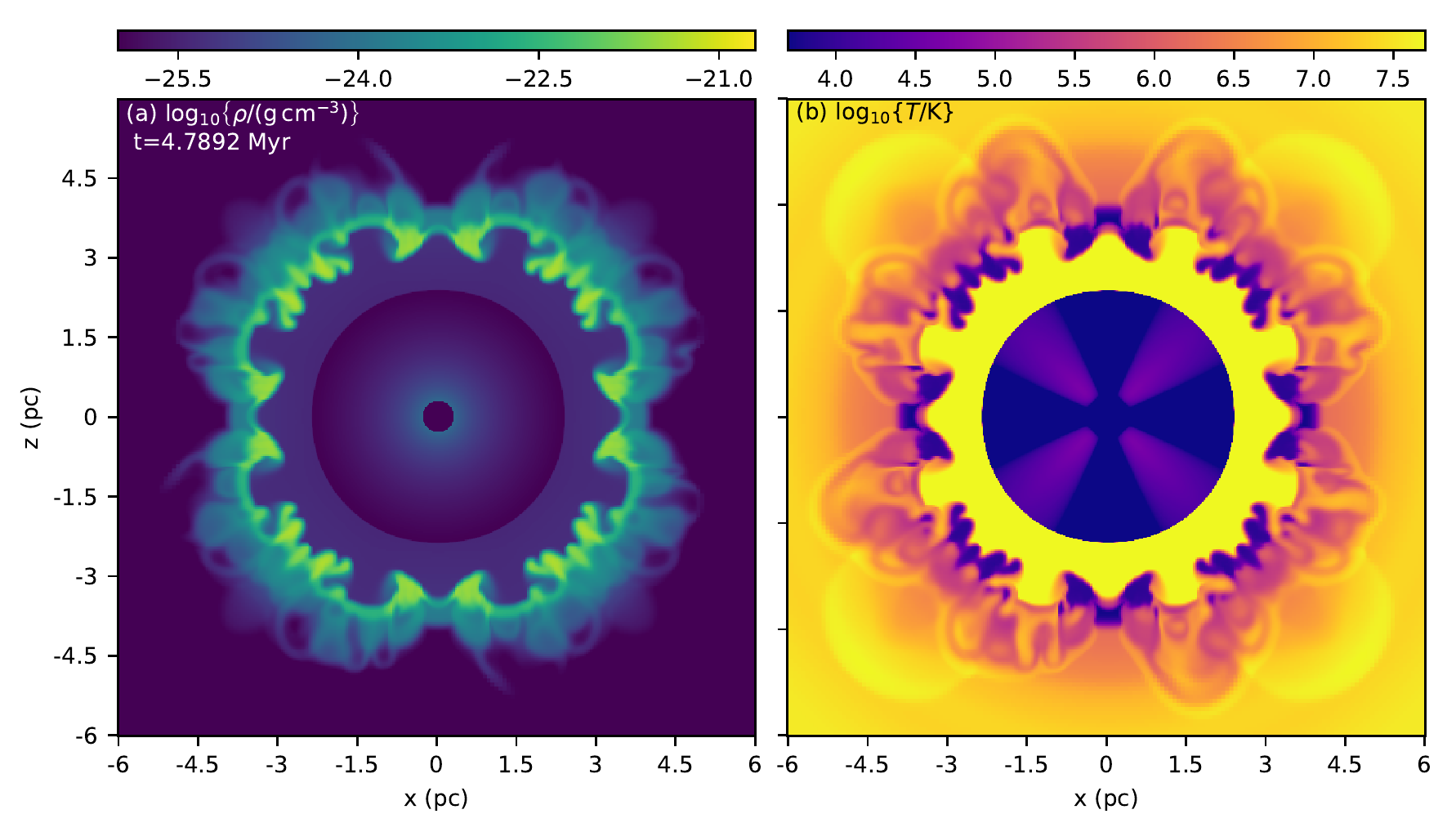}
\includegraphics[trim = 0mm 6mm 0mm 12mm, clip, width=0.8\textwidth]{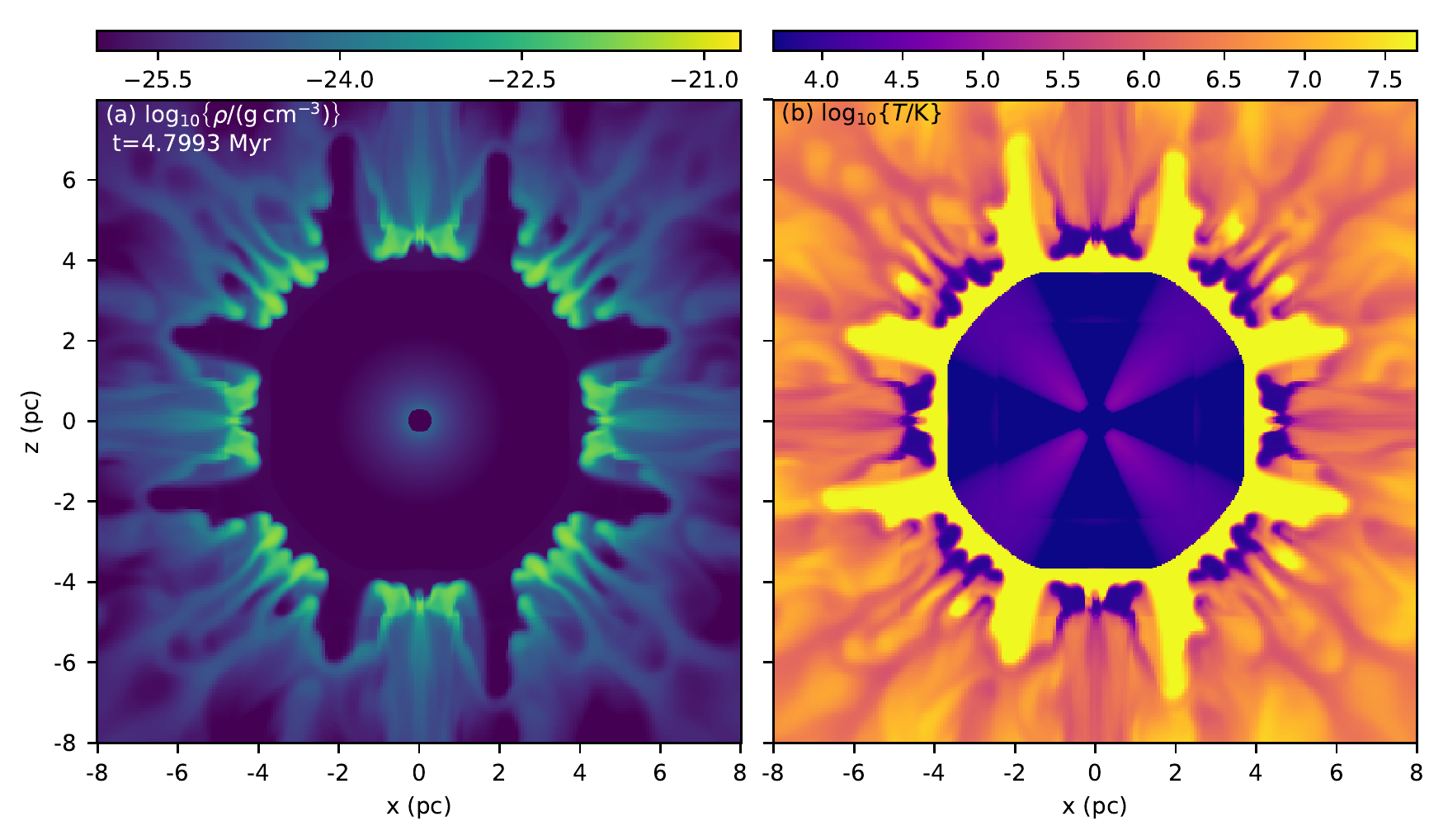}
\caption{
  As Fig.~\ref{fig:wr-3d}, but showing results at 35\,000 and 45\,000 yr after the RSG$\rightarrow$WR transition, and zoomed out to show the expanding nebula.
  }
\label{fig:wr-3d_late}
\end{figure*}

\citet{GarLanMac96} and \citet{FreHenYor06} modelled the circumstellar medium of a 35\,$\mathrm{M}_\odot$ non-rotating star that evolved from main sequence to RSG to WR before explosion, using 2D hydrodynamic and R-HD simulations, respectively.
Here we use the same evolutionary calculation, and focus on the RSG$\rightarrow$WR transition, for which the wind and surface temperature properties are plotted in Fig.~\ref{fig:wr-wind}.
We use the time-varying stellar wind boundary condition (type 1 from section \ref{sec:stellar-wind}) in a 3D R-HD simulation.

\subsubsection{Initial conditions and simulation setup}
We simulate the main sequence and RSG phases using R-HD on a 1D grid with 4096 cells covering 100\,pc, with a background ISM density of $\rho=2.338\times10^{-23}$\,g\,cm$^{-3}$.
This is sufficiently large so that the H\,\textsc{ii} region stays in the simulation domain up to the end of the RSG phase, at $t=4.7542$\,Myr.
This density field is mapped on to a 3D grid with $256^3$ grid cells on each level, and 4 refinement levels, using zero-gradient boundary conditions.
The coarse grid has a range $\{x,y,z\}\in[-30,30]\times10^{18}$\,cm (approximately $\pm10$\,pc) centred on the star.
The finest grid has a cell diameter of $\Delta x \approx 2.93\times10^{16}$\,cm, and the wind boundary region has a radius of 20 cells ($r \approx 5.86\times10^{17}$\,cm, or $\approx 0.2$\,pc).
In comparison with the 2D simulations of \citet{GarLanMac96}, the finest-grid $\Delta x \approx 0.0095$\,pc is comparable to their radial resolution $\Delta r = 0.012$\,pc for the "slow RSG wind" case, and slightly lower resolution than the $\Delta x = 0.00625$\,pc used by \citet{FreHenYor06}.
In terms of the grid used, \citet{FreHenYor06} have comparable geometry to ours, modelling one quadrant of the 2D plane with $125^2$ grid points per level, whereas we model the full 3D space with $256^3$ cells per level, or $128^3$ per octant.

\subsubsection{Results}
The top panel of Fig.~\ref{fig:wr-3d} plots $\log_{10}$ of gas density and temperature 10\,000 yr after the simulation starts.
The RSG wind is being swept up into a thin shocked shell that is dynamically unstable, and a strong reverse shock has formed in the WR wind, heating gas to $\gtrsim10^7$\,K.
Both the RSG and WR winds are spherically symmetric and so the instabilities are seeded by the grid-scale integration errors.
The CSM has already been flash ionized and there is no neutral gas on the domain, except that some of the clumps in the swept-up shell have become optically thick and have partially recombined, but do not cool below $T\sim10^3$\,K.

The hot gas at $r\gtrsim3$\,pc is the relic wind bubble from the main sequence phase, and this provides the external pressure that previously confined the RSG wind.
The RSG wind has been ionized and heated from $T\sim10^2$\,K to $\sim10^4$\,K, however, and is now strongly over-pressurised, expanding into the surrounding hot medium at approximately its sound speed.

The diagonal features in the temperature plot in the freely expanding wind are integration errors arising from the advection of thermal energy in the strongly kinetic-energy-dominated flow (total energy is conserved by the finite-volume scheme).
They do not affect the nebula because the post-shock gas properties are independent of the pre-shock temperature for a large-Mach-number shock.

The second panel shows the gas density and temperature 20\,000\,yr after the WR wind switches on.
The same unstable clumps in the swept-up shell are present, but have grown and have expanded to larger scales.
The expansion velocity of the swept-up shell is approximately 1\,pc per 10\,000 yr ($100\,$km\,s$^{-1}$), and so the forward shock is strongly radiative whereas the reverse shock is adiabatic.
The third panel plots the results after 30\,000 yr, just before the swept-up shell reaches the edge of the RSG wind bubble.
The clumpy nature of the shell means that not all directions will break out of the RSG shell at the same time.
The symmetry in the solution arises because the hydrodynamic solver is almost perfectly symmetric (to roundoff error), and so the integration errors from the grid discretisation are very similar in each octant.

Fig.~\ref{fig:wr-3d_late} shows the later evolution as the swept-up shell breaks out of the RSG wind and into the relic main-sequence bubble.
This is reminiscent of the simulations presented by \citet{RogPit13}, except that their calculations were performed in a dense and turbulent background medium and the WR wind interacted with this interstellar gas and not purely the wind of previous evolutionary phases.
Here the perturbations in the swept-up shell were seeded by the grid geometry rather than a random process, and so the shocked WR wind escapes through eight regularly-spaced channels into the low-density surroundings (in this plane; in 3D there are many channels, but they have a regular spacing), entraining cold RSG wind material as it does so.
A more realistic model would introduce clumpy substructure in both the RSG and WR winds, which would break the symmetry and better reflect the reality that winds of massive stars are strongly clumped \citep{PulVinNaj08}, possibly driven by turbulent sub-surface convection \citep{CanLanBro09, GraFosLan15}.

\subsubsection{Interpretation}
Our results are not comparable to \citet{FreHenYor06} because they used a wind speed for the RSG phase of $v_\infty=75\,\mathrm{km}\,\mathrm{s}^{-1}$ whereas our calculation \citep[based on][see above]{EldGenDai06} gives a wind speed closer to the "slow wind" calculation with $v_\infty=15\,\mathrm{km}\,\mathrm{s}^{-1}$ \citep{GarLanMac96}.
We find that the nebula expands to 1\,pc in 10 kyr, 2\,pc in 20 kyr, 3\,pc in 30 kyr, i.e., expanding at $v_\mathrm{exp}\approx 100\,\mathrm{km}\,\mathrm{s}^{-1}$.
\citet{KooMcK92} calculate the expansion speed of wind-blown bubbles in power-law media, and their equation 3.1, for the case where the reverse shock is adiabatic and the forward shock is radiative, predicts that the shock radius scales as $R\propto t$ for a constant wind expanding into a density profile $\rho\propto r^{-2}$.
The constant prefactor of this equation gives the (constant) expansion velocity as
\begin{equation}
v_\mathrm{exp} = \left( \frac{\dot{M}_\mathrm{wr} v_{\infty,\mathrm{wr}}^{2} v_{\infty,\mathrm{rsg}}}{3 \dot{M}_\mathrm{rsg}} \right)^{1/3}
\end{equation}
For the WR wind parameters shortly after the transition ($\dot{M}_\mathrm{wr}\approx2.75\times10^{-5}\,\mathrm{M}_\odot\,\mathrm{yr}^{-1}$ and $v_{\infty,\mathrm{wr}}\approx1200\,\mathrm{km\,s}^{-1}$) and RSG wind parameters mid-way through the RSG phase ($\dot{M}_\mathrm{rsg}\approx8\times10^{-5}\,\mathrm{M}_\odot\,\mathrm{yr}^{-1}$ and $v_{\infty,\mathrm{rsg}}\approx10\,\mathrm{km\,s}^{-1}$), this corresponds to $v_\mathrm{exp}\approx120\,\mathrm{km\,s}^{-1}$, in good agreement with our numerical results.
\citet{GarLanMac96} found somewhat faster expansion (their fig.~7), but the differences are probably not significant.
The qualitative appearance of the results are very similar, given that the grid geometry is different and the development of instabilities is not expected to be identical.
The formation of clumps in the thin shell, persistent through its expansion, followed by blowout once the edge of the RSG wind bubble is reached, is the same.

This calculation took about 15\,000 core-hours, run on 32 cores for 20 days.
Higher resolution is desirable for better resolving the instability of the swept-up shell, and is required for modelling rotating stars with clumpy winds to resolve the spatial and temporal variations in the wind.
Running with $384^3$ cells on each level would take $\approx75\,000$ core hours, whereas $512^3$ would require $\approx250\,000$ core hours.

\begin{table}
\caption{The properties of the  stars of V444 Cyg system, where primary is a WR star and secondary has spectral type O6, taken from  \citet{SteBloPol92}. 
  The rotation velocities and surface split-monopole magnetic field strengths are notional, for demonstration of the methods only, and both have axis of symmetry $\hat{z}$.
}    
\centering
\begin{tabular}{ p{0.22\textwidth} | c c}
  \hline
  Parameter & Primary & Secondary  \\
  \hline
  Mass Loss rate, $\dot{M}$ [$\mathrm{M}_{\odot}$ yr$^{-1}$]  & $1.4\times10^{-5}$ & $10^{-6}$ \\
  Terminal wind speed, $v_\infty\,  [\mathrm{km}\,\mathrm{s}^{-1}]$  & 2000 & 2000  \\
  Surface rotation speed, $v_\mathrm{rot}\,  [\mathrm{km}\,\mathrm{s}^{-1}]$ & 200 & 200 \\
  Surface split-monopole magnetic field strength, $|\bm{B}|\,$ [G] & 100 & 1 \\
  \hline
  Radius of wind boundary (cm) & $2.6\times10^{11}$ & $2.6\times10^{11}$ \\
  Position of star ($10^{12}$\,cm) & $[-1.84,0,0]$ & $[0.96,0,0]$ \\
  \hline
\end{tabular}
\label{tab:3dwind_param}
\end{table}

\subsection{3D MHD simulation of wind-wind collision}
\label{sec:v444}

\begin{figure*}
\centering
\includegraphics[width=0.9\textwidth]{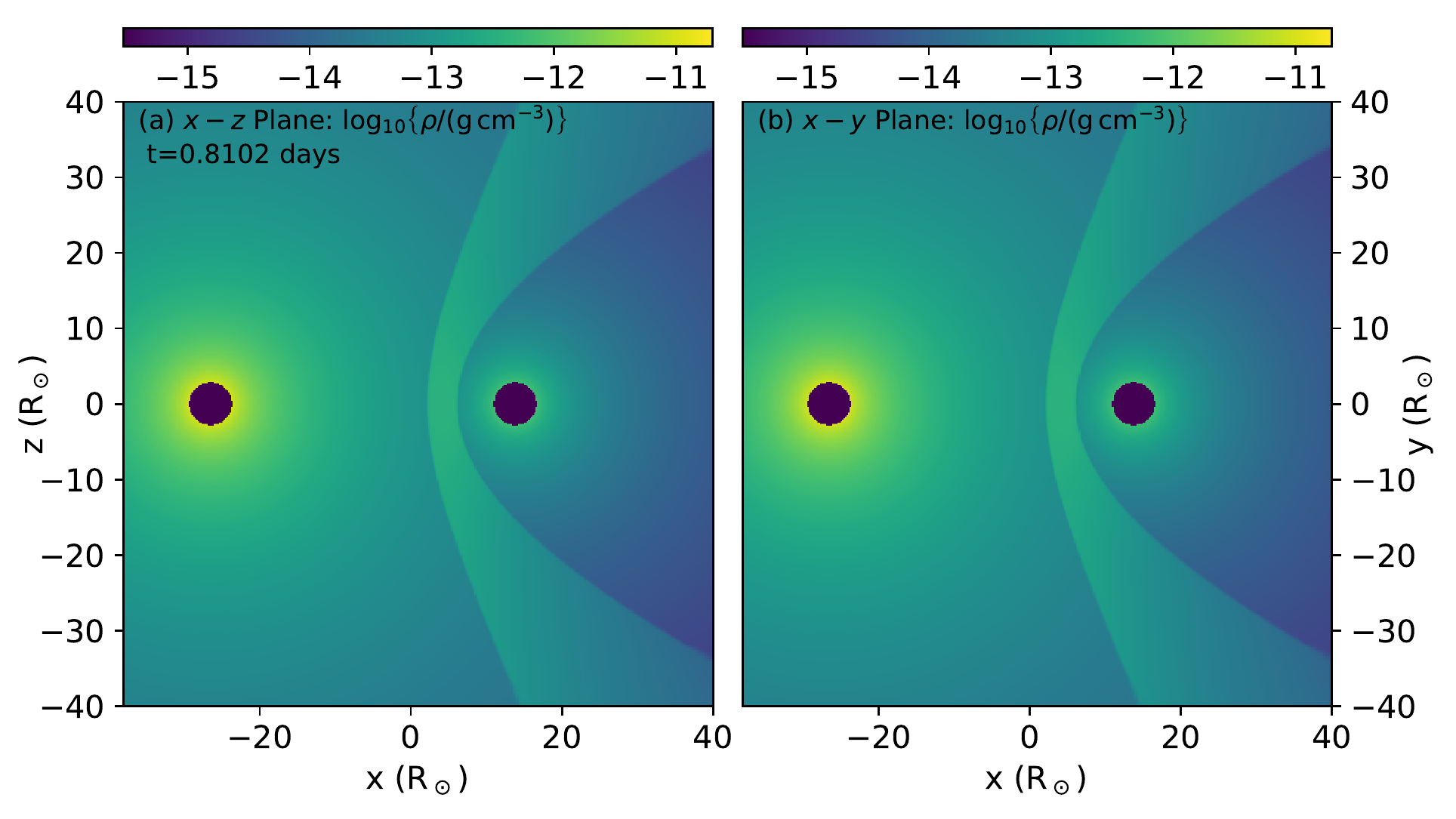}
\includegraphics[width=0.9\textwidth]{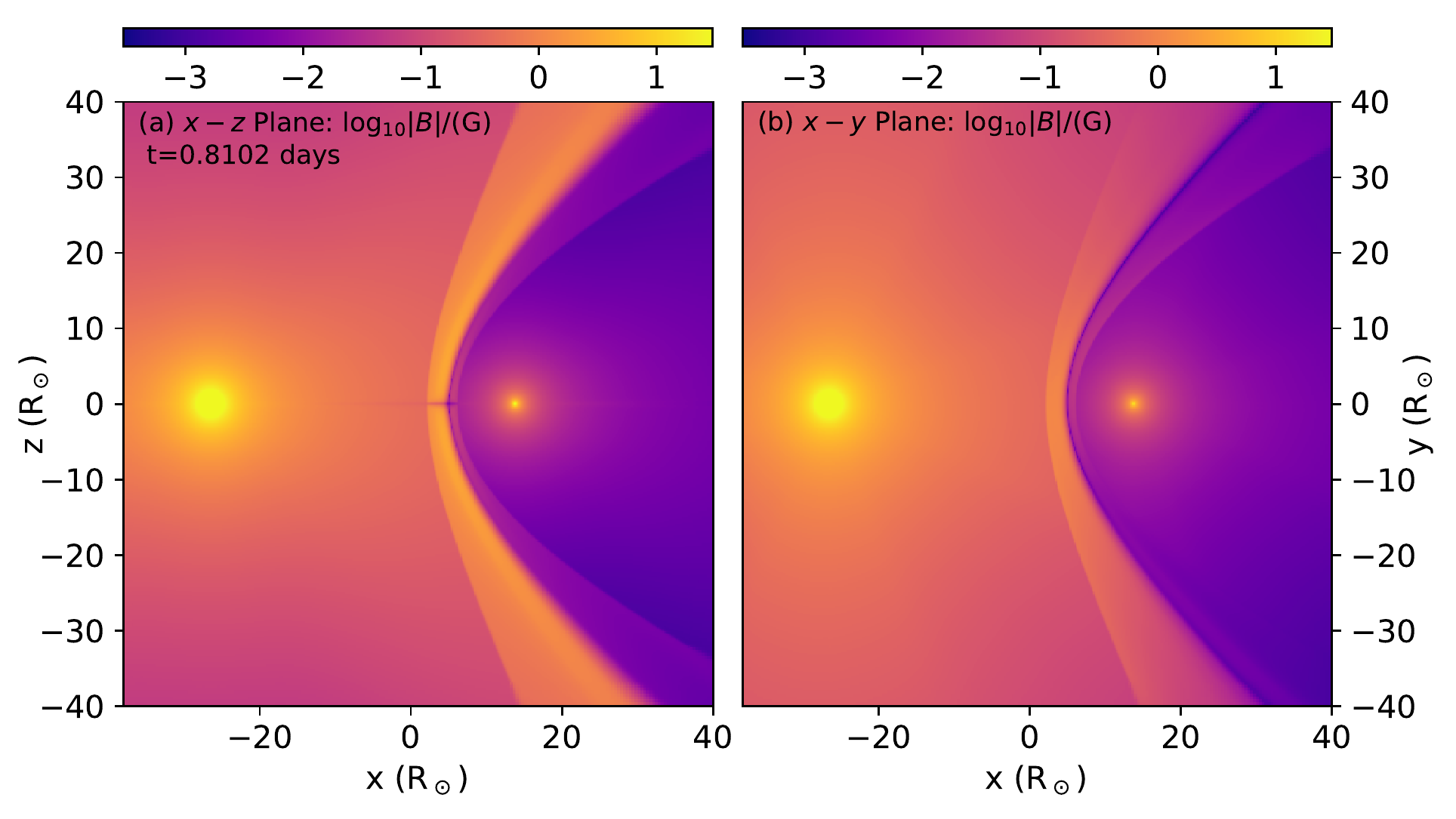}
\caption{
  3D MHD simulation of wind-wind collision from section~\ref{sec:v444}, for the adiabatic case at $t=70\,000$\,s.
  Above: $\log_{10}$ of gas density in the $x$-$z$ plane (left panel) and the $x$-$y$ plane (right panel).
  Below: $\log_{10}$ of magnetic field strength in the $x$-$z$ plane (left panel) and the $x$-$y$ plane (right panel).
  }
\label{fig:v444-3d-ww}
\end{figure*}

Colliding-wind binary systems are fascinating environments to study astrophysical fluid dynamics, shocks and instabilities \citep{SteBloPol92, LamFroDub11, ParPitCor11, MadGulOka13}, magnetism \citep{WalFolMey12, KisReiRei16}, and particle acceleration~\citep{PitDou06, WhiBreKon20, GriReiKis19}.
In principle the nested-grid setup of \textsc{pion} is suitable for modelling such systems, although some code improvements would be required, including implementation of orbital motion for stellar-wind sources, radiative heating/cooling routines that are appropriate for the high densities and chemical abundances encountered, and possibly a wind acceleration region for close binary systems.
Here we present a preliminary 3D MHD simulation of wind-wind collision based on the 2D hydrodynamic calculation of the V444 Cyg system by \citet{SteBloPol92} using the constant wind module (type 0 in section \ref{sec:stellar-wind}).

V444 Cyg is a well-studied eclipsing binary system consisting of two massive stars with powerful winds, a WR primary (spectral type WN5) and an O-type secondary (spectral type O6).
For the modelling of the colliding winds, we took the orbital and stellar wind parameters from \citet{SteBloPol92}, presented in Table~\ref{tab:3dwind_param}.
We make a number of simplifications to the system, partly so the setup is comparable with \citet{SteBloPol92}:
\begin{enumerate}
  \item 
We inject the winds already at the terminal velocity at the wind-boundary radius, and radiation and gravitational forces are neglected.
We set the wind boundary radius for both stars at $r=2.6\times10^{11}$\,cm.
  \item
The orbital period of the system is 4.2 days, nevertheless during this 3D simulation the orbital motion is neglected, because of the extra complications this would introduce (the system geometry will change, and must be simulated for much longer to reach a stationary state).
Thus, we consider that the stars were at rest in an inertial frame with cylindrical symmetry along the $x$-axis, except for the stellar magnetic fields which break the symmetry.
  \item
We use an optically thin cooling function appropriate for photoionized bow shocks \citep{GreMacHaw19}, although this is likely a crude approximation given the large densities and the hydrogen depletion of the WR wind.
\end{enumerate}

The simulation is run with 3 grid levels, each with $384^3$ cells and with the two nested grids centred on the origin at the centre of the domain.
The coarsest grid has $\{x,y,z\}\in [-1.024,1.024]\times10^{13}$\,cm, the next level has $\{x,y,z\}\in [-5.12,5.12]\times10^{12}$\,cm, and the finest level has $\{x,y,z\}\in [-2.56,2.56]\times10^{12}$\,cm, with a cell diameter of $\Delta x = 1.333\times10^{10}$\,cm.
Outflow boundary conditions are employed at all boundaries.

We added stellar rotation and split-monopole magnetic fields for each star, such that the magnetic field in the wind will be swept into a Parker spiral at large radius.
The field strengths were chosen such that the magnetic field is too weak to affect the dynamics of the unshocked wind, i.e., the Alfv\'en Mach number of the wind is large in both cases.
The stars were set rotating well below critical rotation so that latitude-dependent effects are not expected to be strong.

At the first stage we run the simulation with adiabatic hydrodynamics, without taking radiative cooling into account.
The simulation is run for 70\,000 seconds, which is enough time for a stationary shock structure to form around the stagnation point of the flow.
The dynamical time-scale of the wind-wind collision is the stellar separation divided by the wind speed, 14\,000\,s.
The results of the simulation are plotted in Fig.~\ref{fig:v444-3d-ww} for gas density and magnetic field strength.
A stable wind-wind collision has been set up and the weaker wind of the O star gets swept back by the stronger WR wind. 
The collision region, shaped like a bow shock, consist of two shocks and a hot plasma between these shocks.

For the second stage we continue the simulation including radiative cooling.
As noted by \citep{SteBloPol92}, for this system the cooling time is comparable to the advection time for the shocked wind, and so the gas cools and can be compressed to very high densities.
The wind-collision region is unstable, i.e., the shocked region gets narrower and eventually an oscillatory thin-shell instability arises.

\begin{figure*}
\centering
\includegraphics[width=0.9\textwidth]{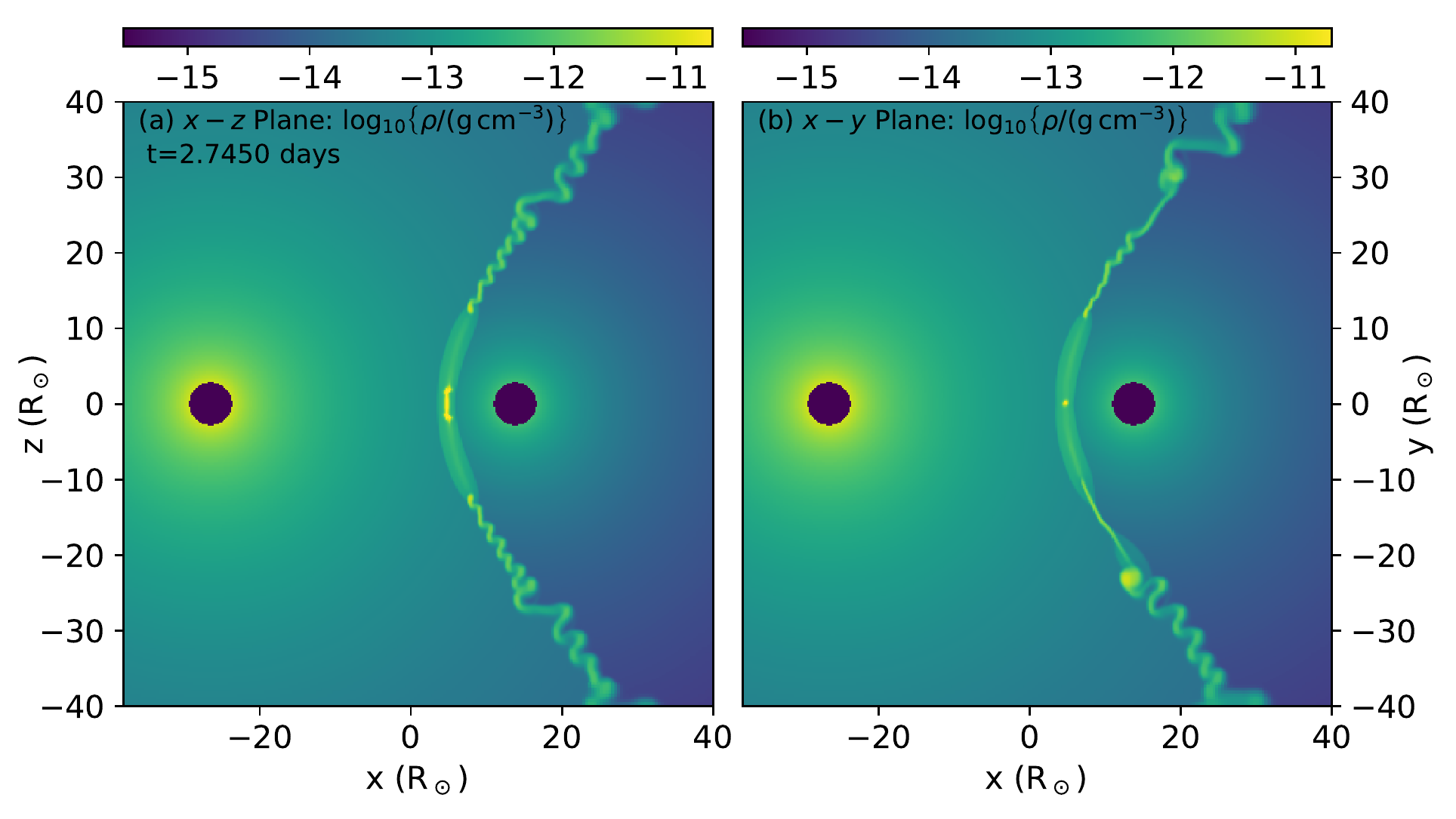}
\includegraphics[width=0.9\textwidth]{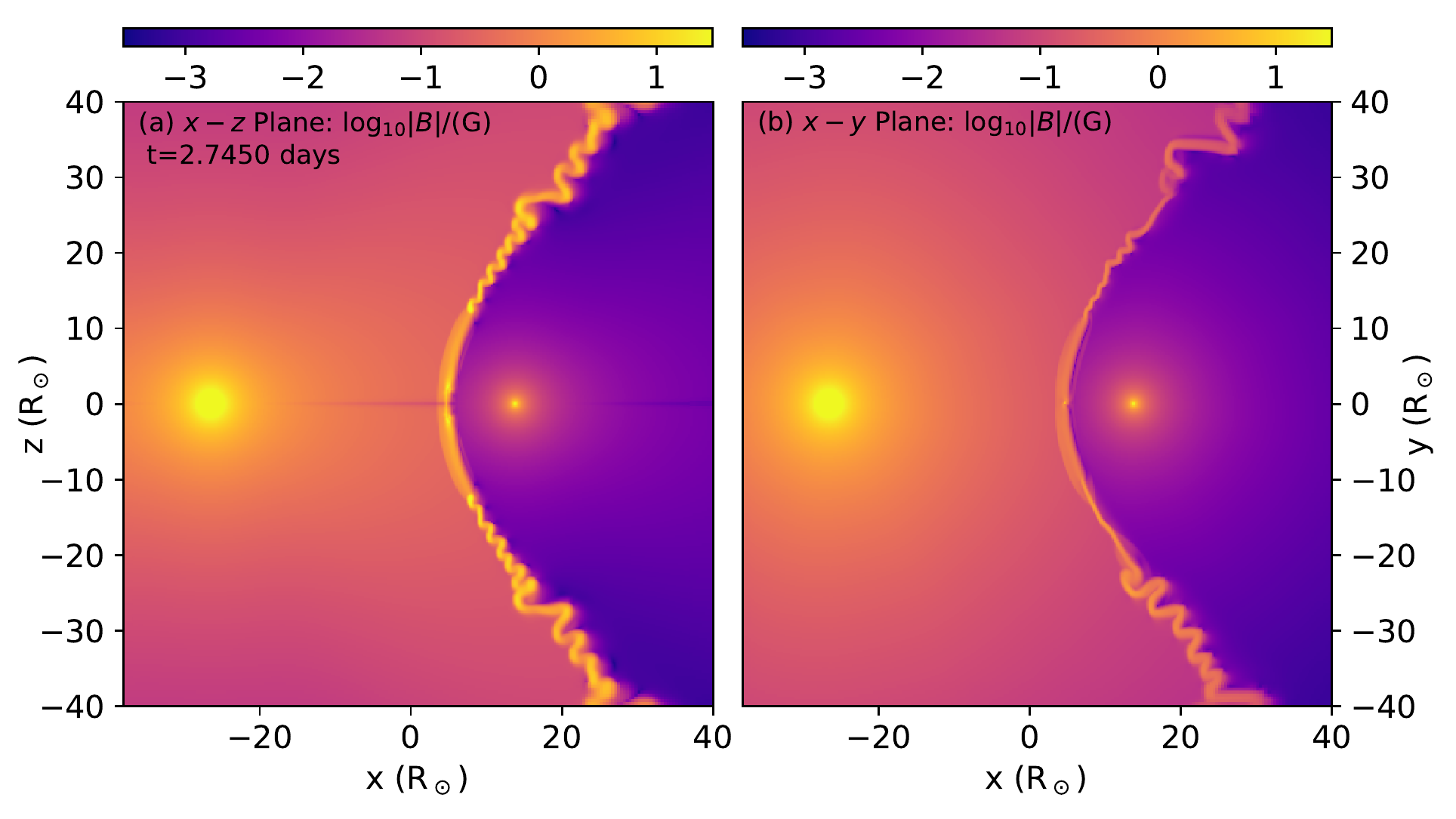}
\caption{
  3D MHD simulation of wind-wind collision from section~\ref{sec:v444}, at $t\approx2.4\times10^{5}$\,s, or $\approx1.7\times10^5$\,s after radiative cooling has been switched on.
  Above: $\log_{10}$ of gas density in the $x$-$z$ plane (left panel) and the $x$-$y$ plane (right panel).
  Below: $\log_{10}$ of magnetic field strength in the $x$-$z$ plane (left panel) and the $x$-$y$ plane (right panel).
  }
\label{fig:v444-3d-cool}
\end{figure*}

Fig.~\ref{fig:v444-3d-cool} shows the same plots as Fig.~\ref{fig:v444-3d-ww}, but at a later time after radiative cooling has been switched on for some time.
The shocks oscillate between strongly radiative and weakly radiative near the stagnation point of the flow.
This appears superficially similar to the case of overstable radiative shocks with velocity $\approx150\,\mathrm{km}\,\mathrm{s}^{-1}$ \citep{InnGidFal87b}, but is actually arising because the advection time and cooling time for the shocked gas are very similar for this setup \citep{SteBloPol92}.
Small perturbations in the hydrodynamics can make the difference as to whether a parcel of gas can cool strongly or not before it is advected away from the stagnation point.
The features seen in all panels near the stagnation point are transitory and unsteady, with knots and rope-like overdensities forming and advecting away to the domain boundaries.
The shocks far from the symmetry axis of the flow are strongly radiative and do not show this oscillatory behaviour.
This is because the shocks are oblique, with smaller Mach number than along the symmetry axis, and so the post-shock temperature is lower and the cooling time is shorter (cooling time has a maximum at $T\approx2\times10^7$\,K and decreases for both lower and higher temperatures).

Gas compression factors of between 10 and 100 are achieved in the radiative shocks (limited by the grid resolution), with somewhat weaker increase in the magnetic field strength because only the component perpendicular to the shock normal is compressed.
For the small separation of the two stars, the magnetic fields of the two winds near the stagnation point are still more radial than toroidal, and only the toroidal component is amplified.
The magnetic field amplitude in the $x$-$y$ plane is less than in the $x$-$z$ plane because the former contains the equatorial current sheet of both stars.

A detailed investigation of the V444 Cyg system would require the inclusion of orbital motion \citep[cf.][]{LamMilLie17}, the finite size of the stars, their wind-acceleration regions, and perhaps also radiative inhibition \citep[e.g.][]{ParPitCor11}.
In future work we will improve the radiative cooling function and implement orbital motion, and use higher-resolution simulations to investigate the MHD properties of shocks in colliding-wind binary systems.
The simulation presented here took 10\,000 core hours, running with 128 MPI processes.

\section{Postprocessing simulation snapshots}
\label{sec:postprocessing}

\subsection{Python library for reading and plotting snapshots}
\label{sec:pypion}
A python library (\texttt{PyPion} \footnote{\href{https://git.dias.ie/massive-stars-software/pypion}{https://git.dias.ie/massive-stars-software/pypion}}) has been developed 
to read \textsc{pion} snapshots into \texttt{Numpy} arrays for plotting and further analysis.
This enables simple post-processing and visualisation on all simulations using python. 

\texttt{PyPion} contains two core scripts that contain modular routines that can be used depending on what type (1D/2D/3D) of \textsc{pion} simulation is run.
The library also works on simulations with refined grids without modification.
An additional script (\texttt{Plotting\_Classes.py}) provides some examples of plots that can be generated from \textsc{pion} data.
Most of the figures in this paper showing multi-dimensional simulations, including e.g., Figs.~\ref{fig:b3d_DB}, \ref{fig:b3d_B100} and \ref{fig:wr-3d}--\ref{fig:v444-3d-cool}, have been produced using this library. 

\begin{enumerate}
    \item \texttt{SiloHeader\_data.py} contains a class with methods to open a \textsc{pion} snapshot in \textsc{silo} format\footnote{\href{https://wci.llnl.gov/simulation/computer-codes/silo}{https://wci.llnl.gov/simulation/computer-codes/silo}} and read metadata from the header directory.
    This data includes the axes dimensions, level dimensions, number of levels, simulation time, number of MPI processes.
    \item \texttt{ReadData.py} contains a class with methods for reading the data for a requested variable (e.g. Density) from a \textsc{silo} file, returning it as a single \texttt{Numpy} array per level.
    When \textsc{pion} is run with multiple MPI processes, each process calculates a subdomain of the grid for each snapshot and saves its data under a Silo directory.
    This class reads each subdomain in turn and adds its data to the correct region in the \texttt{Numpy} array to form an image.
    \item \texttt{Plotting\_Classes.py} contains classes to take data from \texttt{ReadData.py} and make some commonly used plots using \texttt{matplotlib}.
    Several functions have been set up to accommodate different user needs.
\end{enumerate}

\begin{figure*}
\centering
\includegraphics[trim = 0mm 0mm 0mm 2mm, clip, height=0.33\textwidth]{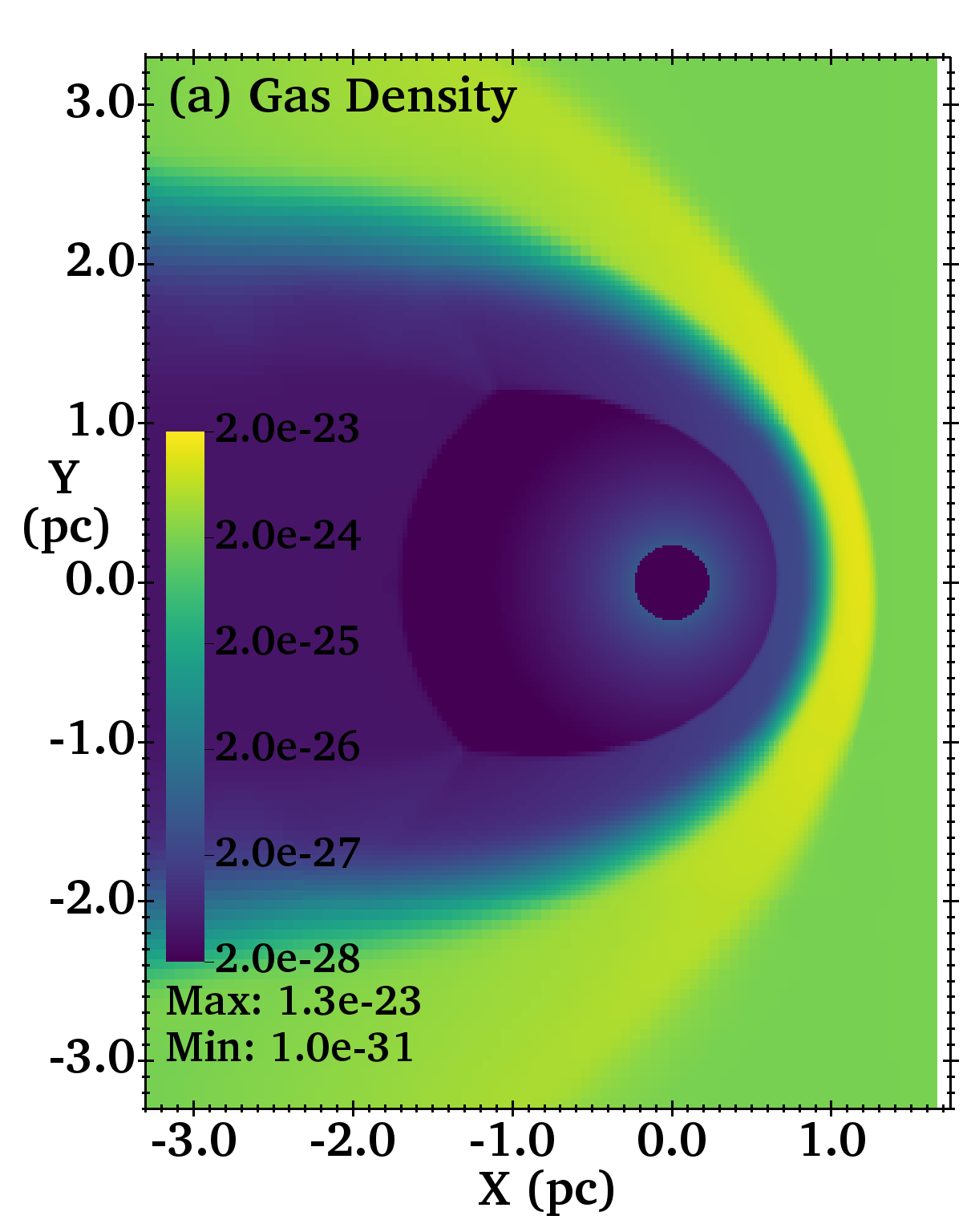}
\includegraphics[trim = 60mm 0mm 0mm 0mm, clip, height=0.33\textwidth]{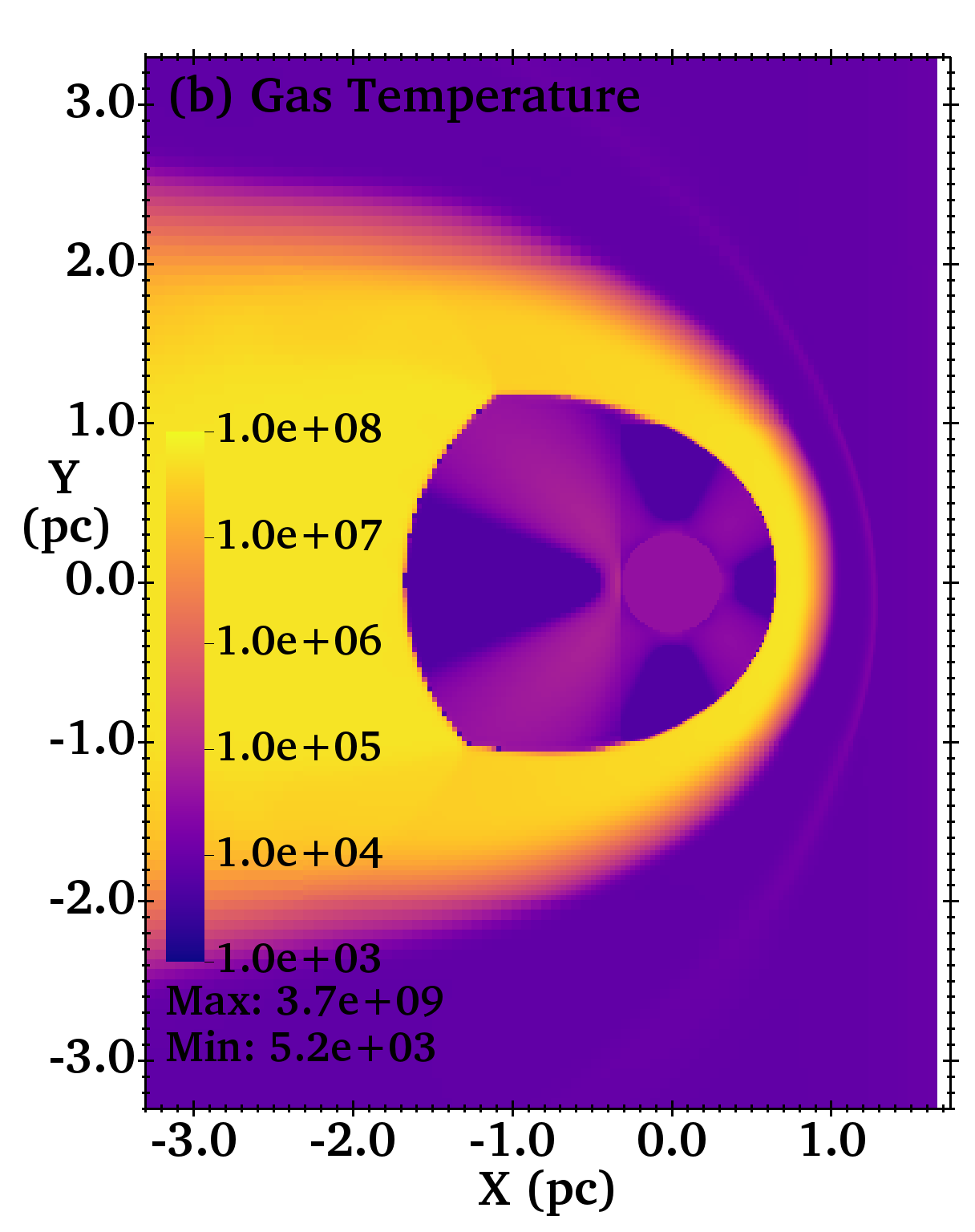}
\includegraphics[trim = 60mm 0mm 0mm 0mm, clip, height=0.33\textwidth]{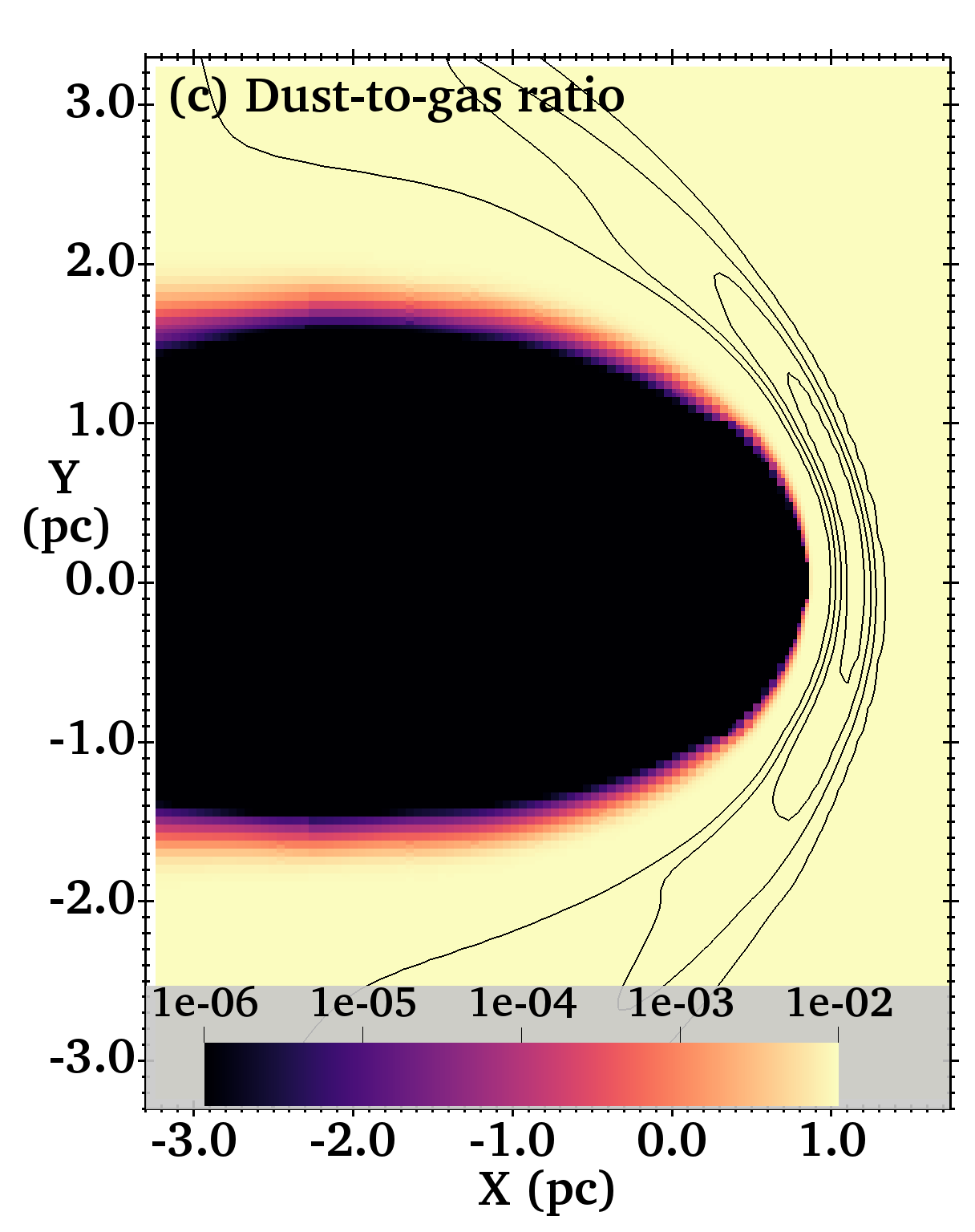}
\includegraphics[trim = 60mm 0mm 0mm 0mm, clip, height=0.33\textwidth]{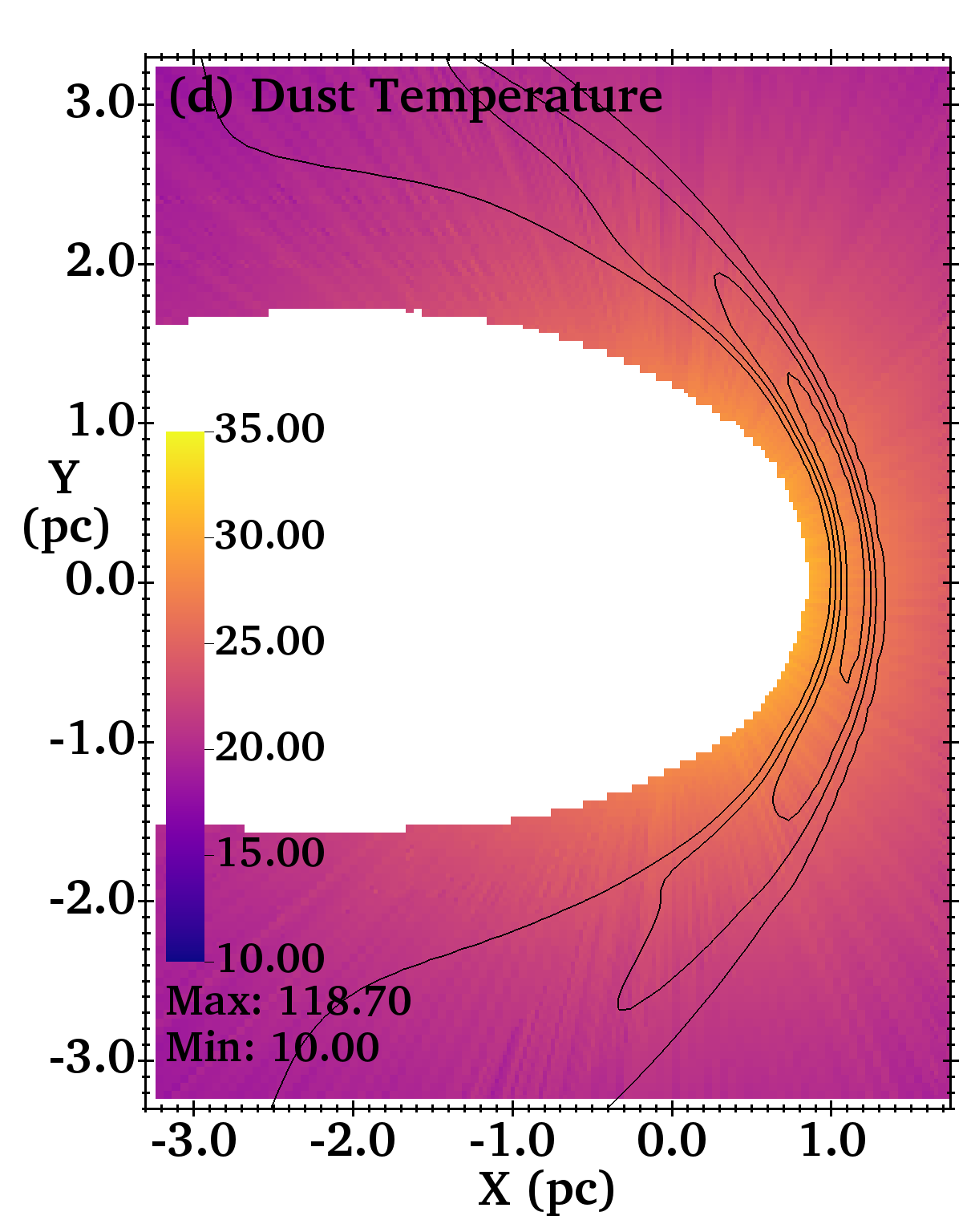}
\caption{
  2D slice through the 3D MHD bow-shock simulation from section~\ref{sec:wind3d} for the star with a 10\,G surface magnetic field showing (a) gas density in g\,cm$^{-3}$, (b) gas temperature in K, (c) dust-to-gas (mass) ratio and (d) dust temperature in K, all on logarithmic colour scales except for dust temperature which is on a linear scale.
  Dust temperature is calculated using \textsc{torus}, and regions with dust-to-gas ratio $<10^{-4}$ are masked for clarity.
  In panels (c) and (d) contours of gas density are plotted, linearly spaced from 0 to $10^{-23}$\,g\,cm$^{-3}$ in steps of $0.25\times10^{-23}$\,g\,cm$^{-3}$.
  }
\label{fig:ostar_dustgas}    
\end{figure*}

\subsection{Radiative transfer with \textsc{torus}}
\label{sec:torus}

We have implemented a method for postprocessing 3D nested-grid \textsc{pion} simulations with the \textsc{torus} Monte Carlo radiative transfer code \citep{HarHawAcr19}.
This builds on the previous implementation that postprocessed 2D uniform-grid \textsc{pion} simulations with \textsc{torus} to make synthetic dust continuum images \citep{MacHawGva16, GvaMacKni17, GreMacHaw19}.
An improvement compared with the method of \citet{GreMacHaw19} is that we now use a passive tracer to distinguish wind from ISM, and this is used to set the dust-to-gas ratio to zero in the wind and to 0.01 in the ISM, with a smoothly varying interface region where wind and ISM are mixed.
Previously we used a simple temperature cut to distinguish wind and ISM, which was effective but not as self-consistent as the new treatment.

\begin{figure*}
\centering
\includegraphics[width=0.33\textwidth]{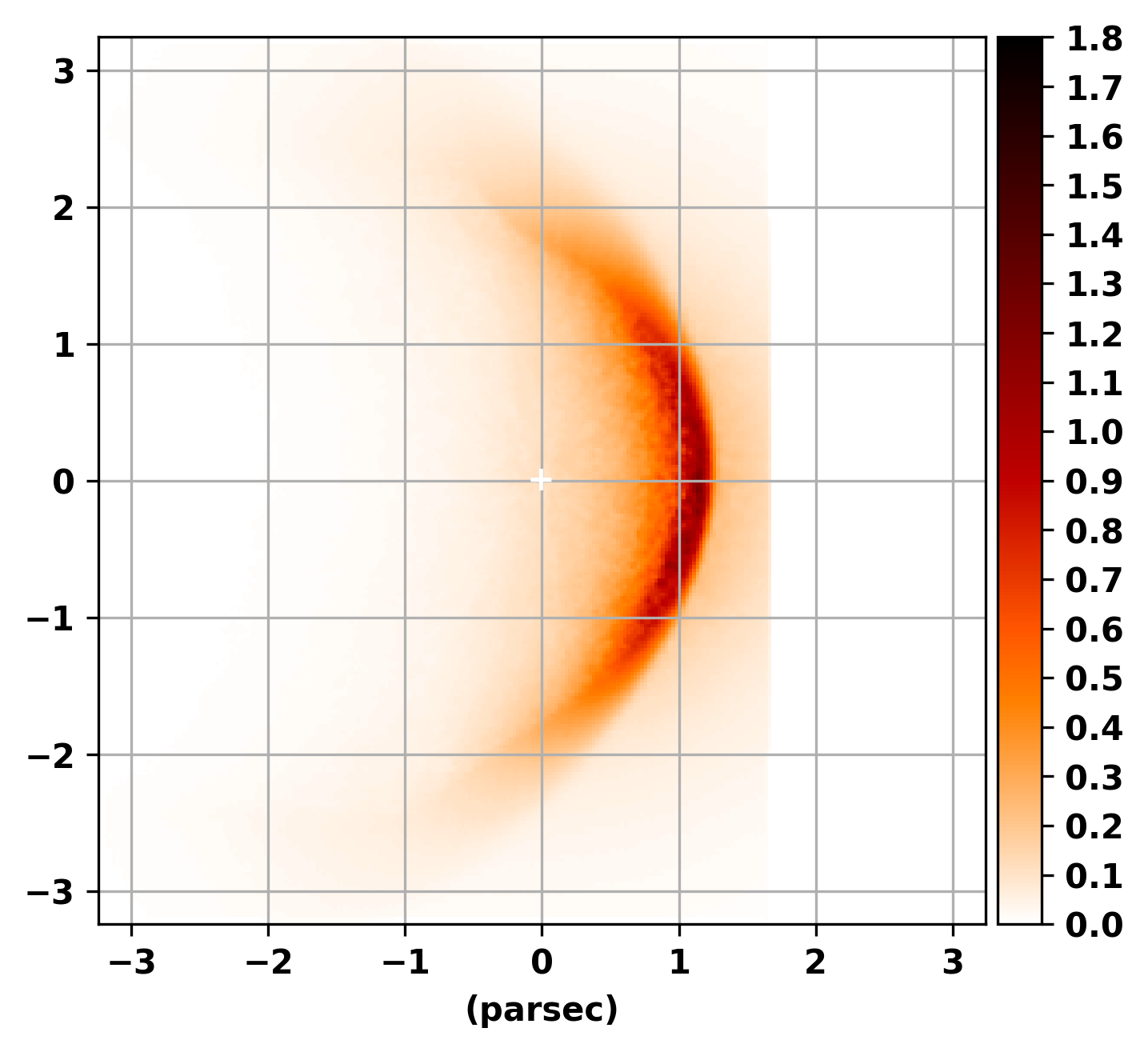}
\includegraphics[width=0.33\textwidth]{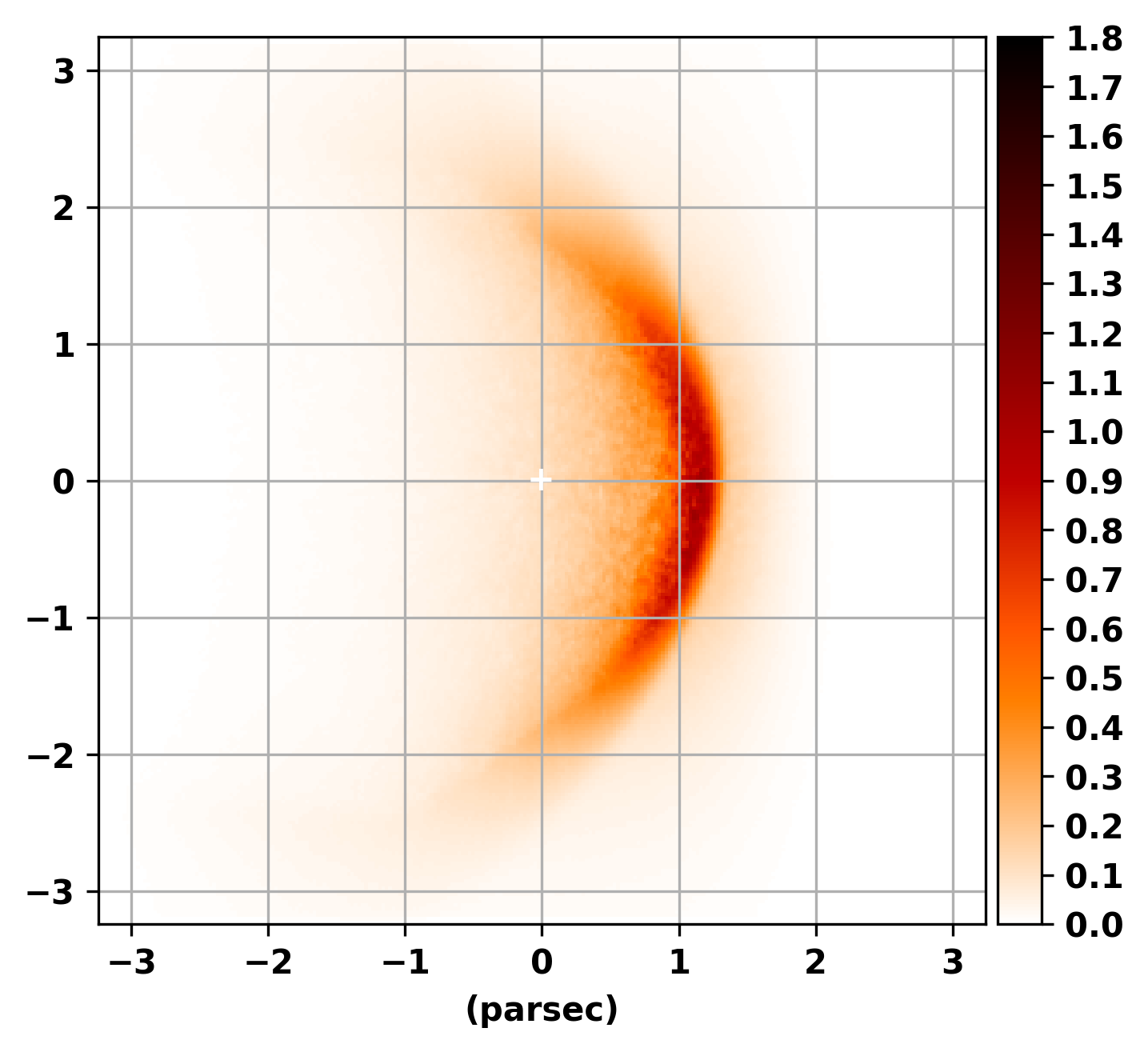}
\includegraphics[width=0.33\textwidth]{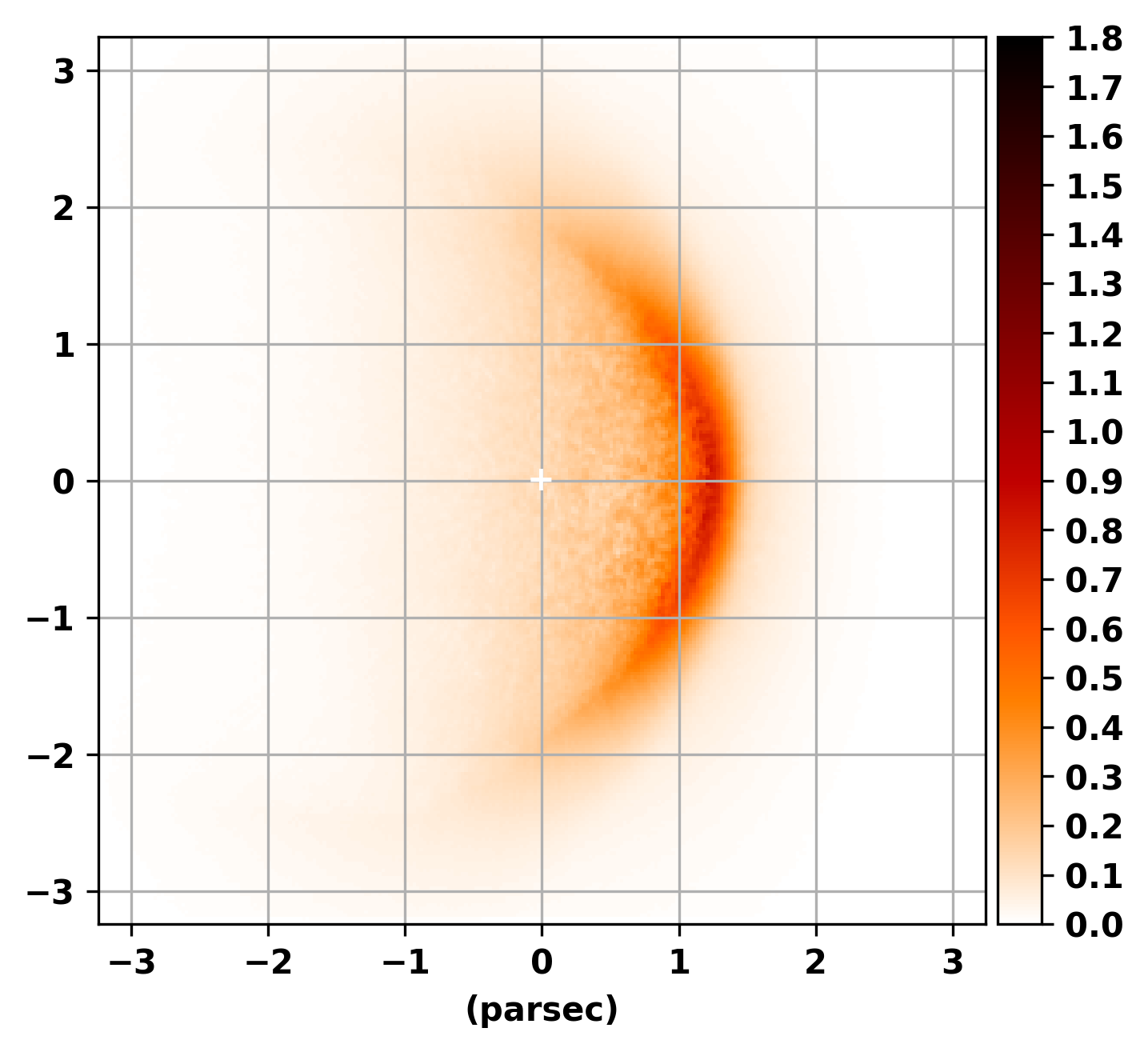}
\includegraphics[width=0.33\textwidth]{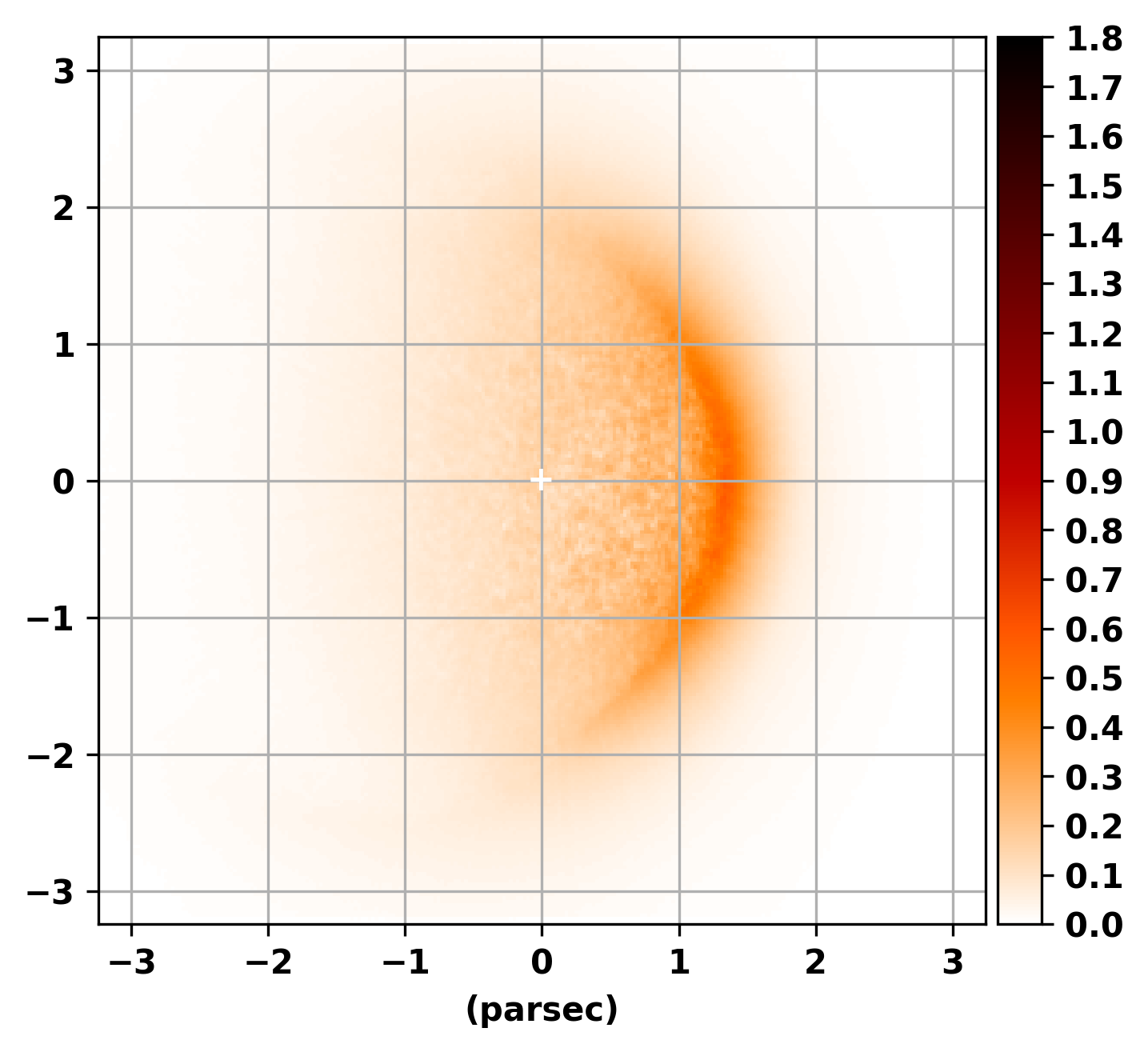}
\includegraphics[width=0.33\textwidth]{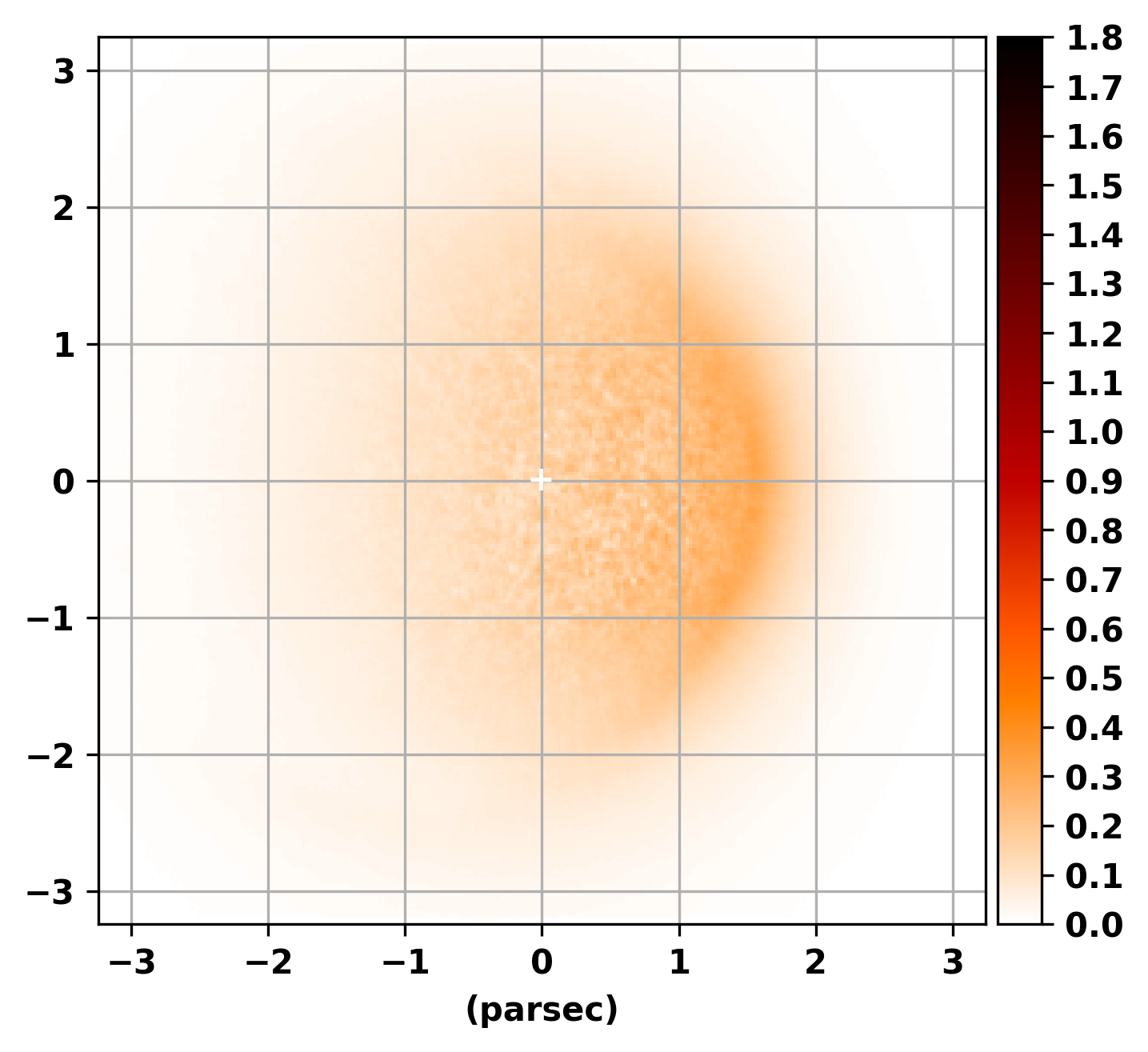}
\includegraphics[width=0.33\textwidth]{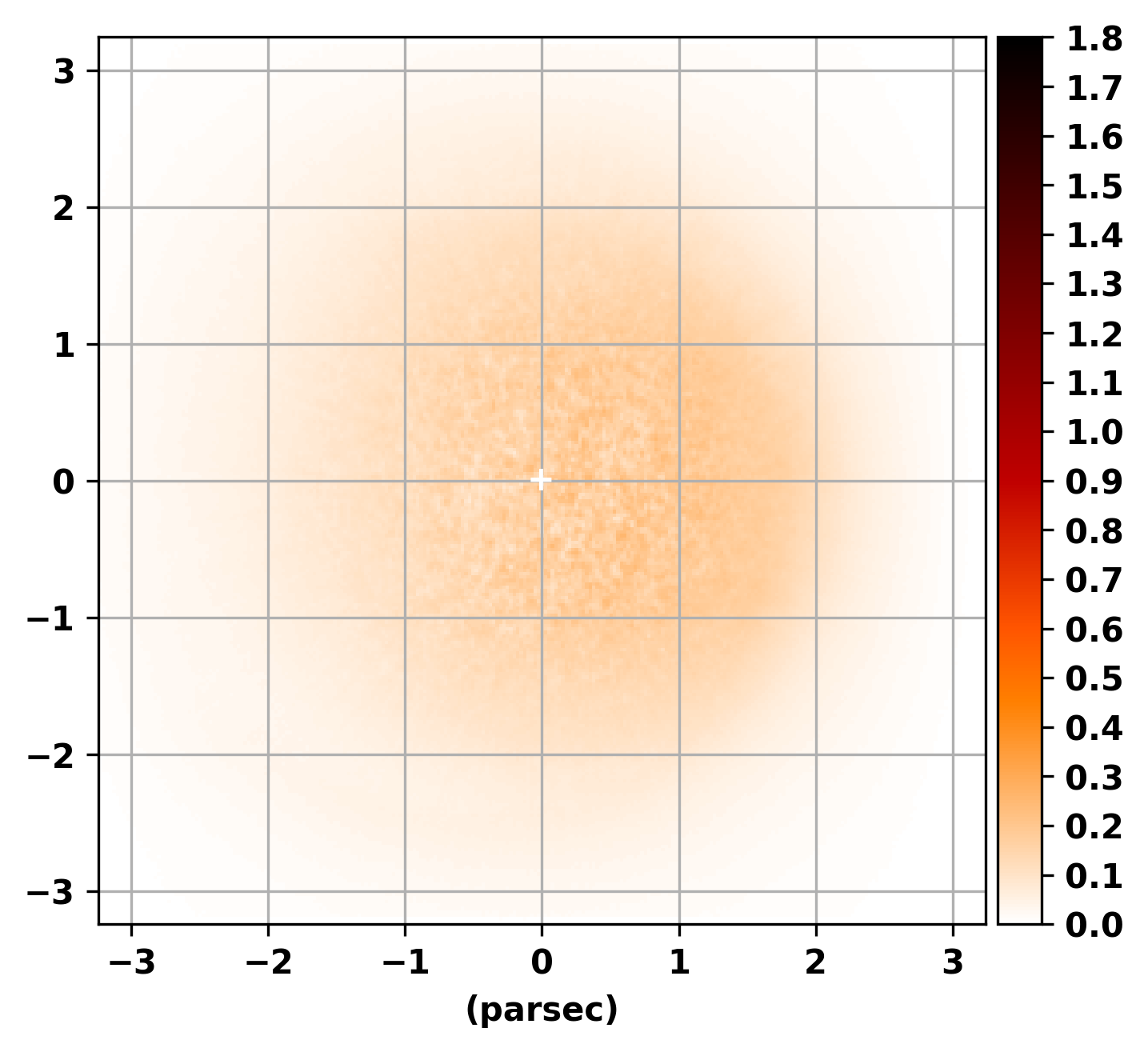}
\caption{
  Dust-emission map of the 3D MHD bow-shock simulation from section~\ref{sec:wind3d} for the star with a 10\,G surface magnetic field, calculated using \textsc{torus}.
  From left to right and top to bottom, the panels show projections with image normal vector at an angle of 90$^\circ$, 75$^\circ$, 60$^\circ$, 45$^\circ$, 30$^\circ$ and 15$^\circ$ 
  with respect to the positive $x$-axis.
  The 50\,$\mu$m intensity is plotted in MJy\,ster$^{-1}$ on a linear scale.
  The empty region upstream from the bow shock in the upper panels is outside the \textsc{pion} simulation domain.
  }
\label{fig:ostar_dust}    
\end{figure*}

\textsc{pion} simulation snapshots are converted to FITS format and read into \textsc{torus} using a C++ programme, \textsc{silo2fits}, provided with the \textsc{pion} source code.
The \textsc{torus} reader maps variables including the density, temperature and dust-to-gas mass ratio on to the \textsc{torus} grid using a bilinear (for 2D models) or trilinear (3D) interpolation.
In static mesh-refinemenet applications, each level of the \textsc{pion} grid is stored in a separate FITS file and these are read into \textsc{torus} sequentially.
Any given cell on the \textsc{torus} grid is populated by the highest resolution \textsc{pion} data available.
The resolution of the \textsc{torus} grid is flexible and can reflect the structure of the grid being read in, or adaptively refine according to its own AMR grid criteria (e.g.\ mass per cell and/or gradients in any quantity), or revert to a uniform mesh.

With the physical parameters of the \textsc{pion} grid read in, appropriate stellar parameters (temperature, radius, location) are added manually to the \textsc{torus} input file.
With the grid and photon sources set up, the full scope of \textsc{torus} functionality is then available \citep{HarHawAcr19}.
For example, this means that the thermally decoupled dust-temperature can be computed in a Monte Carlo radiative equilibrium calculation for arbitrary grain composition and size distribution.
For this test calculation we use silicate grains \citep{Dra03} with a \citet{MatRumNor77} size distribution from 0.01 to 10\,$\mu$m.

Fig.~\ref{fig:ostar_dustgas} shows the \textsc{pion} gas density and temperature in panels (a) and (b) compared with the dust-to-gas ratio (panel c) and the dust temperature, $T_\mathrm{D}$, (panel d) obtained from the \textsc{torus} radiative equilibrium calculation.
The dust density corresponds very closely to the gas density, except that inside the contact discontinuity the dust density decreases to zero in the inner part of the wind bubble, and so it is not shown.
The dust temperature is completely decoupled from the gas temperature because the collisional heating and cooling rates are negligible compared with radiative rates for these diffuse ISM conditions \citep[e.g.][]{MeyMacLan14}.
In the inner part of the bow shock $T_\mathrm{D}\approx30$\,K, whereas further out $T_\mathrm{D}$ decreases to $\approx20$\,K.

The resulting dust emission maps at wavelength $50\,\mu$m are shown in Fig.~\ref{fig:ostar_dust} from a range of viewing angles from edge-on to face-on.
The classic parsec-scale bow-shock morphology \citep[e.g.][]{PerBenBro12, PerBenIse15} is seen in the edge-on panels, where limb-brightening makes the shocked shell much brighter than is seen in the face-on panels.
The relatively low spatial resolution, combined with low-density ISM and small space velocity of the star, ensures that the bow shock is smooth with no apparent instability at either the contact discontinuity or forward shock.
There is a small asymmetry between the upper and lower half-plane, arising because the ISM magnetic field is not parallel to the star's motion, and so the shock compression factor and Mach number is not symmetric about $z=0$.

\section{Performance and Parallel Scaling}
\label{sec:performance}
A number of scaling tests have been performed to assess the performance of \textsc{pion} on HPC systems.
All calculations were run at the Irish Centre for High-End Computing (ICHEC) on the supercomputer \emph{Kay}\footnote{\href{https://www.ichec.ie/about/infrastructure/kay}{https://www.ichec.ie/about/infrastructure/kay}}, using the cluster nodes each consisting of $2\times$ 20-core 2.4 GHz Intel Xeon Gold 6148 (Skylake) processors with 192 GiB of RAM and a 100Gbit OmniPath network adaptor.

\begin{figure} 
\centering
\includegraphics[width=0.45\textwidth]{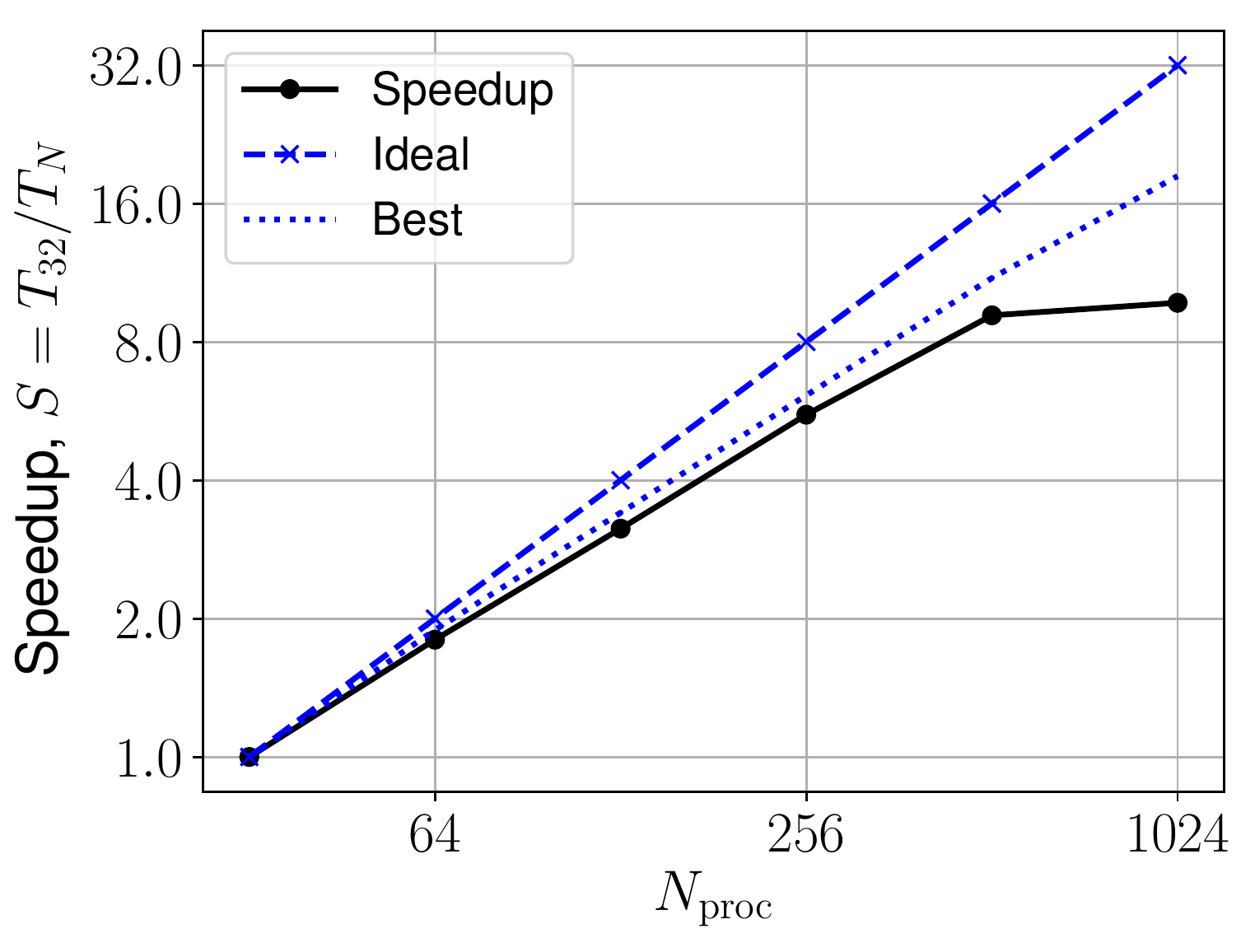}
\caption{
  Speedup of \textsc{pion} for a strong scaling test, running a 3D bow-shock simulation with 3 levels of refinement and $256^3$ grid cells per level, for 2048 timesteps.
  The solid black line shows the attained speedup, the dashed blue line is the ideal case, and the dotted blue line is the best possible speedup taking account of the extra ghost-cell calculations introduced by domain decomposition, but assuming communication overhead is zero.
  The data are from Table~\ref{tab:scaling}.
  }
\label{fig:scaling}    
\end{figure}

\subsection{Strong scaling for 3D MHD}

\begin{table}
\caption{
  Strong scaling of \textsc{pion} for simulation of a bow shock in 3D with 3 levels of refinement and $256^3$ grid cells per level, for 2048 timesteps.
  }
  \centering
  \vspace{2pt}
  \renewcommand{\arraystretch}{1.2}
  \begin{tabular}{| r | r  r | c c |}
    \hline
    $N_\mathrm{proc}$ & Walltime (s) & Core-Hours & Speedup & Efficiency \\
    \hline
    32 & 18\,176 & 161.6 & 1.0 & 1.00  \\
    64 & 10\,101 & 179.6 & 1.8 & 0.90  \\
    128 & 5\,791 & 205.9 & 3.14 & 0.78 \\
    256 & 3\,272 & 232.6 & 5.56 & 0.69 \\
    512 & 1\,989 & 282.9 & 9.14 & 0.57 \\
   1024 & 1\,868 & 531.3 & 9.73 & 0.30 \\
    \hline
  \end{tabular}
  \label{tab:scaling}
\end{table}

To test the strong scaling of \textsc{pion}, a 3D simulation of a bow shock with ideal MHD was run, with a resolution of $256^3$ grid cells on each level, and with 3 levels of refinement.
Note this calculation has no radiative transfer, so there are no long-range interactions and we expect the scaling to be good up to the point where the number of grid cells being communicated in boundary data is comparable with the the number of grid cells being calculated per MPI process.

The star has a mass-loss rate of $\dot{M}=1.74\times10^{-6}$\,M$_\odot$\,yr$^{-1}$, wind velocity $v_\infty=2500\,\mathrm{km}\,\mathrm{s}^{-1}$, and is placed at the origin of the simulation domain.
The star is moving through the ISM at $v_\star=30\,\mathrm{km}\,\mathrm{s}^{-1}$ in the $\hat{x}$ direction, modelled as a flow past the star (which is static on the computational domain), and the uniform ISM has number density is $n_0=100\,\mathrm{cm}^{-3}$.
The stellar magnetic field is taken to be a split monopole (radial field lines) with a surface field strength of $B=10\,\mu$G.
The interstellar magnetic field is oriented perpendicular to the star's space velocity, and has a strength $B_z=25\,\mu$G.
The setup is otherwise similar to the simulation in Section~\ref{sec:wind3d}, just with different star/wind and ISM parameters.

We assume that H and He are both singly ionized by the star's radiation field in the full domain, so radiative photoheating balances radiative cooling at a temperature $T\approx7500$\,K, following the same heating/cooling routines as \citet{GreMacHaw19}.
The simulation was evolved to $t=2.37\times10^{12}$\,s, about 25\% of the time required to reach a stationary state, and a snapshot was saved.
This was taken as a starting point for the scaling test, which consisted of continuing the simulation for 2048 timesteps with MPI process counts between $N_\mathrm{proc}=32$ (run on a single node) to $N_\mathrm{proc}=1024$ (on 26 nodes).
The results are shown in Table~\ref{tab:scaling} and plotted in Fig.~\ref{fig:scaling}, where the speedup, $S$, is defined as the wall-time to run the calculation on 32 MPI processes, $T_{32}$, divided by the wall-time to run on $N$ MPI processes, $T_N$.
The ``ideal'' case is perfect scaling, where doubling the number of MPI processes will decrease the run time by a factor of 2.
The ``best'' case takes account of the extra ghost-cells that must be calculated when the full domain is subdivided into more sub-domains, given that the boundary region is 4 cells thick, but assumes zero communication overhead.
For a $256^3$ grid this is $S=11.02$ for $N=512$, and the ideal value is $S=16$.
Compared with a simulation with 32 MPI processes, a calculation with 512 MPI processes is still 57\% efficient, and the code speeds up by a factor of 9.14 compared with a theoretical best attainable value of 11.02.

Although this strong scaling is good, and allows us to run 3D MHD simulations efficiently on hundreds of cores, we have not tested the scaling to $\gtrsim10^4$ cores.
The ratio of ghost cells to grid cells increases strongly as the number of MPI processes increases and the subdomains assigned to each MPI process get smaller.
From Fig.~\ref{fig:scaling} it is clear that there could be significant gain by switching to a hybrid parallelisation scheme, using multi-threading to reduce the number of MPI processes per node.
We plan to implement this for the next release of \textsc{pion}.

\begin{figure} 
\centering
\includegraphics[width=0.45\textwidth]{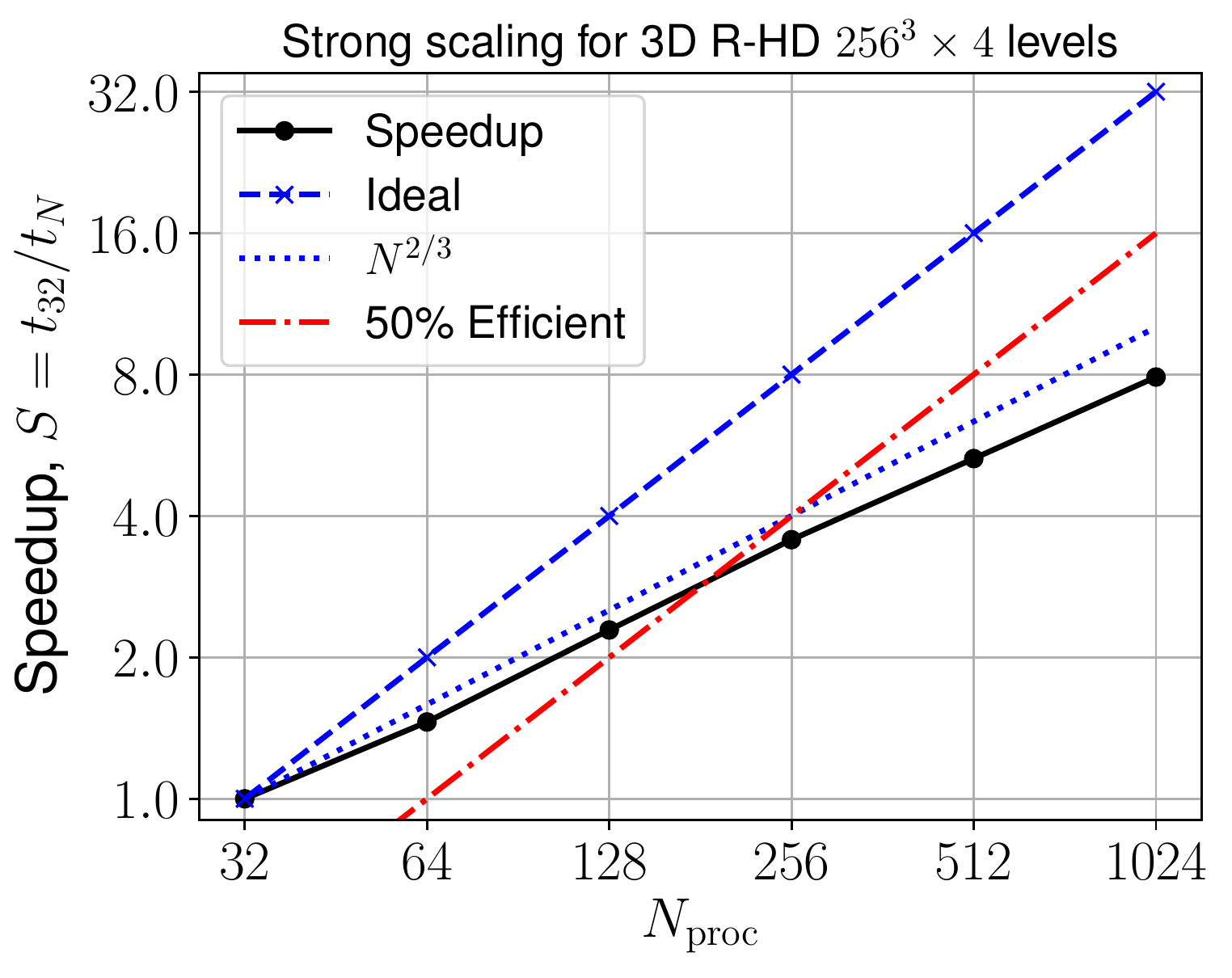}
\caption{
  Strong scaling of \textsc{pion} for 3D R-HD with 4 levels of refinement and $256^3$ grid cells per level, run for 1536 timesteps.
  Curves are as for Fig.~\ref{fig:scaling} except that the blue dotted line now shows a scaling of $N_\mathrm{proc}^{2/3}$, i.e., the expected performance of the 3D radiative transfer algorithm from uniform-grid calulations in \citet{Mac12}.
  }
\label{fig:rhd-strong-scaling}    
\end{figure}

\subsection{Strong scaling for 3D R-HD}

Radiative transfer scales less well than hydrodynamics because we work in the limit where the speed of light is infinite, and so there are long-range interactions between each cell and each radiation source that must be calculated by tracing rays from one sub-domain to the next.
The uniform-grid version of \textsc{pion} had strong-scaling speedup of $S\propto N_\mathrm{proc}^{0.5}$ for 2D calculations and $S\propto N_\mathrm{proc}^{2/3}$ for 3D calculations for the radiative transfer part of the calculation \citep{Mac12}.
For the 3D nested grid we consider an expanding WR nebulae from Section~\ref{sec:GS96}, taking an initial snapshot from $t=4.7638$\,Myr, about 10\,000 yr after the stellar transition from RSG to WR.
The simulation domain has 4 levels of refinement centred on the stellar source at the origin, with $256^3$ grid cells on each level.

The simulation was run for 1536 timesteps (on the finest level), corresponding to about 2400 yr of evolution, and the speedup is plotted in Fig.~\ref{fig:rhd-strong-scaling}.
The scaling is significantly worse for R-HD than for the MHD simulations without radiative transfer and non-equilibrium ionization.
Increasing the core-count by a factor of 8 already reduces the efficiency by more than 50 per cent, and running this simulation on 1024 cores requires $4\times$ more core-hours than running on 32 cores.
The slope of the scaling plot is approximately constant except for the jump from 32 to 64 cores, corresponding to switching from calculating on a single node to multiple nodes with slower communication.
Similar difficulties obtaining good scaling for R-HD have been reported previously \citep[e.g.][]{WisAbe11}, although some innovative algorithms have improved scaling to larger numbers of cores \citep{RosKruOis17}.

\begin{figure} 
\centering
\includegraphics[width=0.45\textwidth]{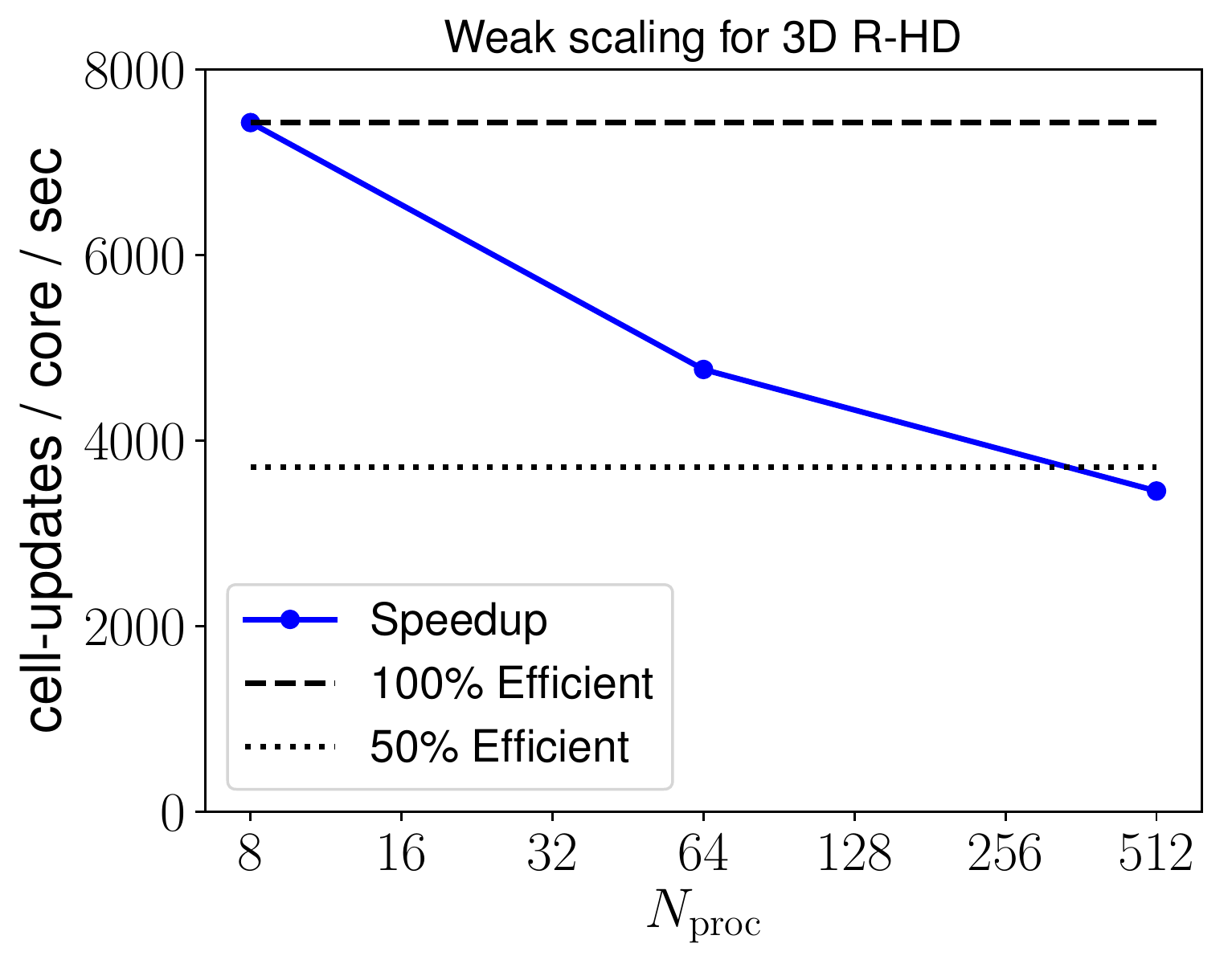}
\caption{
  Weak scaling of \textsc{pion} for 3D radiation-hydrodynamics with 4 levels of refinement and $64^3$ grid cells per MPI process, run for $\approx4.5$ hours walltime.
  }
\label{fig:rhd-weak-scaling}    
\end{figure}

\subsection{Weak scaling for 3D R-HD}

Here we consider the expansion of an H\,\textsc{ii} region and wind bubble from a massive star into a uniform and static ISM.
The medium is dense and the wind is strong, and so the ionization front is trapped by the forward shock driven by the expanding wind bubble.
We take the result from a 1D calculation with PION and map it on to a 3D grid with 4 refinement levels each with $128^3$, $256^3$ or $512^3$ cells.
The weak scaling is tested by running calculations where the number of grid cells per core is constant, so the $128^3$ simulation is run on 8 cores, the $256^3$ on 64 cores, and the $512^3$ on 512 cores.
Each core therefore computes a subdomain of $64^3$ cells in all three cases.

The results are plotted in Fig.~\ref{fig:rhd-weak-scaling}, where we show the number of cell-updates per core per second for simulations with 8, 64 and 512 MPI processes.
The efficiency of the code decreases by about 50 per cent when the number of MPI processes increases by a factor of 64.
The overall performance of \textsc{pion} for this calculation is not optimal, running at nearly $20\times$ slower than simple MHD without any radiative transfer or non-equilibrium-ionization calculation.
There is certainly scope for improving both the overall performance and the parallel scaling of this algorithm, and the size of simulation that can be run is currently limited by the parallel scaling.

\section{Conclusions}
\label{sec:conclusions}

We have presented upgrades to the simulation framework \textsc{pion} for astrophysical fluid dynamics, including the first public release of the source code and associated scripts and postprocessing routines.
The major upgrades are the implementation of static mesh-refinement, the improved robustness of the MHD solver (including improved divergence cleaning), the implementation of the consistent multi-species advection (sCMA method) for advecting elements across the domain, and the addition of latitude-dependent and magnetized winds from rotating stars following \citet{LanGarMac99} and \citet{PogZanOgi04}.

Test calculations showing advection and the expansion of blast waves and ionization fronts across refinement boundaries have been presented.
The advection of a magnetic field loop shows no noticeable artefacts associated with the refinement boundaries.
Blast-wave expansion also works very well for hydrodynamics and for weak magnetic fields, but an artefact appears once the magnetic field becomes dynamically important.
An imprint of the refinement boundary is apparent in the expanding shocked medium, also when run at first-order accuracy and with different methods for divergence cleaning.
Comparing with \textsc{Athena++} \citep{StoTomWhi20} when run on the same problem, the results are very similar, even though \textsc{Athena++} uses a different integration scheme and a different algorithm for dealing with magnetic-field divergence errors.
Looking at the D-type expansion of an ionization front, \textsc{pion} produces results with static mesh-refinement that are at least as accurate as uniform-grid simulations with equivalent resolution, and with a fraction of the computational cost.

Results were presented for a 3D MHD bow shock produced by the wind of a rotating and magnetized O star moving with 30\,km\,s$^{-1}$ through a uniform ISM, a preliminary version of which was presented in \citet{MacGreMou20}.
The classic features of a \citet{Par58} wind were demonstrated: the Parker spiral, equatorial enhancement of the toroidal magnetic field and the equatorial current sheet.
It was shown that for a reasonably strong surface split-monopole magnetic field of 100\,G, the magnetic field strength in the shocked wind bubble can be comparable to that in the shocked ISM.
For bow shocks where synchrotron radiation can be detected \citep[e.g.~BD+43\,3654][]{BenRomMar10}, it may be possible to constrain this magnetic field strength observationally, giving a direct constraint on the stellar magnetic field.

We revisited the calculation of \citet{ChiLanVan08} of the ring nebula produced when a rotating RSG evolves on a blue loop and reaches critical rotation.
Using the latitude-dependent wind prescription of \citet{LanGarMac99} we showed that \textsc{pion} produces results with a 2D nested grid in cylindrical coordinates that are comparable to the 2D spherical-grid computations of \citet{ChiLanVan08}.
We largely confirm their results, although there are some small differences in the details of the hydrodynamic flow.
We also demonstrated how an MHD simulation of such a nebula can be calculated by making a simple assumption about the surface magnetic field strength.
A more realistic calculation would estimate the stellar magnetic field strength from the properties of the stellar envelope.
The code is efficient enough that 3D simulations of ring nebulae are feasible with reasonable computational resources.

Comparing \textsc{pion} with another classic calculation of a circumstellar nebula, we used a 3D nested-grid to simulate the expansion of a spherically symmetric fast wind from a WR star into the slow wind from its previous RSG phase of evolution.
We used the same stellar evolution calculation as the 2D simulations by \citet{GarLanMac96} and \citet{FreHenYor06}, and our 3D simulation has comparable spatial resolution to the previous 2D work.
Again our results are comparable to previous work, although there are some differences in the details.
We showed that the WR wind bubble expands at almost constant speed as predicted by \citet{KooMcK92} for a wind expanding into a $r^{-2}$ density profile.
The symmetry of the winds means that instability is seeded by the integration errors associated with the grid discretisation, and so the solution is artificially symmetric compared with a real nebula.
Implementation of some random or clumpy component to the wind boundary condition would break this symmetry and produce more realistic nebulae.
We plan higher-resolution simulations of this wind-wind interaction for comparison with observations of WR nebulae.

Finally we revisited the 2D simulation by \citet{SteBloPol92} of the wind-wind collision in the WR+O-star binary system V444 Cyg, using 3D MHD simulations including moderate stellar rotation.
We find very similar results for the hydrodynamics of this system, which is marginally in the regime where strong cooling is expected to produce thin-shell instabilities and strong distortions of the bow shock structure.
This proof-of-concept calculation requires a number of enhancements in order to reach state-of-the art, especially the inclusion of orbital motion and a better implementation of radiative cooling for hydrogen-poor plasmas.
It is nevertheless encouraging that one can obtain good results with modest computational resources, and we intend to develop this setup significantly in future work.

The parallel scaling of \textsc{pion} is shown to be very good for MHD calculations without radiative transfer, as long as each MPI process has a local subdomain of $\gtrsim32^3$ grid cells per level.
Beyond this, the ratio of boundary cells to grid cells becomes sufficiently large that the computation and communication overhead is prohibitive.
We anticipate that this scaling can be further improved significantly by implementation of hybrid MPI+OpenMP parallelisation, because of reduced communication overhead and fewer boundary cells with duplicated computation.
Scaling for 3D R-HD simulations is less good, losing $2\times$ in efficiency when the number of MPI processes increases by $8\times$, and significant work is needed to get this running efficiently on large supercomputers.

We introduced the \textsc{PyPion} library of python routines for reading \textsc{pion} snapshots and making various plots of the gas properties as a function of position.
We demonstrate a method for converting \textsc{pion} snapshots to \textsc{fits} images that can be read by the \textsc{torus} Monte Carlo radiative transfer code \citep{HarHawAcr19} and postprocessed to calculate thermal dust emission maps.
This method can also be extended to enable plotting of emission maps from spectral lines, thermal X-rays, or any form of radiation where the emissivity is a simple function of density and temperature.

It is hoped that \textsc{pion} will be a useful tool for the community to model nebulae around evolving massive stars. 
The source code can be obtained via a git repository from \href{https://www.pion.ie/}{https://www.pion.ie/}, and contributed code can be added using a mirrored repository on \emph{GitHub}.
A mailing list is also available for user support and discussion at \href{https://groups.io/g/pion}{https://groups.io/g/pion}.
The methods developed here will be used in future work to study bow shocks and wind-blown nebulae around massive stars with 3D simulations.
Comparing synthetic and real observations will allow us to test the predictions of stellar evolution calculations, to learn more about stellar mass loss, magnetism, and particle acceleration.

\section*{Acknowledgements}

JM is grateful to N.~Langer for advice and discussions on circumstellar medium modelling, and for providing evolutionary calculations used in sections~\ref{sec:chita} and \ref{sec:GS96},
to Jim Stone and Robin Williams for advice and suggestions on mesh refinement, and to Luca Grassitelli for useful discussions on winds from evolved massive stars.
JM acknowledges funding from a Royal Society-Science Foundation Ireland University Research Fellowship (14/RS-URF/3219).
DZ acknowledges funding from an Irish Research Council (IRC) Starting Laureate Award (IRCLA\textbackslash 2017\textbackslash 83).
SG is funded by a Hamilton Scholarship from the Dublin Institute for Advanced Studies.
MM acknowledges funding from a Royal Society Research Fellows Enhancement Award (RGF\textbackslash EA\textbackslash 180214).
TJH is funded by a Royal Society Dorothy Hodgkin Fellowship. 
The authors wish to acknowledge the DJEI/DES/SFI/HEA Irish Centre for High-End Computing (ICHEC) for the provision of computational facilities and support (projects dsast022c, dsast023b).
This research has made use of NASA's Astrophysics Data System Bibliographic Services, the GNU Scientific Library \citep{GSL} and, for visualisation and analysis, VisIt \citep{VisIt}, Numpy \citep{HarMilVan20}, and matplotlib \citep{Hun07}.
This research made use of Astropy, a community-developed core Python package for Astronomy \citep{astropy:2013, astropy:2018}.

\section*{Data Availability Statement}

Instructions for downloading source code for \textsc{pion} and \texttt{PyPion} can be found at \href{https://www.pion.ie/}{https://www.pion.ie/}.
Source code for \textsc{torus} can be obtained from \href{https://bitbucket.org/tjharries/torus/}{https://bitbucket.org/tjharries/torus/}.
The data underlying this article will be shared on reasonable request to the corresponding author.

\bibliographystyle{mnras}
\bibliography{./refs}

\bsp	
\label{lastpage}
\end{document}